\documentclass[a4paper]{aa}  
\usepackage{longtable,lscape}
\usepackage{graphicx}
\usepackage{txfonts}
\usepackage{natbib}
\begin{document}
\title{Results from DROXO. }
\subtitle{III. Observation, source list, and X-ray properties of sources detected in 
the ``Deep Rho Ophiuchi XMM-Newton Observation''.}

\author{I. Pillitteri\inst{1,2,9}
\and S. Sciortino\inst{2}
\and E. Flaccomio\inst{2}
\and B. Stelzer\inst{2}
\and G. Micela\inst{2}
\and F. Damiani\inst{2}
\and L. Testi\inst{3}
\and T. Montmerle\inst{4}
\and N. Grosso\inst{5,6} 
\and F. Favata\inst{7}
\and G. Giardino\inst{8}
}

\offprints{I. Pillitteri}
\institute{DSFA, Universit\`a degli Studi di Palermo, Piazza del Parlamento 1, 90134, Palermo, Italy
\\\email{pilli@astropa.unipa.it}
\and
INAF-- Osservatorio Astronomico di Palermo, Piazza del Parlamento 1, 90134, Palermo, Italy
\and
ESO-- Karl-Scharzschild-Strasse 2, D-85748 Garching~bei~M{\"u}nchen, Germany.
\and 
Laboratoire d'Astrophysique de Grenoble Universit\`e Joseph-Fourier, Grenoble, France
\and
Universit\'e de Strasbourg, Observatoire Astronomique de
Strasbourg, 11 rue de l'universit\'e, 67000 Strasbourg, France
\and
CNRS, UMR 7550, 11 rue de l'universit\'e, 67000 Strasbourg, France
\and
ESA – Planning and Community Coordination Office, Science Programme, Paris, France
\and
Astrophysics Division – RSSD ESA, ESTEC, Noordwijk, The Netherlands
\and
SAO-Harvard Center for Astrophysics, Cambridge MA, USA
}

   \date{Received; accepted }

 \abstract{X-rays from very young stars are powerful probes to investigate 
the mechanisms at work in the very first stages of the star formation and the 
origin of X-ray emission in very young stars.}
{  We present results from a 500 ks long observation of the Rho Ophiuchi 
cloud with a { XMM-Newton} large program named DROXO, 
aiming at studying the X-ray emission of deeply 
embedded young stellar objects (YSOs).}
{The data acquired during the DROXO program were reduced with SAS
software, and filtered in time and energy to improve the signal to noise of detected sources;
light curves and spectra were obtained.}
{ 
We detected $111$ sources, 61 of them associated with $\rho$\,Ophiuchi YSOs as
identified from infrared observations with ISOCAM.  Specifically, we detected
9 out of 11 Class\,I objects, 31 out of 48 Class\,II and 15 out 16 Class\,III objects. 
Six objects out of 21 classified Class\,III candidates are also detected. At the same time 
we suggest that $15$ Class\,III candidates that remain undetected at 
$\log{L_{\rm x}}\,{\rm [erg/s]} < 28.3$ are not related to the cloud population.
The global detection rate is $\sim64\%$. 
We have achieved a flux sensitivity of $\sim 5\cdot 10^{-15}$ erg s$^{-1}$ cm$^{-2}$. 
The $L_\mathrm X$ to $L_\mathrm{bol}$ ratio shows saturation at a value of $\sim-3.5$  
for stars with T$_\mathrm{eff} \le 5000$ K  or 0.7 M$\odot$ as observed in the Orion Nebula. 
The plasma temperatures and the spectrum absorption show a decline with YSO class, 
with Class I YSOs being hotter and more absorbed than Class II and III YSOs. 
In one star (GY~266) with infrared counterpart in 2MASS and Spitzer catalogs 
we have detected a soft excess in the
X-ray spectrum, which is best fitted by a cold thermal component less absorbed than the main 
thermal component of the plasma. This soft component hints at plasma heated by shocks 
due to jets outside the dense circumstellar material.}
{} 

   \maketitle
%

\section{Introduction}
{ Low-mass stars in the pre-Main Sequence (PMS) phase are characterized by intense X-ray emission.
X-ray activity has not been yet firmly confirmed in the very initial protostellar 
cores that accrete material from a surrounding envelope (Class 0 objects), 
{  because X-rays are either} absent or completely absorbed by the dense circumstellar material. 
In Class I, II, and III {  objects} and
until the Zero Age Main Sequence stage the X-ray emission is very strong 
when compared to solar-age stars and also characterized by impulsive variability \citep{Feigelson99}.  
In the case of Class III objects all X-rays are thought to originate from a magnetized stellar 
corona,  which likely resembles a scaled-up version of the corona of late-type Main Sequence 
stars and the Sun.
{  Additional X-ray production mechanisms are possibly at work in} 
Class I and II objects. {  At these young evolutionary stages} 
the magnetic field drives the material in-falling from the disk to the stellar surface {  where the
matter becomes shocked}.  
Furthermore, outflows along the disk axis may interact with the circumstellar 
material and form shocks. {  The plasma in the shocks in both in- and outflows may become heated} 
up to a few million degrees, thus emitting X-rays \citep{Pravdo01,Favata02,Kastner05, Giardino07, Sacco08}.

Significant effort has been devoted to the X-ray observation of star-forming regions, 
which provide natural laboratories where to study the X-ray emission from young stars and its 
implications for the mechanisms of star formation.
The $\rho$ Ophiuchi cloud is among the nearest star-forming regions (120 pc, \citealp{Loinard08})
and has been extensively studied from the infrared (IR) to the X-rays bands. 
The $\rho$ Ophiuchi cloud is shaped as a dense multicore structure,
hosting more than 200 members comprising young stellar objects (YSOs) in all evolutionary 
stages from Class 0 to Class III \citep{Wilking08}. 
While a dispersed population of optically visible stars associated with the 
Upper Scorpius OB association with an age of $\sim 5$ Myr  is present in the region, 
the studies in the IR band have revealed a younger population  of Pre Main Sequence (PMS) stars
and protostellar objects with ages of only 0.3--1 Myr \citep{LR99}, 
which are thus younger than YSOs in other star-forming regions like the Taurus Molecular Cloud 
($\sim 1-5$ Myr) and the Orion Nebula ($\sim 1$ Myr, \citealp{Hillenbrand97}). 
{  The mid-IR survey with ISOCAM on board the ISO satellite} reported by 
\citet{Bontemps01} (hereafter Bo01) allowed the detection of a population of 16 Class I, 
123 Class II and 77 Class III YSOs, adding 71 previously unknown members of Classes I and II. 
X-ray studies carried out with the {\em Einstein} and 
{\em ROSAT} satellites had revealed several tens of embedded Class I, II,
 and III stars that are highly active in X-rays and are characterized by a very strong time 
variability \citep{Montmerle83,Casanova95}.
\citet{Casanova95} tentatively identified several X-ray sources with Class I protostars with
ROSAT/PSPC observations. However, X-ray observations with the
ROSAT/High-Resolution Imager were needed to confirm X-ray emission
from Class I protostars \citep{Grosso97,Grosso01}.
Recent observations with the XMM-Newton \citep{Ozawa05} and Chandra satellites 
\citep{Imanishi01,Flaccomio03} have increased the number of candidate PMS members of the region.
These studies suggested that accreting YSOs have increasing X-ray activity going from
Class~I to Class~III and decreasing plasma temperatures.
The aim of the {\it Deep Rho Ophiuchi XMM-Newton Observation} (DROXO) was to obtain 
a high-sensitivity survey in the Core F of $\rho$~Ophiuchi Cloud by means of a long, 
almost continuous observation. 
We report the data analysis and the X-ray properties of the PMS stars observed
during the DROXO program. The structure of the paper is as follows: 
Sects. 2 and 3  describe the observation and the data analysis, 
Sects. 4 and 5 describe the sensitivity of the survey and the nature of the sources
detected in DROXO. In Sect. 6 we discuss the X-ray emission of the sample of classified
YSOs, in particular we explore the behavior of plasma temperatures, 
absorption, X-ray luminosity and ratio L$_\mathrm X/$L$_\mathrm{bol}$ with respect to
effective temperatures, masses and evolutionary stages.
In Sect. 7 we give a summary. 
Appendix A and B contain tables with the list of detected sources, 
the best-fit parameters of spectra, the upper limits to rate for undetected YSOs in the
ISOCAM sample, the Spitzer counterparts to DROXO sources, and an example of one 
page of the atlas (published electronically only) with the EPIC spectrum and light curve
of each source.  
}

\section{The observation}
The {\em Deep Rho Ophiuchi XMM Observation} is a large program carried with the XMM-Newton 
satellite { pointed toward} core F in the $\rho$ Ophiuchi Cloud for almost eight consecutive days. 
The nominal pointing was at R.A. = 16h27m19.5s and Dec. = $-$24d41m40.9s (J2000), 
and the net exposure time was $\sim$515 ks. 
The pointing approximately coincides with that of the 33~ks XMM-Newton 
exposure studied by \citet{Ozawa05}.
The observation has been carried out in five subsequent satellite orbits 
(0961 $\rightarrow$ 0965), keeping the same position angle in all orbits. 

During the first orbit chip nr. 6 of the MOS 1 camera was damaged apparently by a 
micrometeorite impact \footnote{\em http://xmm.vilspa.esa.es/external/xmm\_news/items/MOS1-CCD6/index.shtml}
and has stopped functioning, so that  only a $\sim$ 28.5 ks exposure is 
available by that chip in DROXO. No damage was registered on the other chips of MOS~1.
During orbit nr. 0964 the $\rho$ Ophiuchi exposure was split in two segments, separated
by $\sim 25$ks, due to a triggered TOO observation. 
During the first three orbits the drift of the satellite was larger
than 6$\arcsec$, and this influenced the computation of exposure maps and 
the subsequent analysis of summed data as explained below.

\begin{figure}
\begin{center}
\includegraphics[width=\columnwidth]{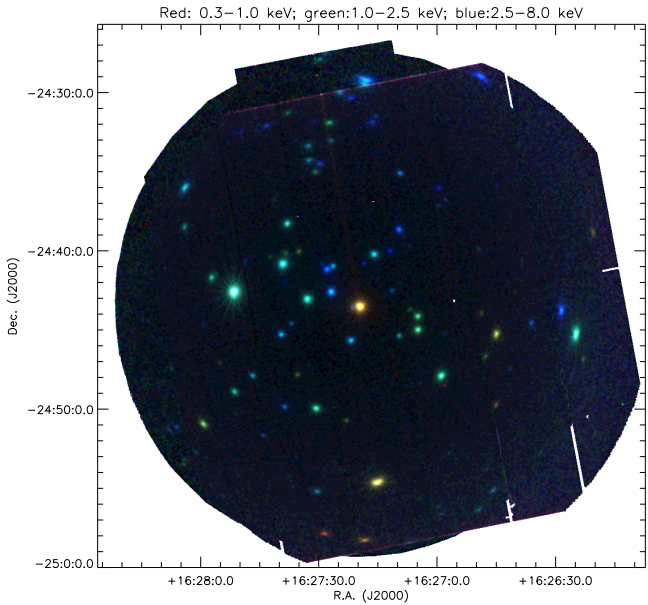}
\caption{\label{img3col} 
{   Merged EPIC image of events recorded in the time-filtered data 
to enhance the signal-to-noise of faint sources. 
Colors encode the following bands: 0.3-1.0~keV (red), 1.0-2.5~keV (green),
2.5-8.0~keV (blue), respectively. The MOS 1, 2, and PN images are normalized by 
effective area and exposure time to reduce instrumental artifacts like CCD gaps.
}
}
\end{center}
\end{figure}

{  In Fig. \ref{img3col} we show the EPIC MOS1, MOS2 and PN images added 
in a pseudo-color Red-Green-Blue image. 
The bands chosen for red, green and blue components are: 0.3-1.0 keV,
1.0-2.5 keV, 2.5-8.0 keV. Sources with a softer/less absorbed spectrum are redder than sources 
with a hard/highly absorbed spectrum. Each CCD image has been divided by the proper 
exposure map and by the average effective area in the energy band
to normalize the different efficiencies of the three cameras.}
 
\section{Data analysis}
The observation data files (ODF) were processed with the SAS software\footnote{See {\em 
http://xmm.vilspa.esa.es/sas}} (version 6.5) to produce full field-of-view 
event lists calibrated in both energy and position.
We subsequently filtered these event files and retained only the events with energy
in the 0.3--10 keV band and those that triggered simultaneously at most
two nearby pixels. This step was executed for each detector (MOS 1, MOS 2, PN) 
and for each of the exposure segments of the five orbits (hereafter step 1). 
The large satellite drift is not taken into account automatically by the SAS software
which results in wrong exposure maps when they are created with the default values. We needed
to reduce the {\sc ATTREBIN} parameter of task {\sc EEXPMAP} to 0.5 $\arcsec$ and to choose
the most accurate algorithm (of which the default was the fastest and less precise one) 
to obtain correctly computed exposure maps.
\subsection{\label{pwxdet} Source detection}
We performed source detection with the PWXDETECT code
developed at INAF-Osservatorio Astronomico di Palermo \citep{Dami97.1,Dami97.2}. 
The code allows the detection of sources starting from unbinned photon positions 
recorded in several datasets from MOS 1, MOS 2 and PN cameras, through a multiscale 
analysis of mexican-hat wavelet convolved images. 
{ All data were properly scaled by time and effective area of each CCD
detector, obtaining a flux image of the sum of all exposures. At the end of the 
process the count rates of detected sources were re-scaled to a reference instrument which, 
in our case, was the EPIC MOS 1. We refer to the count rates derived in this way as 
{\em MOS1 equivalent count rates}. }
\begin{figure}
\begin{center}
\includegraphics[width=\columnwidth]{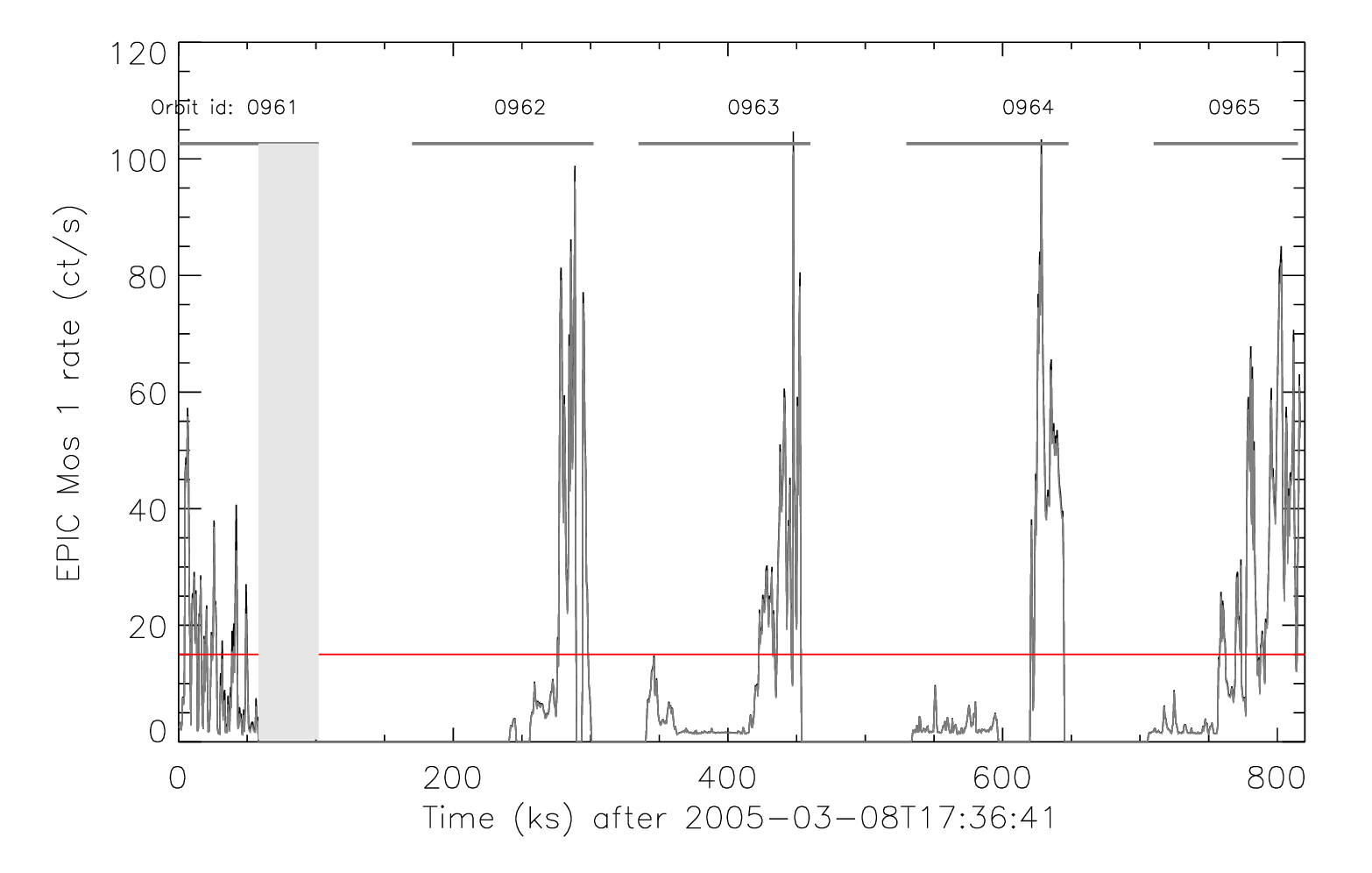}
\includegraphics[width=\columnwidth]{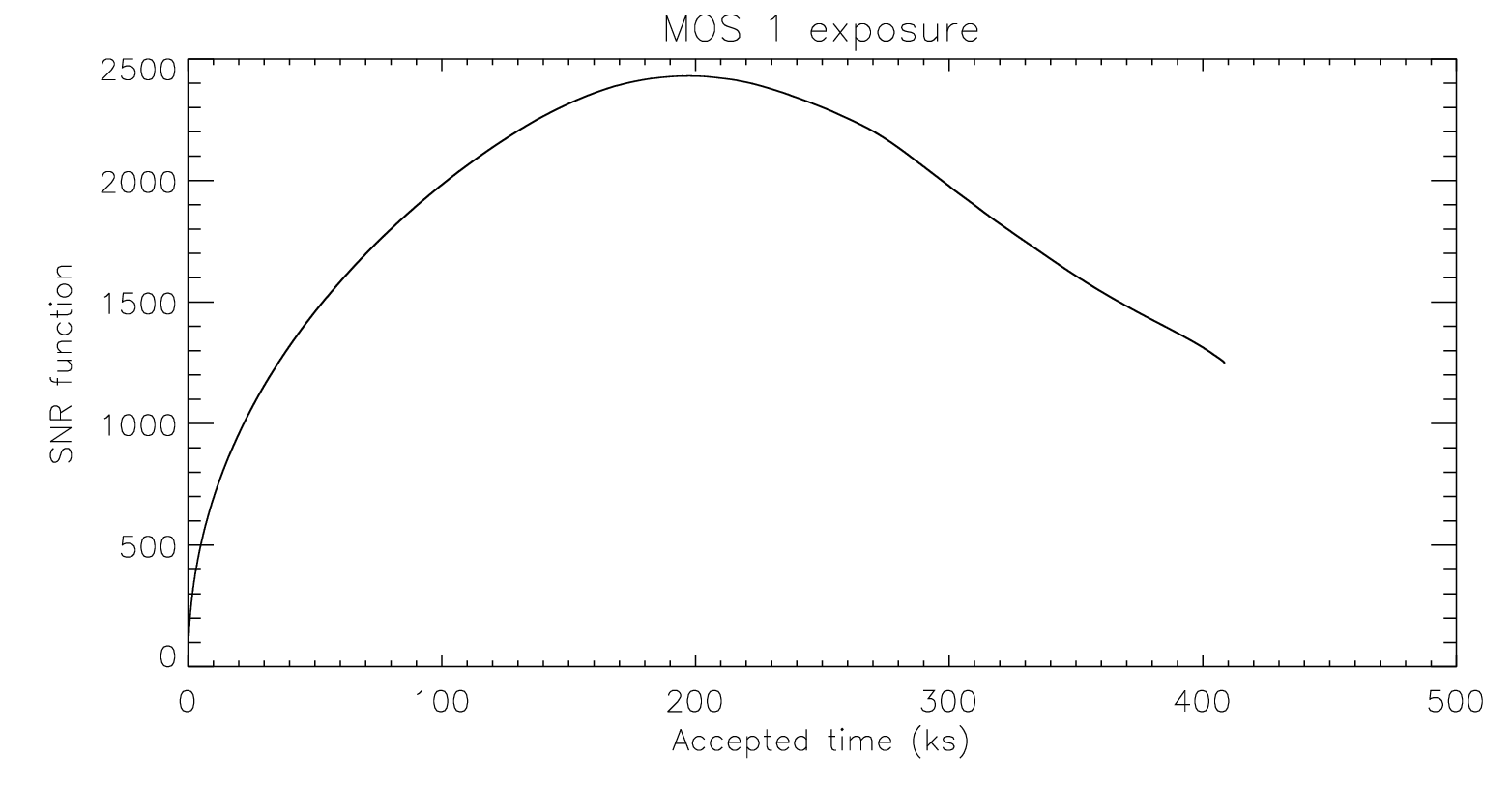}
\caption{\label{highbkg} Top panel: light curve of all {  events} recorded with MOS 1. 
Labels for the individual satellite orbits are given on top of the plot.
The gray shaded area represents the time interval when MOS 1 was
turned off after the micro-meteorite impact. The horizontal line is the {  threshold} count
rate that maximizes the SNR of the full image as described in the text. Bottom panel:
signal-to-noise function vs cumulative time. }
\end{center}
\end{figure}
In order to improve the detection of faint sources, we filtered the data obtained 
at step 1, excluding those time intervals with a high background count rate. 
In fact, the background during DROXO  was highly variable for a significant fraction of the 
exposure time (see Fig. \ref{highbkg}).
{  We excluded all events registered in intervals when the rate on the whole 
image, shown in Fig. \ref{highbkg}, was higher than a given threshold. 
The rate threshold was chosen in a way that maximizes the {\em signal-to-noise-ratio} (SNR) 
for faint sources  \citep[cf.][]{Sciortino01,Damiani2003} and 
improves the source detection process toward faint sources as expected.}
The net cleaned exposure times after this filtering are 198.1~ks (38\%)
273.2~ks (53\%) and 213.2~ks (41\%) for MOS 1, 2 and PN, respectively. 

We detected 111 point sources with a significance threshold corresponding to two spurious 
detections in the whole field-of-view. 
{  The threshold for the detection of faint sources was determined from the analysis of
a large set of simulations of background-only images and then running the detection
code on these images. The simulations provide
a value for the significance threshold to retain at most 1-2 spurious 
sources per field in real data.

Source positions, off-axis distance, exposure times,  and
X-ray count rates are listed in Table~A1 for all detected sources. 
We also report the 2\,MASS designation and other names from the literature for the
optical counterparts identified in Sect.~5.1. The last column
of Table~A1 contains a flag pointing to the source of the previous identifications.
We built an atlas of spectra and light curves for each source 
which is available online\footnote{See: 
{\sc http:www.astropa.unipa.it/~pilli/atlas\_droxo\_sources.pdf}}; 
in Appendix \ref{atlas} we show a page of this 
atlas as an example. }
\subsection{X-ray spectra}
\begin{figure*}
\begin{center}
\includegraphics[width=\textwidth]{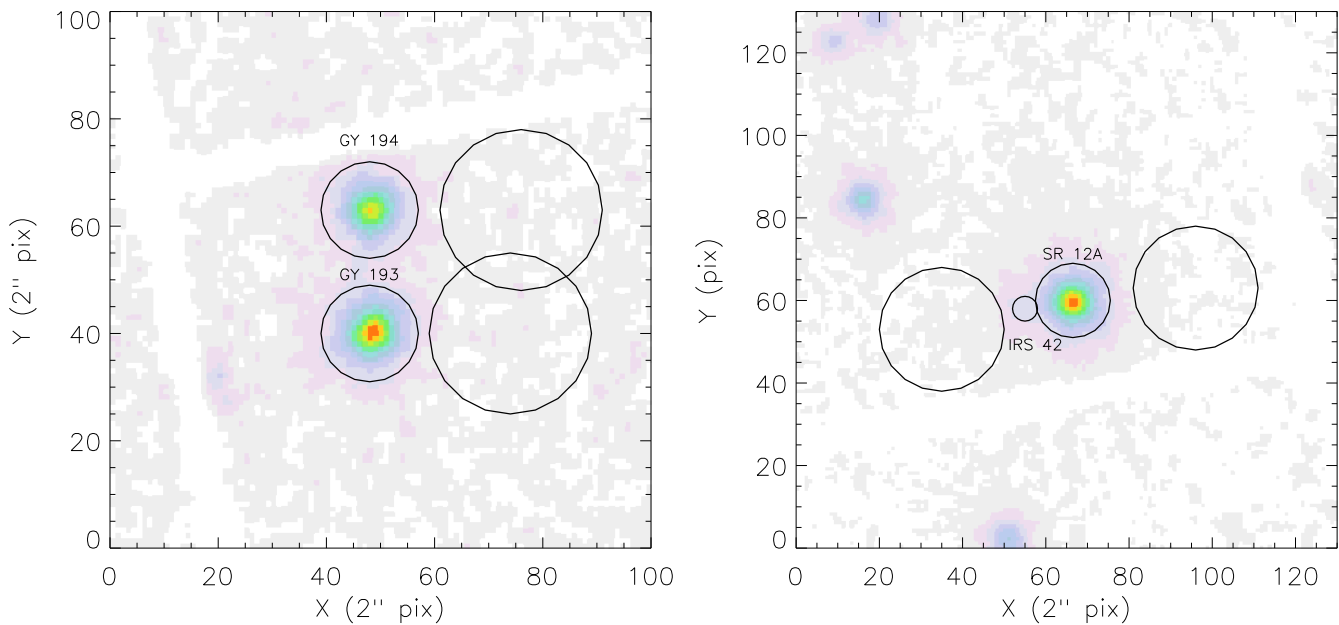}
\caption{\label{bkgsub} Two examples of choices for the source and background extraction regions.
The left panel shows two nearby but distinct sources. The right panel shows an extreme case
in which the region for the faint source is shrunk to minimize the influence of the bright source. 
The background regions are chosen as near as possible to the sources to which 
they refer and at the same distance of the readout node for PN. 
}
\end{center}
\end{figure*}

We produced both light curves and spectra of all sources by selecting
the photons in circular regions around the source positions. 
The radii of the regions depend on the source intensity, on possible 
of nearby sources and the geometry of the CCD. 
We used regions on the same CCD for source and background;
for the background, we avoided to include pixels and columns poorly calibrated in energy.
Figure \ref{bkgsub} shows two examples of the choice of source and background extraction regions.
For PN spectra a further constraint was to have both source and background
extraction regions at {  nearly} the same distance from the CCD readout node. 

To produce the spectra we filtered the photons with respect to {\sc FLAG} 
(chosen to be equal to zero) and {\sc PATTERN} 
(less or equal to four) as indicated in the SAS User's Guide. {  Given that the choice of
GTIs was tailored toward the faintest sources, at this stage we further improved 
the choice of GTIs for bright sources.} 
The time-filtering we adopted for spectra starts from the {\em Good Time Intervals} 
(GTIs) defined initially to perform source detection and iteratively adds temporal bins 
from the light-curve
\footnote{The binning of the light curve for each source is chosen in an 
iterative way, starting from $\Delta T = 1$ ks and progressively enlarging $\Delta T$ 
until the bin with maximum net counts has at least 50 counts or $\Delta T$ 
becomes larger than 10 Ks.} that contribute to increase the total SNR of the spectrum. 
 The procedure starts by adding the time bin with the largest  
individual SNR and is then iterated considering time-bins of  
decreasing SNR until no gain in total SNR is obtained.
As a result, the net exposure time of each spectrum is different.
\subsubsection{Background subtraction \label{bkgsb}} 
We noticed that scaling the background by the geometric areas of the  
source and background extraction regions leads to incorrect estimates of the  
background, especially for faint sources and/or during times of high  
background. 
{  The effect is that background-corrected light curves of faint sources are 
either directly or inversely correlated with the background light curves  
and often lead to negative net count-rates. 
We understood this as the effect of a vignetted background component that is not properly taken  
into account by a purely geometric scaling factor. In order to  
correctly estimate the background contribution to the photons  
extracted in the source region we constructed background maps for each  
instrument (MOS 1, 2 and PN) and for each orbit. These were built by  
removing large regions around detected sources from the images, and  
by subsequently smoothing and interpolating the maps over the source  
extraction areas. Although the resulting scaling factors differ from  
the purely geometric ones by only a few percent, the difference is  
relevant in cases when the background dominates the count rate in the  
source regions and mitigates the above mentioned spurious effects on  
the light curves.}

\subsubsection{\label{xmod} Model fitting of spectra}
We analyzed the spectra of {  all X-ray} sources in the 0.3--10~keV band. 
For a given EPIC detector (MOS~1, MOS~2 or PN) spectra from all orbits were summed up; 
analogously, background spectra were obtained; 
the response matrices and ancillary response files for each spectrum were 
multiplied and then summed by weighting by the exposure time. 
The background was scaled according to the procedure described in Sect. \ref{bkgsb}.

The spectra were grouped prior to fitting with {\sc XSPEC} v.~12.3\ 
to obtain at least a minimum 
SNR in each bin, by considering both source and background photons. 
In order to obtain the largest number of meaningful grouped spectra 
we adopted two schemes of grouping procedure based on high and low SNR of the final 
spectrum. For this purpose we used the same 
procedure as in the {\sc ACIS\_EXTRACT}  package for the analysis of {\em Chandra} 
ACIS data\footnote{See {\sc http://www.astro.psu.edu/xray/docs/TARA/ae\_users\_guide.html} 
for the details of the algorithm.}, adapted to our EPIC spectra to take into account the
background. 

{  We grouped the spectra using two thresholds for the minimum SNR to be obtained
in each bin, i.e. we generated two sets of spectra, one set with a high and one with
a low SNR per bin. The minimum SNR in each spectral bin was imposed on the basis 
of the source count statistics, SNR $\geq 2$ defining our low SNR binning and
SNR $\sim 3-5$ defining our high SNR binning. 
Where possible, we tried to obtain binned spectra with at least eight bins.}

The spectra of all detectors were fitted simultaneously. 
In some cases we had to discard {  the data of one or two of the three EPIC cameras.} 
These cases occur when a source is on CCDs gaps of one or two cameras, 
thus leading to a wrong estimate of the 
point spread function fraction contained in the extraction area. 

The spectra were fitted 
with one-temperature (1-T), two-temperature (2-T) and, 
in some cases, three temperature (3-T) {\sc APEC} models  \citep{Smith2001}
plus absorption  {\sc (WABS)} \citep{Morrison83}, 
the free parameters were the absorption column N$_\mathrm H$, the temperatures, and the 
emission measures. The abundance pattern was fixed
to that found in PMS coronae from the Orion Nebula in the COUP survey \citep{Maggio07}. 
Only in four cases described below (Elias~29, SR~12, IRS~42, src. 61) a more 
complex model was required.

\begin{figure}
\includegraphics[width=\columnwidth]{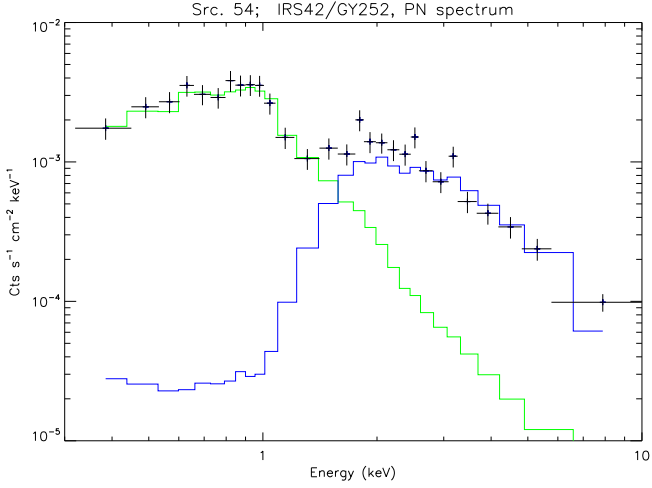}
\caption{\label{irs42spec} Spectrum and best-fit model of IRS42/GY252. 
{  The stepped curves are the contributions of SR~12A (light line) and that of IRS42 
(dark line).  The spectrum of IRS42/GY252 is modeled by an absorbed single temperature APEC model.}
} 
\end{figure}

For Elias~29 (src. 38)  and SR~12A (src 53) the global abundance 
scaling was left as a free parameter to achieve a better fit. 
For Elias~29 we confirm an unusually high abundance ($Z \sim Z_\odot$) 
with respect to those found in PMS coronae ($Z\sim 0.2-0.3 Z_\odot$),
as already reported by \citet{Favata05a}.
The source IRS42\,/GY252 (src. 54) is located in the wings of the much brighter X-ray source
corresponding to SR\,12A (see right panel of Fig.~3). 
{  To take account of the contribution by SR\,12A we added the best fitting 3-T APEC model multiplied
by a constant factor representing the amount of contamination to the model of IRS\,42/GY252. 
This scaling factor was a free fitting parameter. 
The spectrum of IRS\,42/GY\,252 itself can be described with a 1-T APEC model.} 
\label{src54} The spectrum and the best-fit model are shown in Fig. \ref{irs42spec}.
{  As can be seen, IRS\,42/GY\,252 is highly absorbed and the soft emission is attributed
entirely to the contamination by SR\,12A.} 
For the source nr. 61 we used a model given by the sum of two differently absorbed 
{\sc APEC} models to account for the soft excess visible below 1.0 keV, as discussed in Sect. 
\ref{pink} and Fig. \ref{pinkspec}. 

{  Generally, the best fit was chosen on basis of the probability $P(\chi^2 > \chi^2_0)$ 
of obtaining a $\chi^2$ higher than the observed one $\chi^2_0$. 
Our threshold for  acceptable fits was $P(\chi^2 > \chi^2_0) > 0.02$. 
Whenever this criterion was satisfied by the spectra with high SNR per bin 
(typically for medium-high statistics spectra) we chose these; 
otherwise we selected the fits obtained from the spectra with low SNR per bin.
We always selected the best-fit model with the smallest number of free parameters. 
In seven cases, even for some bright sources, no formally acceptable fit was found and we
allowed for a lower $P(\chi^2 > \chi^2_0)$. However, we accepted the best fit results 
by visually checking that the overall shape of the spectrum is well reproduced by the model. 

With this procedure we obtained spectral fits for $91 / 111$ sources.}   
Table \ref{t1fit} summarizes the results. 
The columns report the source number,  the available EPIC datasets, 
the data sets we chose for fitting, a flag indicating the binning of the spectra we used 
(high or low SNR), the type of model, the model best-fit parameters, the unabsorbed flux and luminosity 
in the 0.3--10~keV band, the $\chi^2$ statistics, degrees of freedom and the probability $P(\chi^2 > \chi_0^2)$.
{  For those weak sources without a spectral analysis, 
we calculated fluxes and X-ray luminosities in the 0.3--10.0 keV band 
by using their count rates and PIMMS 
software\footnote{see http://heasarc.gsfc.nasa.gov/Tools/w3pimms.html} 
assuming a 1-T model with kT and N$_H$ equal to the median of values derived from 
the best-fit procedure of spectra (see Sect. \ref{sens}) and a distance of 120 pc.}
\subsection{\label{photos} Photospheric parameters}
{  In Sects. 5 and 6 we will focus our discussion on the X-ray properties of 
the sample of PMS stars classified by Bo01.}
The photospheric stellar parameters ($T_\mathrm{eff}$, $L_\mathrm{bol}$ and masses) 
for the Class~II and III objects from the list of Bo01 were estimated from the near IR 
(2MASS) photometry. The procedure we used closely follows that adopted by Bo01 and
improved by \citet{Natta06}. We assumed that the J-band emission
from these sources is dominated by the stellar photosphere and is only marginally 
contaminated by the emission from circumstellar material and also that the IR colors 
of Class~II sources can be described as the emission from a passive circumstellar disk as 
described by \citet{Meyer97}. These assumptions obviously do not apply 
to Class~I sources and for this reason photospheric parameters for these objects 
were not estimated. 

We used the \citet{Cardelli89} extinction law with R$_v=4.4$, which we think is
appropriate for Ophiuchus. A small number of sources ($\sim$15\%)
have colors slightly bluer than those of reddened main sequence stars, presumably due to
photometric uncertainties, which are on the order of 0.1 mag, while the offsets of these
objects with respect to the reddened sequence range between 0 and 0.15 mag.
Dereddening these sources by extrapolating the colors
of Class~II and III sources would produce an overestimate of the extinction. 
For these objects we assigned the colors of the closest photosphere model on the
reddened main sequence. 
Table \ref{photprop} lists the effective temperatures, masses and bolometric luminosities
of ISOCAM objects.
 
The values of the $J$-band extinction we derive are very similar to ones from 
\citet{Natta06}, with only the significant exception of WL~16, for which our 
procedure produces a significantly higher extinction.

\section{Sensitivity of the survey \label{sens}}
\begin{figure}
\includegraphics[width=\columnwidth]{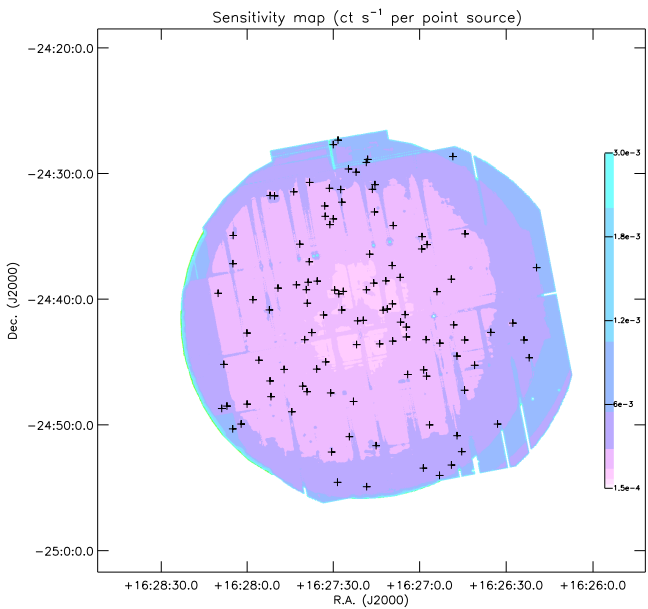}
\caption{\label{smap}  Sensitivity map of the DROXO field in units of ct / s per point source 
(logarithmic scale).  The map refers to the sum of PN and MOS instruments.
The intensity scale is indicated in the vertical strip on the right side of the map. 
Crosses indicate the positions of detected sources.}
\end{figure}

The faintest detected source in DROXO has a net count rate of 
$8.3\cdot 10^{-2}$ ct ks$^{-1}$ in MOS equivalent units. 
{  In Fig. \ref{smap}  we plot a sensitivity map in units of count rate per point source 
of the DROXO field of view. 
It is built starting from a smoothed background map, after removing the contributions of 
the sources, and taking into account the vignetted exposure map and the threshold used 
for detection. The map shows that the sensitivity varies by a factor 2.5\ in the area covered
by the three EPIC cameras, and it is quite constant across a 60\% fraction of 
the field of view because of the characteristics  of the detector. 
To translate the limit count rate in a limit flux we needed a conversion factor that
depends on the spectrum temperature and, critically, on the absorption.
From the analysis of the spectra we know that the absorption is also strongly variable 
from source to source by more than a factor 100, which can be due to local 
material around the sources or dense cloud material in front of the objects, or both.
The plasma temperatures that we find from the spectra vary from $kT \la 1$~keV to a few keV.
The combined action of local absorption and plasma temperatures of undetected sources
causes the limit flux to vary more than the range observed in the sensitivity 
map showed in Fig. \ref{smap}. 

{  For example, we can estimate a conversion factor from count rates to fluxes
by using the median of the column absorption ($N_\mathrm H = 2.3 \cdot10^{22}$  cm$^{-2}$) 
and of the plasma  temperatures ($kT = 3.1$ keV) from the spectral fits, and this gives us  
$cf = 6.1\cdot 10^{-11}$ erg cts$^{-1}$ cm$^{-2}$. 
This yields a limiting unabsorbed flux in the 0.3--10 keV band of $\sim5\cdot10^{-15}$ 
erg s$^{-1}$ cm$^{-2}$ and a luminosity of $\sim8.7 \cdot 10^{27}$~erg~s$^{-1}$.
The $cf$s derived from all sources are in the range $cf = 2.3-12.7\cdot 
10^{-11}$ erg cm$^{-2}$ cts$^{-1}$, yielding a limit flux comprised in the range 
$F_\mathrm X = 5-22 \cdot10-15$ erg cm$^{-2}$ s$^{-1}$ and luminosities $L_\mathrm X = 7-38 
\cdot 10^{27}$ erg s$^{-1}$, respectively. These values are indicative of the sensitivity
we achieved across the field of view, but local strong absorption can 
significantly lower the actual sensitivity in that position.} 
}

\begin{figure*}
\begin{center}
\includegraphics[width=\textwidth]{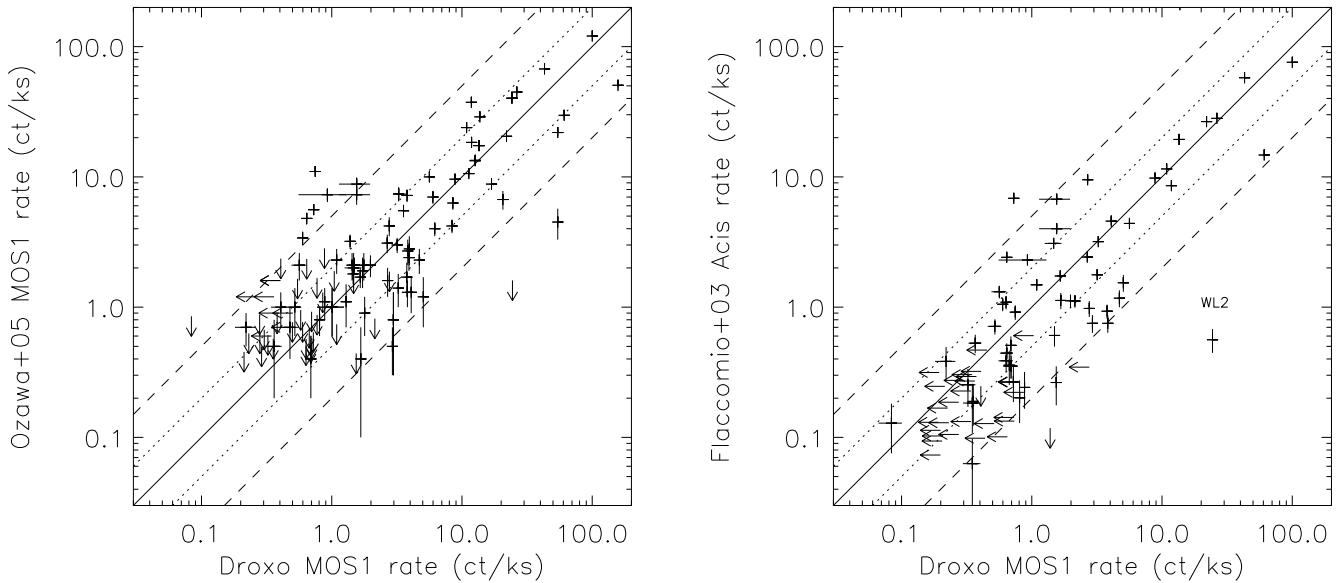}
\caption{\label{confozawaacis} Left panel: MOS 1 rate from \citet{Ozawa05}  
vs. DROXO MOS 1 rate. The lines mark the region for rate variations of factors 2 and 5.
Right panel: the same for the ACIS rate from \citet{Flaccomio03}. 
  Upper limits to count rate of sources undetected in the compared samples are 
indicated by vertical and horizontal arrows.  Errors are quoted at 1$\sigma$ level.} 
\end{center}
\end{figure*}

\subsection{\label{confx} Comparison with other X-ray surveys of $\rho$ Ophiuchi} 

In Fig. \ref{confozawaacis} we show the scatter plots of the MOS 1 count rate from \citet{Ozawa05} 
and Chandra-ACIS rate (which has an effective area comparable to the EPIC MOS) from \citet{Flaccomio03}
vs. the DROXO MOS equivalent rate.
In DROXO we detected 80 of the X-ray sources reported by \citet{Ozawa05}, while 7 of the sources 
found by Ozawa and collaborators remain undetected in DROXO; analogously we identified 62 sources 
from \citet{Flaccomio03}, while 30 of those sources have not been detected. 
We computed an upper limit to the count rates for those sources that are 
detected by \citet{Flaccomio03} and  \citet{Ozawa05} but are undetected in DROXO 
(indicated as arrows in Fig. \ref{confozawaacis}).
{  We also re-analyzed the previous XMM-Newton observation reported by \citet{Ozawa05}
with the same detection procedure as in DROXO, and we calculated the upper limits to count rates for 
DROXO sources undetected in the Ozawa survey. Analogously we calculated upper limits to count
rates for two sources detected in DROXO but undetected by \citet{Flaccomio03}.} 
The count rates of sources detected in DROXO and in the two other surveys 
globally agree within a factor $\sim$5. Because there is no systematic trend in the scatter 
we conclude that it is likely to be attributed to
variability. Although a detailed study of the time variability is beyond the scope of this paper,
we report here that 52\% of the sources have variable rates at more than 90\% significance level 
when compared with the Ozawa et al. survey, and 79\% are found variable to be compared with the  
survey of Flaccomio et al.

{ In the $33$\,ks {\em XMM-Newton} survey of \citet{Ozawa05} the faintest detected source has a count
rate of 0.5~ct~ks$^{-1}$. Using the conversion factor derived by us for DROXO the corresponding 
luminosity is 5.3$\cdot10^{28}$ erg s$^{-1}$, which scales reasonably well with the exposure times 
of both surveys.} Flaccomio et al.'s survey reaches a limiting rate similar to DROXO,  
likely due to the lower Chandra-ACIS background with respect to that found in our EPIC observation. 
Most of the ACIS sources undetected in DROXO are near the sensitivity limit of DROXO, and a variability 
of factor 2  can easily explain their missed detections.

Sources 33 (GY~304) and 43 are brighter in DROXO than in Ozawa et al. survey 
by more than a factor five.
DoAr~25 suffered of strong pile-up in the Ozawa et al. survey. These authors 
do not report a direct measurement of the count rate for this object, but they derived
a luminosity from the spectral analysis. From its X-ray luminosity we inferred a count rate 
greater than 200 ct/ks. We checked that in DROXO DoAr~25 and the other two brightest sources
in the field, SR~12A and IRS~55, are not affected by pile-up. 
The variability of DoAr~25 with respect to the XMM observation detailed in \citet{Ozawa05} 
is at least a factor 4. 

\begin{figure}
\includegraphics[width=\columnwidth]{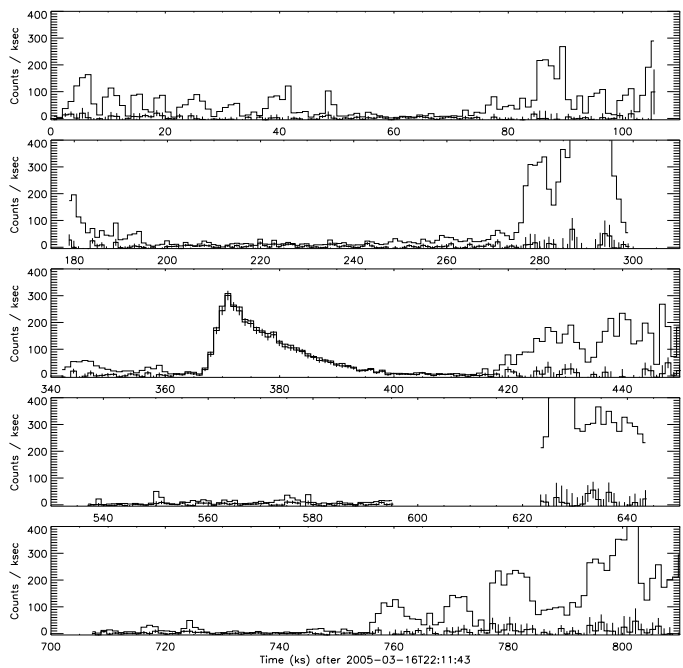}
\caption{\label{lcwl2} PN Light curve of WL2 (source nr. 15). 
The bold light curve is the net rate, while  the thin line is the total rate
(source + background). A large flare lasting $\sim$35 ks is well visible. The 
source quiescent rate is comparable to the background rate.}
\end{figure}

Among the most variable sources, WL~2 (labeled in Fig. \ref{confozawaacis}) shows a big flare 
in the DROXO light curve (Fig. \ref{lcwl2}). The quiescent rate of WL2 is $\sim6$ ct ks$^{-1}$ 
and the peak rate is $\sim300$ ct~ks$^{-1}$ explaining its large offset position 
in the scatter plot of the right panel in Fig.\ref{confozawaacis}. 
The quiescent rate is consistent with the one measured by \citet{Ozawa05}. 
The long duration of this flare on WL\,2 ($\sim 35$\,ks) underlines the need
for long observations of YSOs to properly assess their quiescent emission.

\section{The nature of the X-ray sources}
\subsection{Optical and IR counterparts of the DROXO sources}
We searched for optical and IR counterparts of DROXO sources in the 2MASS, Spitzer C2D 
\citep{Evans2003}\footnote{See also http://ssc.spitzer.caltech.edu/legacy/c2dhistory.html} and
ISOCAM (Bo01) surveys as well as in the optical and IR surveys of \citet{Natta06}, \citet{Wilking05}, 
\citet{Barsony05} and \citet{LR99}. 
The match radius takes into account the uncertainties on the X-ray source position 
in DROXO, {  which is generally on the order of $\sim1\arcsec$ },  and other catalogs. The
mean value for the match radius is $\sim1.0\arcsec$ with a 0.1-0.9 quantile range of 
$\sim 0.5--5.8\arcsec$. As anticipated in Sect.~3.1, Table \ref{tabdet} reports the literature names of 
DROXO counterparts and a flag indicating in which catalogs the source is identified.
Sources with counterparts in the ISOCAM survey are indicated with letter ``I'' in 
the ``Flag'' column; sources identified only in Spitzer are indicated with letter ``S'' ; 
sources which have only a X-ray identification in the surveys by \citet{Flaccomio06} 
and/or \citet{Ozawa05} are indicated with letter ``X''. 
Among the $111$ X-ray sources $78$ have a counterpart in 2\,MASS and
$61$ have a counterpart in ISOCAM data. Two objects have no 2\,MASS
and no ISOCAM counterpart, but have been identified in the literature
(IRS50 and WSB46).
 
An additional $16$ X-ray sources have been newly identified with Spitzer. 
This leaves $16$ DROXO sources without known optical/IR counterpart. Of these, $6$ were detected
in previous X-ray surveys and $10$ are presented here for the first time.
 We calculated upper limits to the X-ray fluxes and luminosities for the sample of undetected
ISOCAM objects falling in the field of view of DROXO {  using the conversion factor
derived in Sect.~\ref{sens}}.  

\subsection{Unidentified X-ray sources}
The inspection of light curves and spectra can give some
indication on the nature of sources with unknown optical/IR
counterpart, designated `U' or `X' in Table \ref{tabdet}.
Five unidentified sources (Src.~\# 5, 7, 16, 26, 58)
are too faint for spectral analysis, but the other five unidentified X-ray sources
(Src.~\# 12, 19, 88, 96, 110) have spectral parameters compatible with the expectation for
a YSO in $\rho$\,Oph, i.e. $kT$ of a few keV and 
$N_{\rm H} > 10^{22}\,{\rm cm^{-2}}$. Furthermore, Src.~110
shows impulsive time variability similar to flares typical for PMS stars.
There are also six sources (Src.~\# 20, 29, 41, 45, 48, 95) 
without optical/IR counterpart that have been detected in previous X-ray surveys. 
For Src. \# 20, 29 and 41 we had too few counts to obtain meaningful spectral fits, 
and their light curves show some variability, but no clearly identifiable flare. 
For Src. \# 45, 48 and 95 the spectral analysis gives a high absorption 
(above 10$^{22}$ cm$^{-2}$). While the temperature of Src. \# 95 is not
constrained, Src. \# 45 and 48 have plasma temperatures of 5.4 and 4.5 keV. 
 For these two sources a power law best fit to their spectra is also acceptable.
These three objects could have characteristics 
consistent with those of highly embedded YSOs, although they are undetected on the 
millimetric surveys of \citet{Motte98,Johnstone00,Jorgensen08}. 
The lack of 2MASS counterparts suggests that these sources  could 
have an extragalactic nature. On the other hand, it is not ruled out tthat they are
very low mass PMS stars or even brown dwarfs.

\begin{figure*}
\begin{center}
 \includegraphics[width=8cm,angle=90]{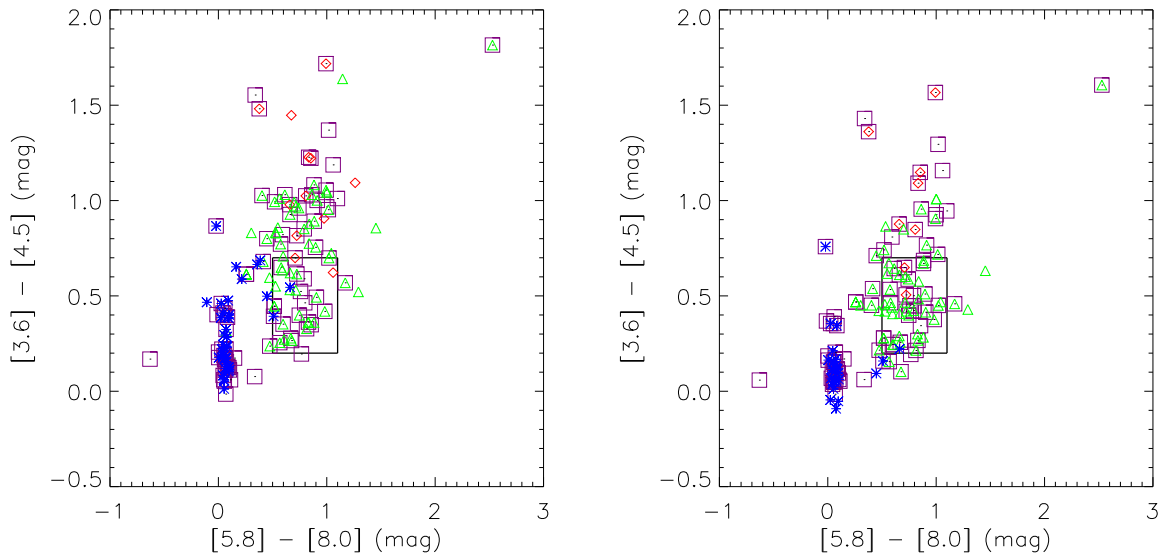}
\caption{\label{irac} IRAC color-color diagrams of objects in the DROXO field 
with information on the reddening from J or V band photometry or from $N_\mathrm H$.
The left panel shows colors before correction for the reddening, the right panel 
shows  dereddened IRAC colors. Purple squares mark objects with X-ray emission in DROXO. 
Asterisks are Class III YSOs, 
triangles are Class II YSOs and diamonds are Class I YSOs. 
The rectangle marks the region where Class II YSO colors are expected.
}
\end{center}
\end{figure*}

\subsection{Mid-IR photometry from {\em Spitzer}} 
{  The Spitzer photometry for the X-ray sources identified in the C2D survey
are listed in Table A4.}  
We derived magnitudes in the 3.6, 4.5, 5.8, 8.0 $\mu$m IRAC bands and constructed 
the color-color diagram shown in Fig. \ref{irac}
(color index $[3.6]-[4.5]$ vs color index $[5.8]-[8.0]$).
This diagram provides a rough classification of YSOs according to \citet{Allen04} and \citet{Hartmann05}. 
Normal, unreddened stars or Class~III / Weak~T-Tauri stars with very low reddening 
should occupy a region centered around (0,0). 
Reddening due to matter along the line of sight tends to disperse vertically data points along 
the [3.6]$-$[4.5] IRAC color index, while the [5.8]$-$[8.0] color index is not affected by this source of
reddening \citep[see][]{Flaherty2007}. 
Infrared emission from a circumstellar disk in Class I and II sources moves the objects 
both vertically and horizontally toward redder values of both indexes.
We expect to find Class II YSOs in the region marked with the black box or above it 
and very embedded protostars (Class 0/I) are expected to lie at $[3.6]-[4.5] \ga$ 1.1 and 
at  $[5.8]-[8.0] > 1$ \citep{Allen04}.  

We plotted only the objects for which we have information on the reddening from J and V band 
photometry or from $N_\mathrm H$.
In the left panel of Fig. \ref{irac} 
the symbols represents the IRAC colors before correction for reddening.
Objects detected in DROXO are marked with squares, diamonds are Class~I YSOs as
classified by Bo01, triangles are Class~II YSOs, asterisks are Class~III YSOs.
The points with ISOCAM classification follow the 
distribution outlined above, with Class~III~YSOs around or dispersed vertically above the (0,0)
point, and Class II YSOs located in the rectangle or dispersed vertically above this region.

In order to investigate the nature of these latter ``dispersed'' objects we dereddened 
the colors of those objects for which we had information on the {  extinction. If available
we used the reddening $A_\mathrm J$  or $A_\mathrm V$  
from the literature, otherwise the column absorption  $N_\mathrm H$ from the X-ray spectra
with the conversion factor of $A_V= 1.51\cdot 10^{-22}\cdot\mathrm N_h$ \citep{Mathis1990}. 
To convert these extinctions to the Spitzer bands we followed   
the calibrations by \citet{Flaherty2007} and \citet{Carpenter01}.}
The relation between reddening in 2MASS K$_\mathrm s$ and IRAC bands provided by
 \citet{Flaherty2007} for $\rho$ Ophiuchi indicates that no reddening is present 
in the $[5.8]-[8.0]$ color index. 
{  Therefore, dereddening shifts the objects vertically downwards in Fig. \ref{irac}. 
The right hand panel of Fig.\ref{irac} shows the dereddened IRAC colors. 
Evidently, most objects with $[3.6]-[4.5] > 0.7$
in the left hand side of Fig. \ref{irac}, fall in the ``canonical" IRAC 
Class\,II area after dereddening. Therefore, the majority of objects
with apparently protostellar IRAC colors are probably strongly reddened 
Class\,II sources. 
We note also that the ISOCAM sub-sample without X-ray counterparts has the same color distribution 
as the ISOCAM/DROXO sample. This indicates that no bias is present against X-ray properties.
}

\citet{Prisinzano08} found that Class~I protostars in the Orion Nebula can be separated into
two distinct subclasses: the first, Ia, with a rising SED from $K$ to 8$\mu$m and
lower X-ray emission level than the second  one, Ib, characterized by rising SED up to 4.5$\mu$m.  
This latter group shows IRAC colors more similar to those of Class II objects. 
The first subclass, Class Ia, is instead populated by more embedded 
protostars with lower and, perhaps, more absorbed X-ray emission. 
{  They find 23 Class 0-Ia YSOs, 22 Class Ib YSOs
and 148 Class II YSOs. In COUP the fraction of Class Ia on Class II YSOs is $\sim16\%$, 
the fraction of Class I versus Class II YSOs in Orion is $\sim30\%$.
In DROXO the fraction of Class~I objects with  [3.6]-[4.5] $> 1$ (after dereddening)
is $\sim7\%$ with respect to Class II objects. 
This sample should be composed by very embedded objects similar to 
Class Ia defined by \citet{Prisinzano08}.}
The number of very embedded protostars in Core F of $\rho$ Ophiuchi is thus very low
with respect to Class II objects. The same evidence has been obtained by 
{  \citet{Jorgensen08}. In that paper the fraction of Class I to Class II 
is reported to be lower in $\rho$ Ophiuchi than that present in 
other star-forming regions like Perseus 
(10\% in $\rho$~Ophiuchi vs 90 \% in Perseus,  respectively)}, 
hinting for a different age or formation time-scales for these two star-forming regions. 

{  Three objects (GY92, GY289, GY203) which have been classified as Class III by Bo01
are found with Spitzer colors similar to Class II YSOs. 
Furthermore, we find three objects classified as Class II YSOs (GY~146, GY~240, GY~450) that
are bluer than 0.4~mag in ([5.8] - [8.0]) IRAC color. These objects may have transition disks 
\citep[see][and references therein]{Kim2009}.
The star with the reddest IRAC colors located in the top right corner of Fig. \ref{irac} 
is WL22/GY174. It is classified
as a Class II object and probably it suffers of strong foreground extinction as 
pointed out by \citet{Wilking01}. This is supported also from our spectral fit to its 
X-ray spectrum: we find a N$_\mathrm H$ absorption value of $1.3\cdot10^{23}$ cm$^{-2}$, 
a factor of 5 higher than the average of the DROXO sample. }

\section{Results}
\subsection{Coronal temperatures and absorption \label{xfit}}
Table \ref{t1fit} lists the temperatures obtained through a thermal model 
fit to the X-ray spectra {  as described in Sect.~\ref{xmod}.}  
Most of the spectra are reasonably well described by a model with a single thermal component; usually 
we find temperatures higher than 1.5 keV and absorption column N$_\mathrm H$ higher 
than $10^{22}$ cm$^{-2}$. 
In a few cases two or three thermal components are needed to fit the spectra, 
depending on the characteristics of the spectrum and its count statistics.
For these cases we calculated the average of the two or three plasma temperatures 
weighted by their emission measures to obtain a representative mean temperature for these spectra.
The median of the representative plasma temperatures is $\sim $3.1 keV with a 
1$\sigma$ range between 1.6 and 8 keV (the 10\%--90\% quantile range is 
0.8--13 keV).  
The 10\%--90\% quantile range of the N$_\mathrm H$ column is 
$2.8\cdot 10^{21} - 7.2\cdot 10^{22}$ cm$^{-2}$  and the median is 
$2.3\cdot 10^{22}$ cm$^{-2}$.

\subsection{X-ray luminosities and stellar parameters}
\begin{figure*}
\begin{center}
 \includegraphics[width=6.5cm,angle=-90]{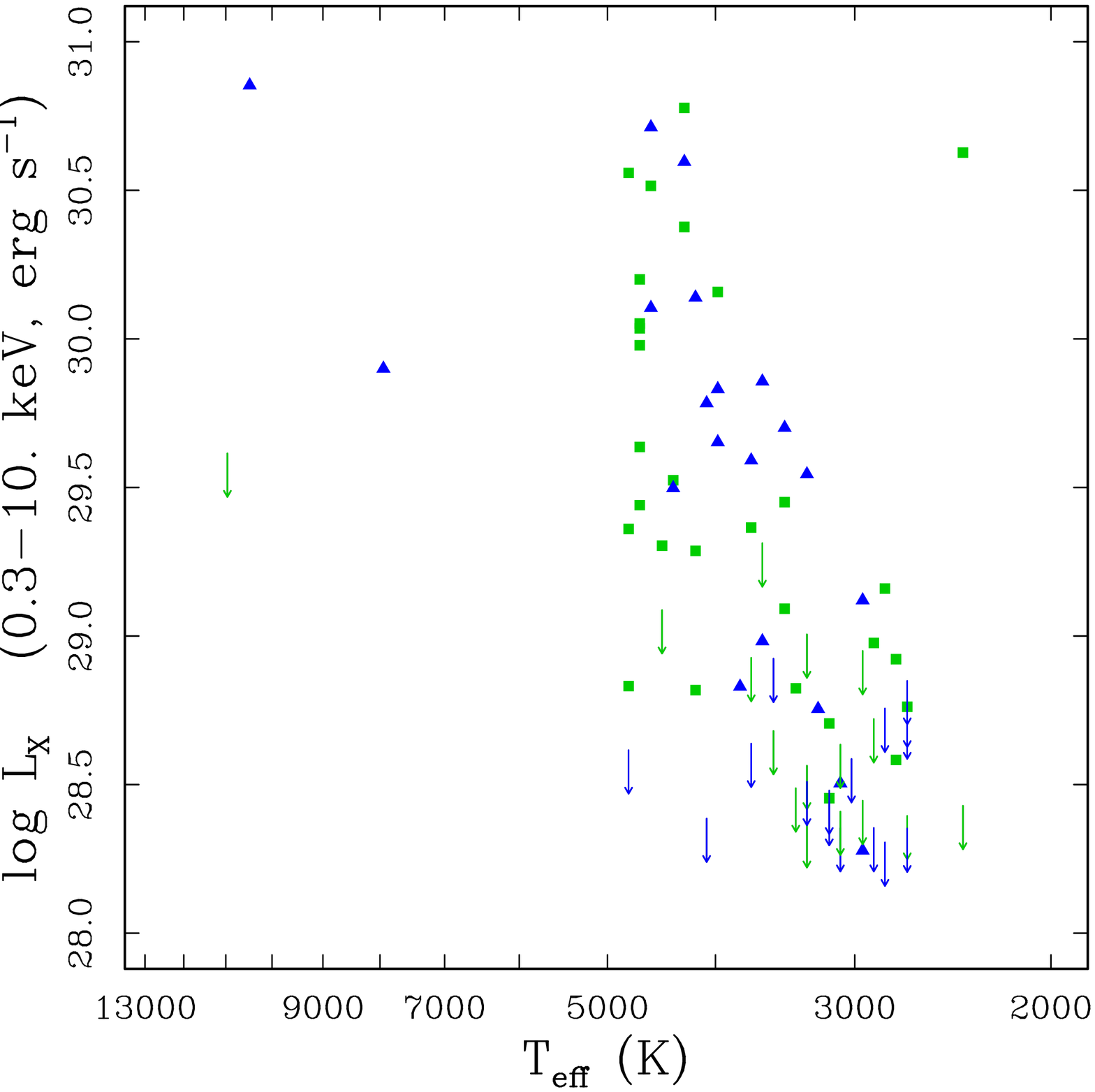}
 \includegraphics[width=6.5cm,angle=-90]{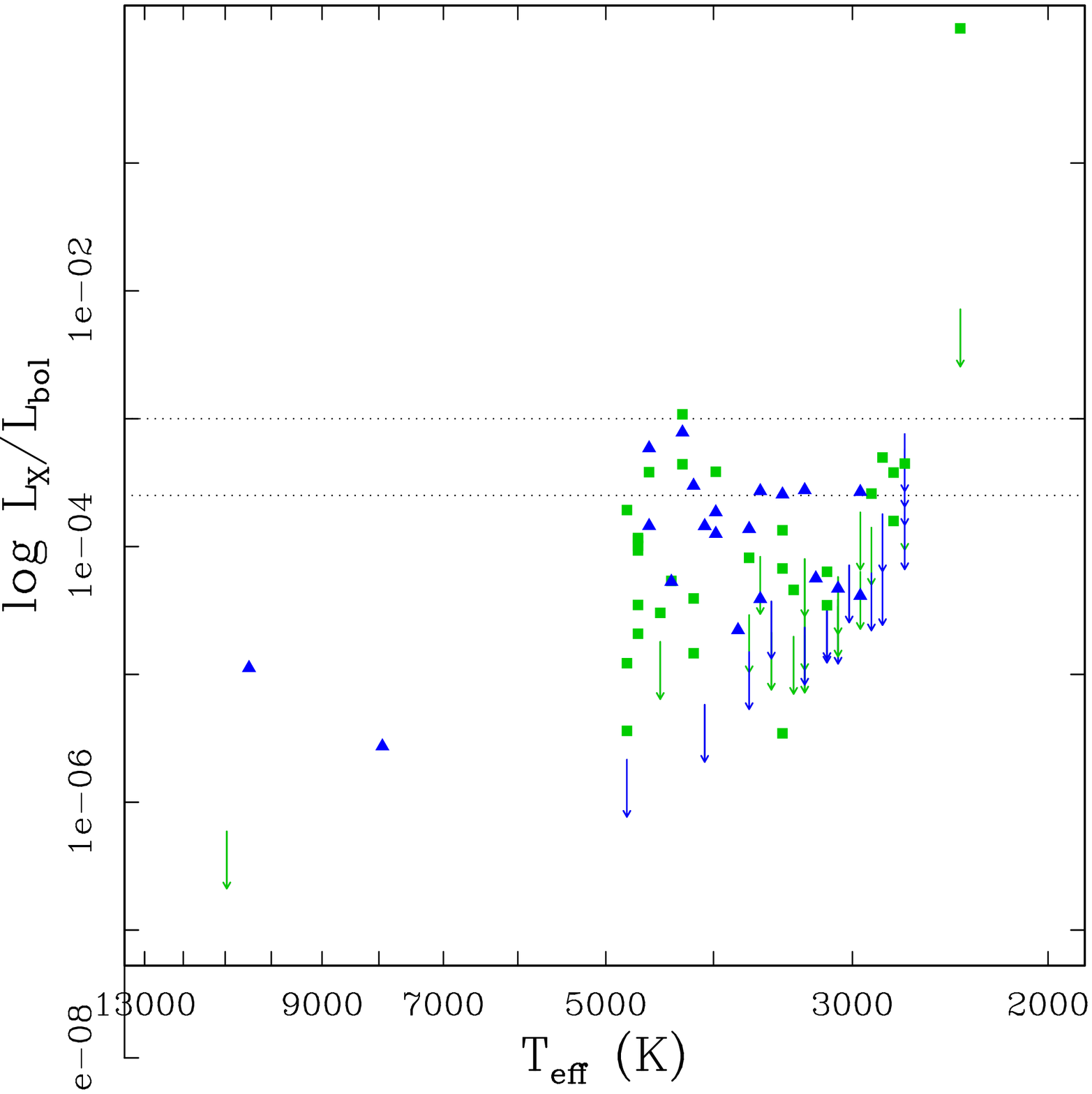} 
 \includegraphics[width=6.5cm,angle=-90]{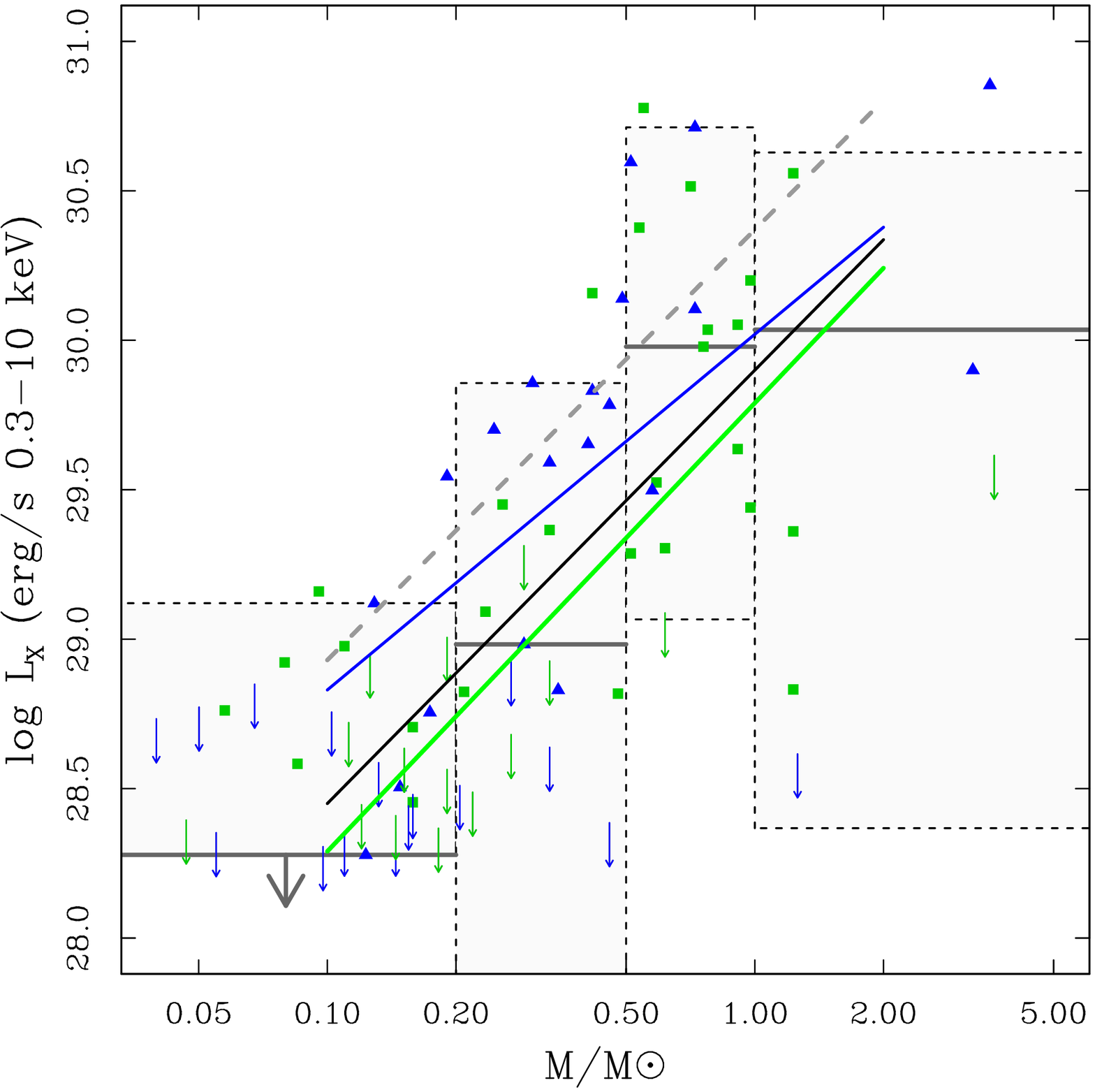}
 \includegraphics[width=6.5cm,angle=-90]{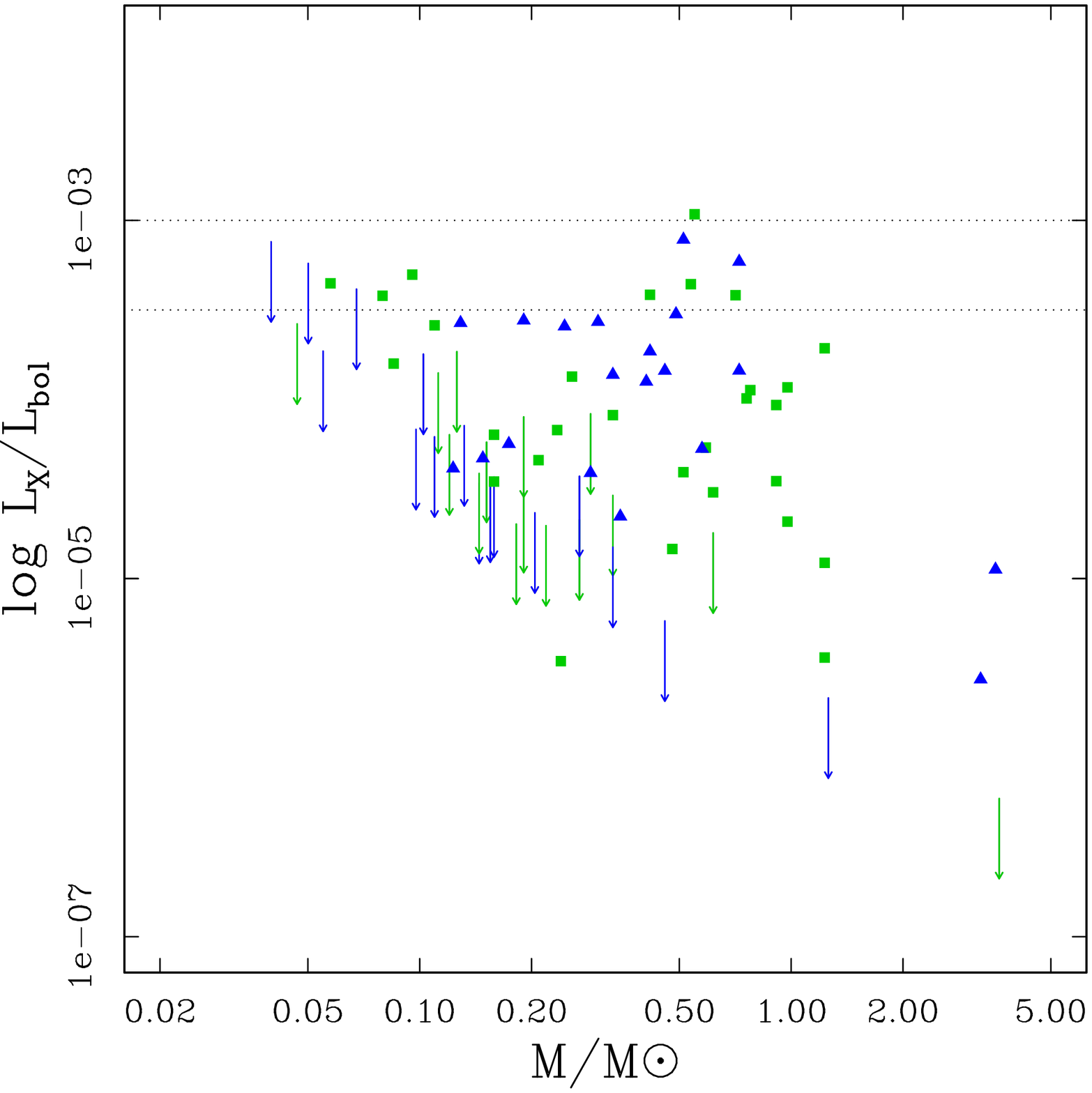}
\caption{\label{t_lx} Top left panel: X-ray luminosities (0.3-10. keV) 
versus effective temperatures of YSOs identified with ISOCAM objects (Bo01). 
Symbols are:  squares Class II YSOs, triangles Class III YSOs.
Upper limits to luminosities were calculated for ISOCAM objects not 
detected in DROXO (vertical arrows). 
Top right panel: $L_\mathrm X/L_{bol}$ ratios vs $T_\mathrm{eff}$. 
The upper limit with highest $T_\mathrm{eff}$ and mass is relative to WL~16 and
calculated taking into account an  $A_\mathrm V \sim 30$ mag and a $kT = 1$ keV 
(see text). The detection point at the lowest $T_\mathrm{eff}$ is from WL~2 that 
exhibited the large flare shown in Fig. \ref{lcwl2}. 
Bottom left and right panels: X-ray luminosities and $L_\mathrm X/L_{bol}$ vs. mass
(in units of $M_\odot$). The boxes in the left panel are the 10\%$-$90\% quantile ranges of 
luminosities derived from the Kaplan-Meier estimator of XLFs in four mass ranges 
($\le 0.2$, 0.2--0.5, 0.5--1.0, $\ge 1.0 M_\odot$). 
Horizontal segments mark the median of $L_\mathrm X$ in each mass interval, 
in the lowest mass bin the median is an upper limit. 
The solid black line is the relation between mass and luminosity fitted 
between 0.1 and 2 M$_\odot$, blue and green lines are the best fits for Class III 
and Class II YSOs respectively. 
The dashed gray line is the analogous obtained by \citet{Preibisch05} to Orion 
young stars. 
Horizontal lines in the bottom right panel are the values of saturation of 
$\log L_\mathrm X/ L_\mathrm{bol}$ equal to -3 and -3.5.
} 
\end{center}
\end{figure*}

{  We examine the relation between X-ray luminosity and X-ray to bolometric 
luminosity ratio and mass and  effective temperature for all X-ray 
sources for which stellar parameters could be determined in 
Sect. \ref{photos} (Fig. \ref{t_lx}). }
{  The X-ray luminosities for the sample with known stellar parameters} increase from  
$\sim2\cdot10^{28}$ ergs s$^{-1}$  up to $\sim7 \cdot10^{30}$ ergs s$^{-1}$ in the 
3000--5000 K range (Fig. \ref{t_lx}, top left panel) or  $\sim 0.1 - 0.7$ M$_\odot$ (left bottom panel).  
Most of the YSOs have effective temperatures around  3000--5000 K (K type stars), with the
exception of the three hot stars WL~5, WL~16 and WL~19.
The fraction of X-ray detections among ISOCAM YSOs increases from $\sim0.1 M_\odot$  
to $\sim 1 M_\odot$. 
The boxes in the bottom left panel of Fig. \ref{t_lx} are the 10--90\% quantiles obtained with 
ASURV software \citep{Schmitt85,Lavalley92} 
for $\log L_\mathrm X$ values in four ranges of mass ($<0.2$, $0.2-0.5$, $0.5-1.0,> 1.0$ $M_\odot$). 
For these mass ranges the median values of $\log L_\mathrm X$ are 
$\leq$28.3, 29.0, 30.0 and 30.0, respectively. 
The COUP sample contains very few upper limits in each mass range  whereas in DROXO 
the fraction of upper limits is $\sim 30$\,\% reducing the $L_{\rm x}$ medians in each mass range. 
As discussed below in Sect. \ref{ysoclasses}, we suggest that a fraction of Class III YSOS 
are likely spurious cloud members. 
{ We fitted the relation  between X-ray luminosity and mass excluding
these suspect members in the Class III sample obtaining 
$\log L_\mathrm X \sim 29.9\pm0.2 + \log M^{(1.45 \pm 0.25)}$ with the same procedure used by \citet{Preibisch05}
for COUP sample.  
We also fitted  X-ray luminosity and mass relation
for Class II and Class III samples separately, finding similar slopes but different normalizations.
The relations are \\
$\log L_\mathrm X \sim 29.8\pm0.2 + \log M^{(1.5 \pm 0.3)}$ and \\
$\log L_\mathrm X \sim 30.0\pm0.3 + \log M^{(1.2 \pm 0.5)}$ \\
for Class II and Class III YSOs respectively. 
The slopes of these relations are very similar to those found for ONC in COUP 
program (1.44, \citealp{Preibisch05}). The lower normalizations that 
we find in the $\rho$~Ophiuchi sample with respect to ONC
(up to 0.6 dex for DROXO Class~II objects with respect to COUP)
can be due to systematic effects like distance estimate 
(for Orion it was used a value of 450~pc, 
more recent estimates place ONC to $\sim$400~pc, 
\citealp[see Sect. 8.1.1 in ][and references therein]{Mayne08}) and correction for 
absorption, but an intrinsic difference could be present. Thus 
PMS stars in $\rho$~Ophiuchi are on average less luminous than PMS stars
in Orion. Because there are upper limits in DROXO, this difference should be more marked.
Also in the Taurus Molecular Cloud (TMC) \citet{Telleschi07} find a power law
index of $\sim1.5$ between $\log L_\mathrm X$ and mass.

The $\log L_\mathrm X / L_\mathrm{bol}$ ratio increases with decreasing stellar masses (Fig. \ref{t_lx}, 
bottom and top right panels) saturating at $\sim-3.5$  for stars cooler than 5000 K and 
with masses of $\sim 0.7$ M$_\odot$, slightly  below the ``canonical'' value of ${-3}$ observed 
in MS young stars, but very similar to the value reported by \citet{Preibisch05} for Orion YSOs
(${-3.6}$). 

From the relation between mass and X-ray luminosity and the hypothesis that
the ratio  $\log L_\mathrm X / L_\mathrm{bol}$  for PMS stars is saturated at a level of $-3.5$,
\citet{Telleschi07} derived an empirical mass -- bolometric luminosity relation for PMS stars
$\log L_\mathrm{bol} \sim \log (M/M_\odot)^{1.49}$ which is shallower than the 
relation that yields for MS stars. By considering that the saturation limit of  
$\log L_\mathrm X / L_\mathrm{bol}$ for our sample of YSOs of $\rho$~Ophiuchi 
is similar to that of TMC and ONC ($\sim -3.5$, see fig. \ref{t_lx}) and that the slope of 
the relation between mass and  $L_\mathrm{X}$ is quite similar for $\rho$~Ophiuchi, ONC and TMC, 
it is suggested that the relation between mass and $L_\mathrm{bol}$ found from \citet{Telleschi07} 
for coeval PMS stars should apply also for our sample.
}

Among the three hottest stars, characterized by effective temperatures around 10000 K, 
we detected the Class III YSOs WL~19 and WL~5, while WL\,16 (Class II) remains undetected. 
X-ray spectra of WL~19 and WL~5 show quite hot plasma temperatures 
(kT= 3.7 keV and 4.5 keV, respectively) and 
high absorption (N$_H = 10^{23}$ and 6.5$\cdot10^{22}$ cm$^{-2}$, respectively).
The undetected Class II YSO WL~16  is a peculiar object: it consists of a massive star 
(L$=250$L$_\odot$, $M \sim 4$M$_\odot$) that illuminates a 
circumstellar disk visible only at mid-IR wavelengths \citep{Ressler03}. 
It suffers of strong absorption ($A_V \ge 30$) likely due to a foreground screen of cloud material.
Taking into account this high absorption, the upper limit to luminosity  is inversely correlated with 
the plasma temperature: for $kT = 4$~keV we obtain  $\log L_\mathrm X \la 28.8$, for $kT = 1$~keV   
$\log L_\mathrm X \la 29.6$,  for $kT = 0.5$~keV $\log L_\mathrm X \la 30.7$.
According to their position in the HR diagram these three objects are
intermediate-mass pre-MS stars and their X-ray luminosities and $L_\mathrm X/L_\mathrm{bol}$ ratios 
are in the range typically observed for Herbig Ae/Be stars  \citep{Stelzer09}.

The coolest object is the binary system WL2/GY128 \citep{Barsony05}. 
A discrepancy between its spectral type and the effective temperature is present in the literature. 
While its spectral type is comprised between K and M as reported by 
\citet{LR99}, its temperature, estimated by \citet{Natta06}, is very low ($\sim$2300 K).  
 As discussed in Sect. \ref{confx}, WL\,2 has undergone a huge flare during DROXO. 
Its quiescent X-ray luminosity is $\sim10^{29}$ erg s$^{-1}$ which is typical for young K--M 
type stars but unexpectedly high for a low mass brown dwarf. 
Likely the photospheric temperature of this object is more similar to that of late K or M-type stars.

Only two bona fide brown dwarfs are in the field of view. We detected GY~310 
(log $L_\mathrm X = 28.92$ erg s$^{-1}$), but not GY~141, which was detected by \citet{Ozawa05} 
during a flare with a flux 90 times higher than in a previous Chandra observation.

\subsection{\em X-ray emission of different YSO classes.\label{ysoclasses}} 

We examined the X-ray detection rates for YSOs in different evolutionary states
referring to the YSO classification of Bo01. 
Their catalog comprises 16 Class\,I, 123 Class\,II, and 38 bona-fide 
Class\,III. The latter are classified as Class III YSOs
on the basis of absence of IR excess and detection in X-ray images (ROSAT) and/or radio
band (VLA). Given the low ROSAT sensitivity at $L_\mathrm X \sim 3\cdot 10^{29}$ erg s$^{-1}$ 
the sample of Class III YSOs is presumably incomplete.
\citet{Bontemps01} report also a list of 39 candidate Class\,III stars which are 
objects with stellar-like colors. They are found in the dense part 
of the cloud and have no X-ray detection at the ROSAT sensitivity limit. 
\citet{Bontemps01} conclude that these objects are in excess with respect 
to the number of expected field stars because of the high extinction 
in the region where they lie, and for this reason these objects were 
assumed Class III star candidates.
{  For the detection rates we considered only the fraction 
of $\rho$\,Oph members that are within the DROXO field-of-view
(col.~2 of Table~\ref{tab:yso_detectionrate}). 
The number of X-ray-detected stars from each YSO group is given in Col.~3, 
and the detection fraction is summarized
in Col.~4. The last two columns of Table~\ref{tab:yso_detectionrate} provide
the medians of the plasma temperatures and of X-ray luminosities for each sub-sample. 
These last values were obtained with ASURV and take into account the upper limits
for non-detected objects.
Before proceeding to a discussion of the numbers given in 
Table~\ref{tab:yso_detectionrate} an assessment of biases and completeness 
is in order. The ISOCAM sample is not well suited to select Class\,III
sources. The evidently low X-ray detection rate of candidate Class\,III
stars from Bo01 suggests that most of them are spurious
members. We detected six Class\,III candidates with $\log L_\mathrm X > 28.3$, 
and found that they have {\em Spitzer} photometry consistent with Class III objects 
\footnote{Four of six are classified as `YSO'  and two of six  are classified as `star' in
the {\em Spitzer} catalog, \citet[see][]{Evans03,Padgett2008,Evans09} 
for this scheme of classification}. There remain 15 out of 21 Class III  candidates 
undetected at $\log L_\mathrm X < 28.3$. 
Their masses are comprised between 0.1 and 0.5 $M_\odot$, and their dereddened IRAC colors [3.6] -- [4.5]
are comprised between 0.0 and 0.2 mag, whereas the six X-ray-detected Class III candidates have 
[3.6] -- [4.5] colors up to 0.4 mag. In the following discussion we will also consider a 
sub-sample of Class~III  objects without the 15 X-ray-undetected Class~III candidates.}

\begin{figure*}
\begin{center}
\includegraphics[width=6.5cm,angle=-90]{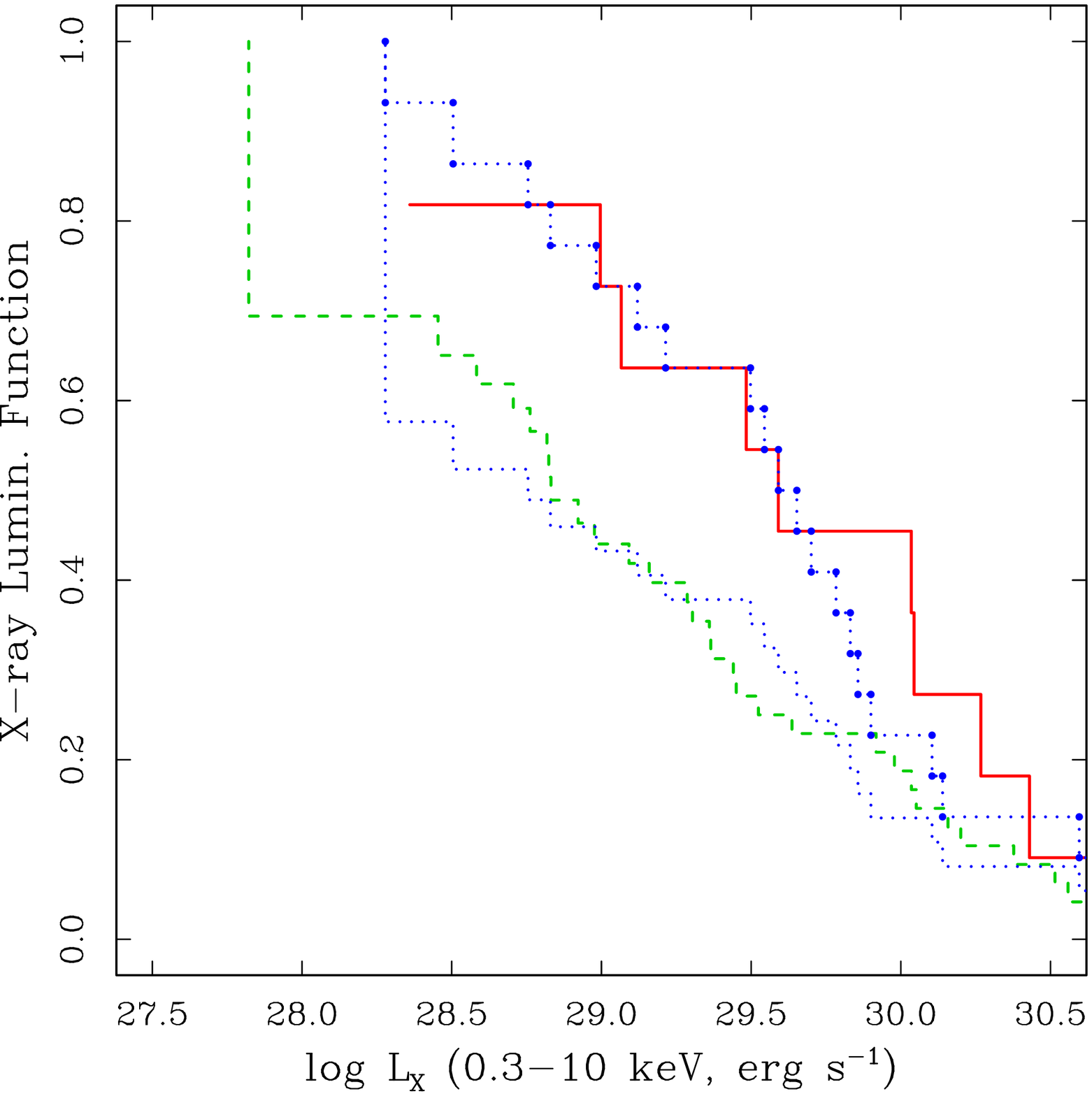}
\includegraphics[width=6.5cm,angle=-90]{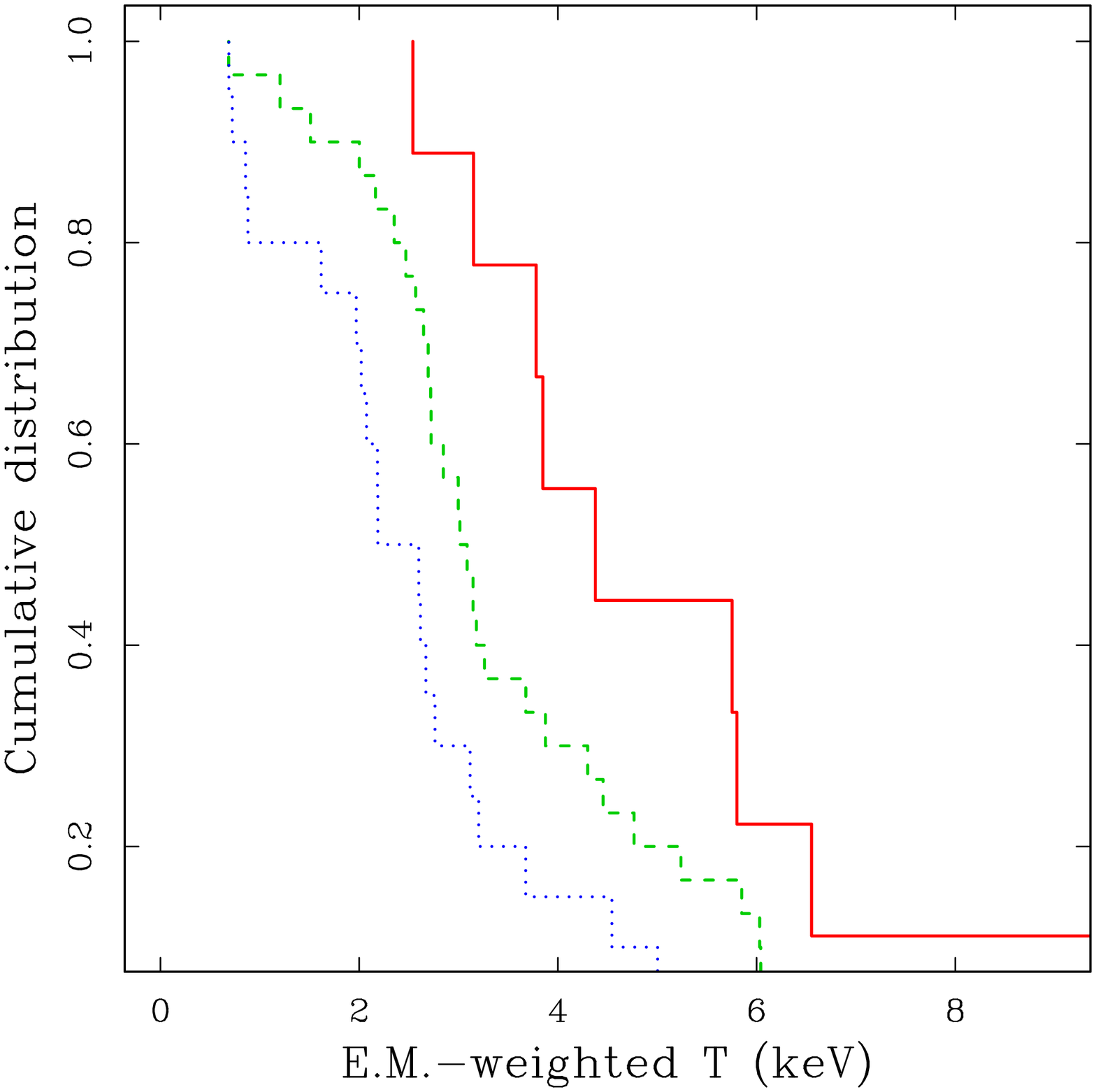}
\includegraphics[width=6.5cm,angle=-90]{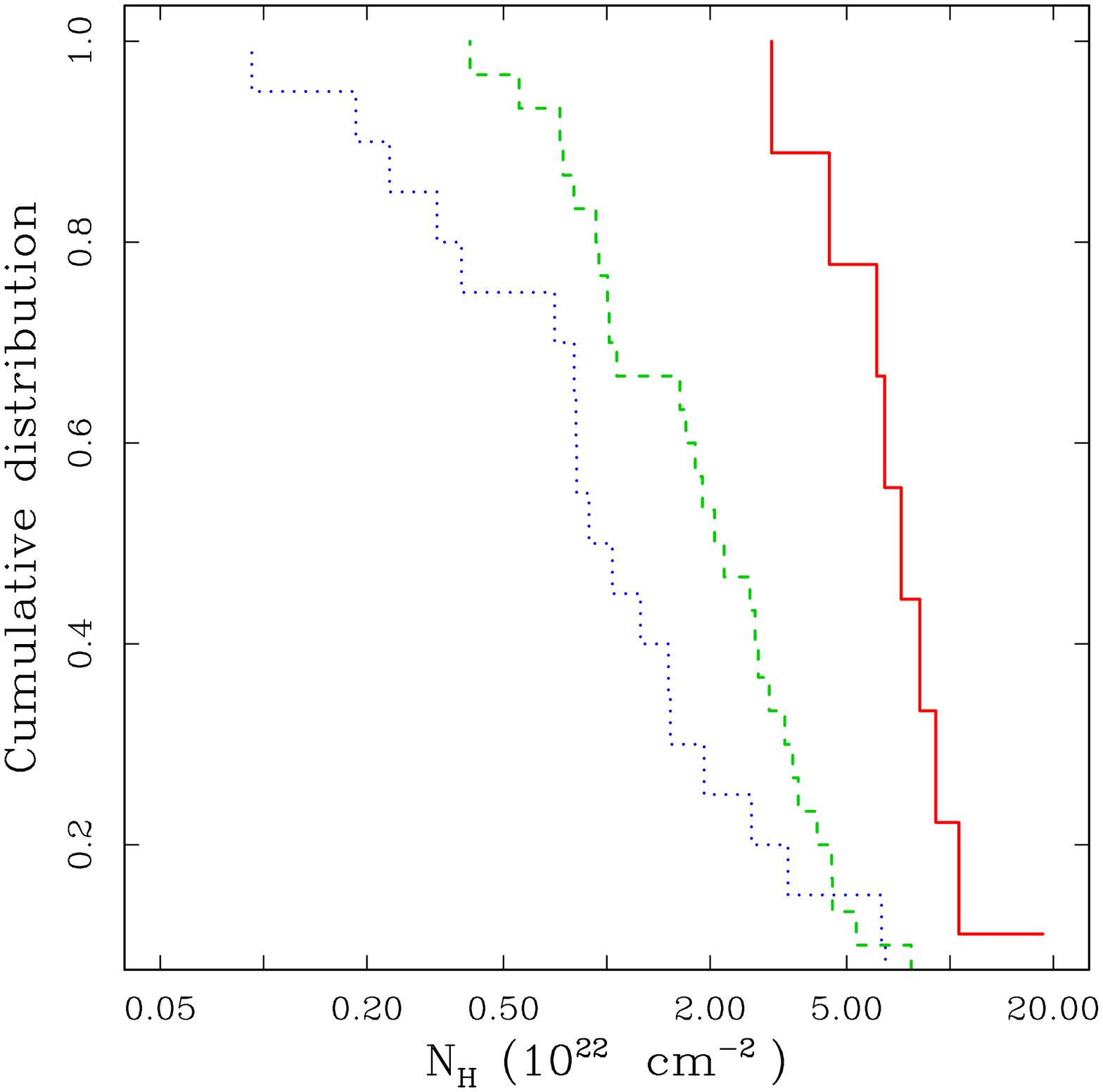}
\includegraphics[width=6.5cm,angle=-90]{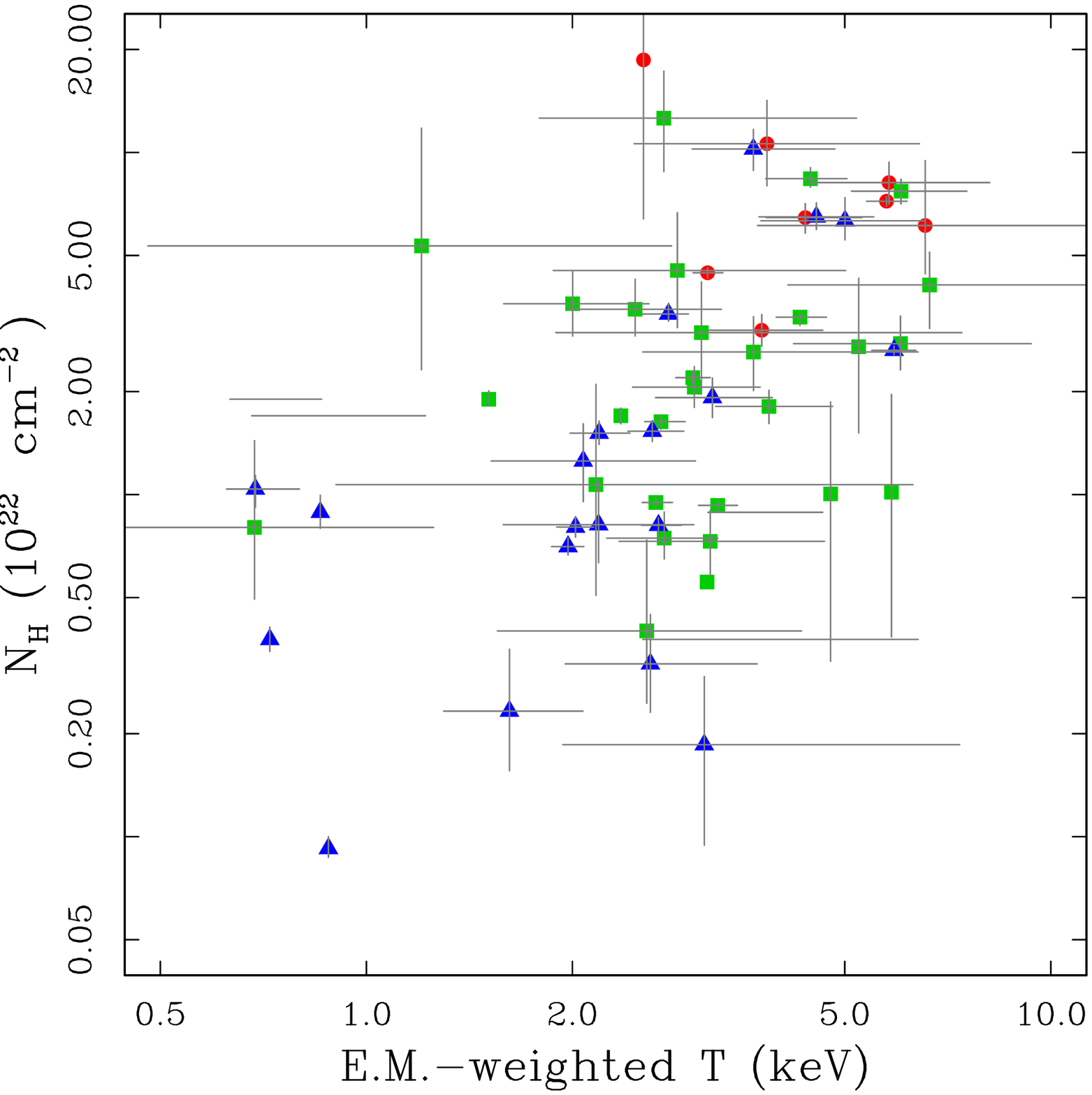}
\caption{Top left panel: Kaplan-Meier estimators of cumulative distributions 
of $L_\mathrm X$ for Class I YSOs (solid line), Class II YSOs (dashed line), 
Class III YSOs (dotted line) and {\em corrected} XLF 
(dotted line + points) of Class III YSOs from Bo01. \label{wttctt}
Top right panel: cumulative distribution of mean plasma temperatures obtained from fit 
to spectra for Class I (solid line), Class II (dashed line), Class III (dotted line) objects. 
Bottom left and right panels: cumulative distributions of $N_\mathrm H$ column absorption 
among the different classes (left) and scatter plot of mean plasma temperatures  
vs. $N_\mathrm H$ (right). 
Circles are Class I YSOs, squares are Class II YSOs, and triangles are Class III YSOs.} 
\end{center}
\end{figure*}

Previous studies of star-forming regions have shown that for each given mass range,  
Class I and II YSOs show lower  X-ray luminosities than Class III YSOs
\citep{Neuhau95,Flaccomio03,Preibisch05,Telleschi07,Prisinzano08}.
Figure \ref{wttctt} (top left panel) shows the Kaplan-Meier estimators of X-ray 
luminosity function (XLF) of Class I, II, and III objects following the Bo01 classification. 
Upper limits are mostly concentrated below the lowest detection values for Class II and III YSOs. 
We plot also the XLF of Class III YSOs (dots and dashed line) after excluding 15  
upper limits of Class~III YSO candidates from Bo01. 
In this way we try to correct the bias introduced by spurious members
that very likely contaminate the Class~III YSO sample as discussed above.
This {\em corrected} Class III XLF is similar to the XLF of Class~I YSOs.
The {\em corrected} Class III XLF shows higher luminosity levels with respect to Class II XLF.
Two sample tests yield a probability of $\ge99\%$ for the two distributions of being different, 
thus suggesting that in $\rho$~Ophiuchi Class III~YSOs are more luminous than Class II YSOs.
We observe also that Class I YSOs have similar X-ray luminosity levels compared with Class~III objects.
The high X-ray detection rate and high X-ray luminosities among Class I objects are 
surprising.
For comparison, in ONC Prisinzano et al. (2008) have found that the X-ray 
luminosities of Class\,I are lower than those of more evolved Class\,II YSOs.
Our X-ray bright Class~I YSOs could be explained as an effect of different 
mass distributions in different samples but, 
given that we cannot estimate masses of Class I objects with our method, 
we cannot test this hypothesis.

\begin{table}
\begin{center}
\caption{X-ray detection rates and median of logarithm of X-ray luminosities  and plasma
temperatures for the different YSO groups following the classification given by 
Bontemps et al. (2001).}
\label{tab:yso_detectionrate}
\begin{tabular}{lccccc}\hline \hline
YSO Class & $N_{\rm DROXO,FOV}$ & $N_{\rm det}$ & detection & $\log \rm L_{\rm x, med}$ & median kT  \\ 
          &                     &               & fraction  & [erg/s]    &  keV     \\ \hline
I         & 11     & 9        & 82\,\% & 29.6  & 4.4\\
II        & 48     & 31       & 77\,\% & 28.8  & 3.1 \\
bona-fide III & 16     & 15   & 94\,\% & 29.7  & 2.4 \\
candidate III & 21     & 6    & 29\,\% & $\le28.3$ & -- \\ \hline 
Total      &  96 & 61 & 64\% & \\
\hline
\end{tabular} 
\end{center}
\end{table}

The distribution of mean plasma temperatures in Fig.~\ref{wttctt} (top right panel) 
shows that on average the emitting plasma in Class I YSOs is hotter than in Class II 
and Class III YSOs. The median 
plasma temperatures for Class I, II and III are reported in Table \ref{tab:yso_detectionrate}.
Two-sample tests give a probability of $99\%$ to reject the null hypothesis that the distributions 
of temperature for Class III and Class I YSOs are drawn from the same distribution, the probability
is  $\sim$98\% when comparing Class II and Class I temperatures, and $\sim90\%$
when comparing Class II and Class III YSOs temperature distributions.
As expected, the distributions of $N_\mathrm H$ show that the absorption is higher
on average in Class I YSOs than in Class II and III (see Fig. \ref{wttctt} bottom left 
and right panels), indicating circumstellar matter in Class~I objects.

\subsection{Source with a soft X-ray excess \label{pink}}
\begin{figure}
\begin{center}
\includegraphics[width=\columnwidth]{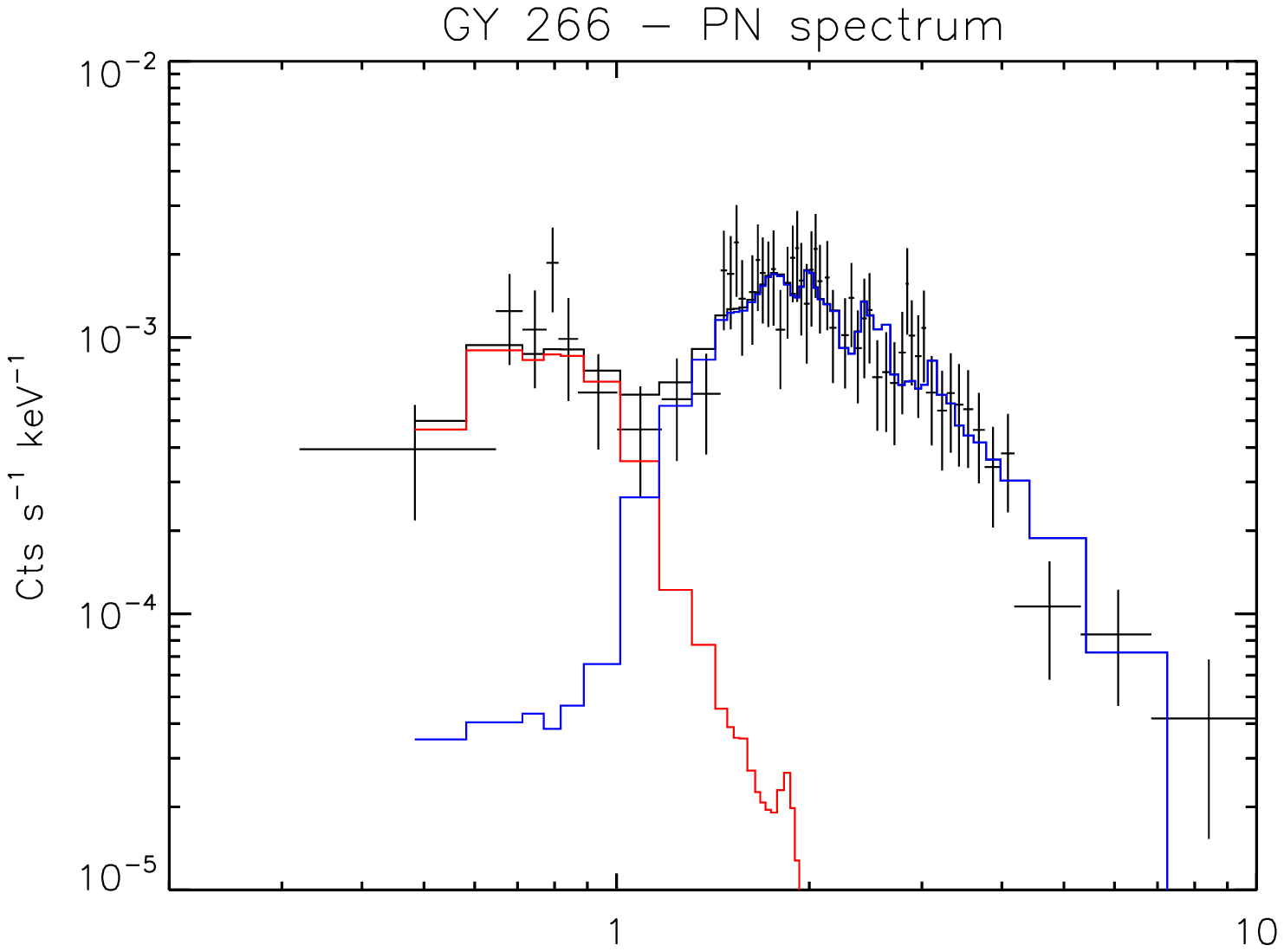}
\caption{\label{pinkspec} X-ray spectrum of source \# 61 (GY~266) with a soft component 
in the 0.3--1 keV energy range, less absorbed than the hot component.}
\end{center}
\end{figure}

{  We discuss the case of the source nr.~61, which shows
an excess of soft photons in the 0.3--1.0 keV band significantly different
from a typical coronal singly-absorbed multi-temperature plasma. 
This object has no counterpart in the ISOCAM survey, although in the literature is identified
with GY~266 and is classified as a variable star by \citet{deOliveira08}.
Furthermore, it has an IR counterpart in {\em 2MASS} and {\em Spitzer} catalogs.
From IRAC and MIPS photometry this object is cataloged as a normal star \citep{Evans03}. }
The DROXO spectrum of this object is shown in Fig. \ref{pinkspec}. 
We modeled the spectrum with two thermal components, which were differently absorbed. 
The soft component has kT = $0.47_{0.23}^{0.67}$ keV ($\sim5.5$ MK) and is 
absorbed by a N$_\mathrm H$ column of $\sim 3_{0.}^{34} \cdot 10^{20}$  cm$^{-2}$. 
This is one of the softest thermal components we have determined in the whole sample of DROXO spectra.
The hot component has kT = 1.8$_{1.4}^{2.4}$ keV absorbed by 
N$_\mathrm H \sim 2.5_{2.1}^{3.2}\cdot 10^{22}$  cm$^{-2}$. 
The emission measure ratio of soft to hot component is $\sim 1\%$. 
{  The hot component has the typical temperature and column density
of a YSO in $\rho$\,Oph, whereas the weakly absorbed component is unusually
soft.} 

\citet{Guedel07a} have reported similar examples of X-ray spectra
with soft excess from the XEST survey. They modeled them with a 2-T plasma
with individual absorptions and hypothesized that the soft component be
associated to the emission from shocks in jets. 
\citet{Bonito07} have studied the soft X-ray emission that could arise from
bipolar jets in YSOs through extensive MHD simulations. They find that a 
jet less dense than the ambient medium in which it propagates can emit soft X-rays 
with plasma temperatures of 2-3 MK, very similar to that of 
the soft component we found. 
{  Moreover, \citet{Bally03} and \citet{Favata06} observed that 
the X-ray emission of HH~154 jet is located at the base of the shock 
and that morphological changes are detected on a time scale of four years. 
The X-ray luminosity of that jet is $4\cdot10^{29}$ erg s$^{-1}$. In our case the unabsorbed 
luminosity of the soft component in the spectrum of GY~266 is much lower than that
of HH~154 (3.4$\cdot10^{27}$ erg s$^{-1}$). Given the angular resolution of EPIC we
cannot spatially resolve the soft component from the hot component.}
Thanks to angular resolution of Chandra  \citep{Guedel08} resolved separated 
soft emission from the coronal emission in the star DG Tau belonging to the Taurus Molecular Cloud. 
The coincidence of the elongated soft X-ray emission with the direction of 
the optical jet represents a clear case for the jet scenario in DG~Tau. 
In our case the lack of IR excess is difficult to reconcile with  
the circumstellar disk, which is expected along with the jet.
Explaining the origin of the X-ray soft component of GY~266 requires further 
investigation of this star.

\section{Discussion and conclusions}
{   We described the {\em Deep Rho Ophiuchi XMM-Newton Observation} (DROXO),
aimed at exploring at high sensitivity the X-ray emission of YSOs in the $\rho$ Ophiuchi 
core F region. We detected 111 X-rays sources, and for 91 of them we obtained a model fit to their
X-ray spectra. By using optical and IR data we estimated the photospheric parameters of
most of the sample of X-ray detected YSOs. 

We find 10 unidentified sources and a further 6 sources already detected in previous X-ray surveys,
but without optical/IR counterpart. 
Three of them (src. num. 45, 48 and 95) show light curves with some impulsive variability. The
spectra of src. num. 45 and 48 have a good fit with an absorbed power law, while
src. num. 95 has a spectrum compatible with a thermal model. 
The lack of 2MASS counterparts could be a hint that they are of extragalactic nature,
but, given the sensitivity of 2MASS catalog, it is not possible to rule out they 
are very low mass PMS stars or even brown dwarfs, especially src. num. 95.

The sensitivity of the survey is $f_\mathrm X \sim 5\cdot10^{-15}$ erg s$^{-1}$ cm$^{-2}$ 
in flux and log $L_\mathrm{X}$ (erg s$^{-1}) \sim 27.9$ in luminosity, but it strongly depends
on the local absorption, which is largely variable in the field-of-view.
The sample of 96 classified YSOs given by Bo01 in the DROXO field of view has allowed 
us to explore the X-ray emission from Class I, II, and III YSOs of
$\rho$~Ophiuchi. The X-ray properties of $\rho$~Ophiuchi PMS stars obtained from DROXO
were compared with those of the Orion Nebula Cloud PMS stars studied in the COUP survey.
When fitting the relation between mass and X-ray luminosity with a power law, the index 
that we find for DROXO sample is quite similar to that found for ONC and TMC \citep{Preibisch05,Telleschi07}.
With respect to COUP we find a lower normalization, suggesting that on average the PMS stars
in the range of masses between 0.1--2 $M_\odot$ in our sample are less luminous 
than their analogs in COUP.
 Young stellar objects in $\rho$~Ophiuchi exhibit a saturation of the ratio 
L$_\mathrm X / $L$_\mathrm{bol}$
near the ``canonical'' value of  $\sim 10^{-3}$ for masses between 0.5 and 1 M$_\odot$, while 
stars with masses below 0.5 M$\odot$ show a lower limit of saturation  
L$_\mathrm X / $L$_\mathrm{bol}\sim10^{-3.5}$.  This is consistent with what was found by 
\citet{Preibisch05} for the COUP sample.

We detected a large fraction of YSOs in the field of view, 94\% of which are
bona fide Class III stars, 77\% of Class~II and 82\% of Class~I YSOs, respectively. 
We confirm the high detection rate among Class~I YSOs found by \citet{Imanishi01}.
The detection rate in our Class I sample is higher than the analog rate found in the COUP 
survey by \citet{Prisinzano08} (82\% vs. 62\%) despite the higher sensitivity of 
COUP with respect to DROXO. 
\citet{Prisinzano08} make a distinction between 
Class Ia (characterized by a rising SED from $K$ to 8$\mu$m) and Class Ib (rising SED
up to 4.5$\mu$m). The detection rates in these two subclasses are 44\% and 82\% 
respectively. X-ray luminosities of Class Ib objects are higher 
than those of Class Ia objects.
The rate of X-ray detection of Class~Ib objects in Orion is 
identical to that of our sample of Class~I objects in $\rho$~Ophiuchi. 
With support of Spitzer photometry we suggest that only a small fraction of our Class I stars
are deeply embedded objects. From X-ray data the high detection rate suggests that 
our Class I sample  could be mainly formed by Class Ib YSOs as defined by Prisinzano et al. (2008).
These objects in Orion have X-ray luminosities more similar to Class II objects. 
In $\rho$~Oph we find that our nine detected Class I stars 
are on average more luminous than Class II stars. 

\citet{Bontemps01} list 39 stars defined as Class III candidates, 21 of which are surveyed in DROXO. 
Six out of 21 are detected in DROXO, providing  support for their PMS nature.
The other 15 are undetected below $\log L_\mathrm X = 28.3$.  Likely they are not PMS stars 
and thus not members of the $\rho$~Oph cloud.
By means of XLFs we evaluated the level of emission of Class~I, II, and III YSOs.
After excluding the suspect Class~III YSOs, we find that  Class I YSOs emit at the same 
level of Class III YSOs (median $\log L_\mathrm X \sim {29.6}$ 
and {29.7}, respectively) and both samples are more luminous than 
Class II YSOs  (median $\log L_\mathrm X$ $ \leq {28.8}$). 
In COUP Class III objects are more luminous than Class I and II 
for masses between 0.5 and 1.2 M$\odot$, while for masses in 0.1-0.5 M$\odot$ Class Ib 
sources are slightly more luminous than Class II and III 
(see Fig. 8 in \citealp{Prisinzano08}).

The analysis of X-ray spectra leads us to conclude that the mean plasma temperatures of 
Class~I protostars are higher than in Class~II and Class~III YSOs. 
We find a clear trend of decreasing plasma temperatures passing from Class I to Class III objects,
differently from what is observed in 
ONC where \citet{Prisinzano08} do not find a significant evolution of temperatures 
from Class~I to Class~III objects.
Four Class III stars and one Class II star  (GY~112, SR~12A, GY~296, GY~380 and GY~3)
have spectra noticeably softer than the rest of YSOs with mean temperatures around 0.7-0.8 keV. 

The absorption derived from X-ray spectra is higher in Class~I than in Class~II and III. 
Given the high absorption among Class~I YSOs, it is impossible to
determine if a soft thermal component in their spectra is completely absorbed or
is not present at all. This soft component would reduce the intrinsic $kT$ for
Class\,I objects, making them more similar to the ones in the ONC. However, higher
absorption should lead to an underestimate of the luminosities of heavily absorbed spectra 
like those of Class~I objects, while we observe
a luminosity for them higher than that of in Class~II and Class~III YSOs.
Furthermore, Prisinzano et al. (2008) have found in extensive simulations that 
varying $N_\mathrm H$ does not significantly influence the XLF.

The star GY~266 shows a peculiar X-ray spectrum composed by two thermal components 
that are differently absorbed. The hot and heavily absorbed component is similar to 
that found in other PMS stars, while the soft one, less absorbed, could arise from 
unresolved jet shocks.

\begin{acknowledgements}
The authors acknowledge an anonymous referee for the useful comments that improved
the paper.
IP, SS, EF, BS, GM, and FD acknowledge financial support from ASI/INAF contract nr. I/023/050.
LT acknowledges support from ASI-INAF I/016/07/0.
\end{acknowledgements}


\Online

\begin{appendix}
\section{Online tables}
\longtab{1}{\small
\begin{landscape}
\begin{longtable}{l l l l l l l l l l l l}
\caption{\label{tabdet}  List of sources detected in the DROXO EPIC images. 
The columns refer to X-ray positions, position accuracy, 
summed exposure times from the three EPIC cameras, count rates scaled to MOS detector units 
(see Sect. \ref{pwxdet})  with 1$\sigma$ errors, 
2MASS identifier, ISOCAM identifier, SED classification given by \citet{Bontemps01}, 
literature names, flag column for identifications (see footnote). 
Coordinate errors take into account position uncertainties calculated by the detection code 
and the best-match radius from the optical/IR objects. 
Class III candidates from Bo01 are indicated by ''III?".}\\
\hline\hline
DROXO source number & R.A.         & DEC         & Pos. Error       &   Off-axis   & Exp. Time     & Rate        & 2MASS id. & ISOCAM id.$^a$ & CLASS  & Name & Flag$^b$ \\ 
  & J2000        &      J2000  & $\arcsec$        &  $\arcmin$   &  ks           & ct/ks       &         &                &        &      &          \\
\hline
\endfirsthead
\caption{continued.}\\
\hline\hline
DROXO source number & R.A.         & DEC         & Pos. Error       &   Off-axis    & Exp. Time     & Rate        & 2MASS id. & ISOCAM id.$^a$ & CLASS  & Name & Flag$^b$ \\ 
  & J2000        &      J2000  & $\arcsec$        & $\arcmin$      &  ks           & ct/ks        &               &        &      &          \\
\hline
\endhead
\hline
\endfoot 
1&	    16:26:19.4   	&	   $-24$:37:29.0   	&3.5&13.3&137.6&  2.6  $\pm$    0.2&	   16261949$-24$37275   	&	 -- 	&	               	&	            	&	   S     \\
2&	    16:26:21.9   	&	   $-24$:44:39.1   	&3.1&13.2&155.8&  1.0  $\pm$    0.1&	   16262189$-24$44397   	&	 32 	&	          II   	&	            GY3  	&	  I \\
3&	    16:26:23.7   	&	   $-24$:43:14.3   	&1.3&12.4&207.7& 54.7  $\pm$    0.6&	   16262367$-24$43138   	&	 38 	&	          II   	&	    DoAr25/GY17  	&	  I \\
4&	    16:26:27.6   	&	   $-24$:41:53.9   	&1.2&11.3&249.9& 20.6  $\pm$    0.3&	   16262753$-24$41535   	&	 43 	&	          II   	&	           GY33   	&	  I\\
5&	    16:26:32.9   	&	   $-24$:49:55.9   	&7.9&14.0&199.4&  0.55  $\pm$    0.1&	                      	&	 -- 	&	               	&	          	&	  U       \\
6&	    16:26:35.3   	&	   $-24$:42:38.5   	&3.8& 9.8&297.6&  3.0  $\pm$    0.1&	                      	&	 -- 	&	               	&	             	&	  S     \\
7&	    16:26:40.9   	&	   $-24$:45:15.5   	&9.3& 9.6&299.3&  0.77  $\pm$    0.1&	                      	&	 -- 	&	               	&	          	&	  U       \\
8&	    16:26:44.3   	&	   $-24$:34:48.3   	&3.0& 9.1&305.3&  0.56  $\pm$    0.07&	   16264419$-24$34483   	&	 65 	&	           I   	&	     WL12/GY111  	&	  I \\
9&	    16:26:44.3   	&	   $-24$:43:14.9   	&1.2& 8.0&359.5& 11.8  $\pm$    0.2&	   16264429$-24$43141   	&	 66 	&	         III   	&	          GY112  	&	  I \\
10&	   16:26:44.4   	&	   $-24$:47:14.7   	&2.1&10.2&271.0&  1.8  $\pm$    0.1&	   16264441$-24$47138   	&	 -- 	&	               	&	           	&	  S       \\
11&	   16:26:45.3   	&	   $-24$:52:07.8   	&6.5&14.0&193.4&  2.8  $\pm$    0.2&	                      	&	 -- 	&	               	&	           	&	  S       \\
12&	   16:26:47.0   	&	   $-24$:50:51.5   	&9.7&12.7&233.5&  0.85  $\pm$    0.1&	                      	&	 -- 	&	               	&	            	&	  U       \\
13&	   16:26:47.0   	&	   $-24$:44:31.6   	&3.2& 8.1&257.3&  1.7  $\pm$    0.1&	   16264705$-24$44298   	&	 69 	&	         III   	&	          GY122   	&	  I\\
14&	   16:26:48.2   	&	   $-24$:42:02.8   	&3.1& 6.8&393.0&  1.3  $\pm$    0.09&	   16264810$-24$42033   	&	 -- 	&	               	&	          	&	  S      \\
15&	   16:26:48.5   	&	   $-24$:28:39.6   	&1.9&13.2&120.4& 24.4  $\pm$    0.6&	   16264848$-24$28389   	&	 70 	&	          II   	&	      WL2/GY128  	&	  I  \\
16&	   16:26:48.9   	&	   $-24$:53:11.7   	&9.6&14.5&222.6&  2.3  $\pm$    0.2&	                      	&	 -- 	&	               	&	           	&	  U       \\
17&	   16:26:49.0   	&	   $-24$:38:25.0   	&2.6& 6.5&396.8&  0.67  $\pm$    0.06&	   16264897$-24$38252   	&	 72 	&	          II   	&	     WL18/GY129  	&	  I \\
18&	   16:26:53.0   	&	   $-24$:43:29.9   	&8.3& 6.4&301.4&  0.3  $\pm$    0.06&	                      	&	 -- 	&	               	&	            	&	  S       \\
19&	   16:26:53.0   	&	   $-24$:54:01.4   	&2.3&14.9&218.9&  0.48  $\pm$    0.06&	                      	&	 -- 	&	               	&	           	&	  U       \\
20&	   16:26:54.0   	&	   $-24$:39:24.1   	&4.4& 5.2&443.3&  0.32  $\pm$    0.05&	                      	&	 -- 	&	               	&	           	&	  X       \\
21&	   16:26:56.5   	&	   $-24$:50:00.0   	&5.0&10.9&316.9&  0.38  $\pm$    0.06&	                      	&	 -- 	&	               	&	           	&	  S        \\
22&	   16:26:57.4   	&	   $-24$:35:39.7   	&2.5& 6.3&399.7&  0.72  $\pm$    0.06&	   16265733$-24$35388   	&	 84 	&	          II   	&	     WL21/GY164  	&	  I  \\
23&	   16:26:57.6   	&	   $-24$:46:07.6   	&1.7& 7.4&324.6&  0.28  $\pm$    0.04&	   16265752$-24$46060   	&	 -- 	&	               	&	         	&	  S          \\
24&	   16:26:57.7   	&	   $-24$:43:13.4   	&7.9& 5.3&483.2&  0.7  $\pm$    0.07&	                      	&	 -- 	&	               	&	          	&	  S         \\
25&	   16:26:58.6   	&	   $-24$:45:37.7   	&1.0& 6.9&443.7& 24.2  $\pm$    0.3&	   16265850$-24$45368   	&	 88 	&	          II   	&	           SR24N$-$SR24S  	&	  I  \\
26&	   16:26:58.7   	&	   $-24$:53:26.4   	&7.1&13.9&171.7&  0.77  $\pm$    0.1&	                      	&	 -- 	&	               	&	          	&	  U       \\
27&	   16:26:59.2   	&	   $-24$:35:01.7   	&1.9& 6.5&355.3&  2.0  $\pm$    0.1&	   16265916$-24$34588   	&	 90 	&	          II   	&	     WL22/GY174  	&	  I  \\
28&	   16:26:59.3   	&	   $-24$:36:00.6   	&4.2& 5.7&324.2&  0.69  $\pm$    0.09&	   16265904$-24$35568   	&	 89 	&	          II   	&	     WL14/GY172  	&	  I  \\
29&	   16:27:04.2   	&	   $-24$:45:59.7   	&4.9& 6.5&293.4&  0.36  $\pm$    0.05&	                      	&	 -- 	&	               	&	        	&	  X         \\
30&	   16:27:04.6   	&	   $-24$:43:00.6   	&1.1& 4.0&539.4& 12.6  $\pm$    0.2&	   16270451$-24$42596   	&	 96 	&	         III   	&	          GY193  	&	  I  \\
31&	   16:27:04.6   	&	   $-24$:42:13.6   	&1.2& 3.5&547.0&  8.38  $\pm$    0.1&	   16270456$-24$42140   	&	 97 	&	         III   	&	         GY194  	&	  I \\
32&	   16:27:05.0   	&	   $-24$:41:14.0   	&4.7& 2.9&376.3&  0.22  $\pm$    0.04&	                      	&	 -- 	&	               	&	             	&	  S    \\
33&	   16:27:06.6   	&	   $-24$:41:49.8   	&1.8& 2.9&304.4&  0.69  $\pm$    0.06&	   16270659$-24$41488   	&	 102 	&	          II   	&	          GY204  	&	  I \\
34&	   16:27:06.8   	&	   $-24$:38:16.4   	&2.3& 2.9&531.6&  0.75  $\pm$    0.05&	   16270677$-24$38149   	&	 103 	&	          II   	&	     WL17/GY205  	&	  I \\
35&	   16:27:09.2   	&	   $-24$:34:09.0   	&1.6& 6.2&338.4&  3.8  $\pm$    0.1&	   16270910$-24$34081   	&	 105 	&	          II   	&	     WL10/GY211  	&	  I \\
36&	   16:27:09.4   	&	   $-24$:43:20.3   	&1.6& 3.6&259.4&  3.8  $\pm$    0.1&	   16270931$-24$43196   	&	 -- 	&	               	&	              	&	  S     \\
37&	   16:27:09.4   	&	   $-24$:40:22.6   	&2.8& 1.7&592.4&  0.37  $\pm$    0.04&	   16270935$-24$40224   	&	 107 	&	          II   	&	          GY213  	&	  I  \\
38&	   16:27:09.5   	&	   $-24$:37:19.8   	&1.2& 3.2&455.1& 10.9  $\pm$    0.2&	   16270943$-24$37187   	&	 108 	&	           I   	&	     EL29/GY214  	&	  I \\
39&	   16:27:11.2   	&	   $-24$:40:47.0   	&1.7& 1.4&446.0&  2.1  $\pm$    0.09&	   16271117$-24$40466   	&	 112 	&	          II   	&	          GY224  	&	  I \\
40&	   16:27:11.7   	&	   $-24$:38:32.9   	&1.5& 2.0&572.4&  3.18  $\pm$    0.09&	   16271171$-24$38320   	&	 114 	&	         III   	&	     WL19/GY227  	&	  I \\
41&	   16:27:12.7   	&	   $-24$:40:52.9   	&2.9& 1.2&611.5&  0.084  $\pm$    0.02&	                      	&	 -- 	&	               	&	         	&	  X        \\
42&	   16:27:13.9   	&	   $-24$:43:33.6   	&3.0& 3.5&592.5&  0.29  $\pm$    0.04&	   16271382$-24$43316   	&	 117 	&	          II   	&	          GY235  	&	  I \\
43&	   16:27:15.1   	&	   $-24$:51:39.3   	&1.2&11.5&324.5& 54.4  $\pm$    0.5&	   16271513$-24$51388   	&	 -- 	&	               	&	      WSB 46  	&	  S      \\
44&	   16:27:15.5   	&	   $-24$:30:53.6   	&1.8& 9.2&269.8&  0.88  $\pm$    0.08&	   16271551$-24$30536   	&	 119 	&	          II   	&	    IRS35/GY238  	&	  I \\
45&	   16:27:15.7   	&	   $-24$:33:04.0   	&4.3& 7.1&329.9&  0.64  $\pm$    0.08&	                      	&	 -- 	&	               	&	           	&	  X       \\
46&	   16:27:15.9   	&	   $-24$:38:43.7   	&1.1& 1.4&597.1& 13.5  $\pm$    0.2&	   16271569$-24$38434   	&	 121 	&	          II   	&	     WL20/GY240  	&	  I  \\
47&	   16:27:16.4   	&	   $-24$:31:15.8   	&2.4& 8.9&281.2&  3.8  $\pm$    0.2&	   16271643$-24$31145   	&	 -- 	&	               	&	          	&	  S        \\
48&	   16:27:17.4   	&	   $-24$:36:25.6   	&4.7& 3.7&523.3&  0.35  $\pm$    0.05&	                      	&	 -- 	&	               	&	        	&	  X         \\
49&	   16:27:18.0   	&	   $-24$:28:53.4   	&2.0&11.2& 63.7& 60.9  $\pm$    2.&	   16271817$-24$28526   	&	 125 	&	         III   	&	      WL5/GY246  	&	  I \\
50&	   16:27:18.4   	&	   $-24$:54:55.2   	&2.3&14.8&225.9&  6.1  $\pm$    0.2&	   16271836$-24$54537   	&	 -- 	&	               	&	          	&	  S        \\
51&	   16:27:18.5   	&	   $-24$:39:15.4   	&1.8& 1.0&616.5&  1.4  $\pm$    0.06&	   16271838$-24$39146   	&	 127 	&	           I   	&	          GY245  	&	  I \\
52&	   16:27:18.5   	&	   $-24$:29:07.7   	&2.3&11.0& 58.4&  1.6  $\pm$    0.4&	   16271848$-24$29059	 &	 128$-$129 &	       II$-$II   	&	      WL3/GY249$-$WL4/GY247  	&	  I \\
53&	   16:27:19.6   	&	   $-24$:41:41.4   	&0.8& 1.7&628.2&100.0  $\pm$    0.5&	   16271951$-24$41403   	&	 130 	&	         III   	&	     SR12/GY250  	&	  I \\
54&	   16:27:21.5   	&	   $-24$:41:43.6   	&1.3& 1.9&609.0&  0.6  $\pm$    0.05&	   16272146$-24$41430   	&	 132 	&	          II   	&	    IRS42/GY252   	&	  I \\
55&	   16:27:21.9   	&	   $-24$:43:37.0   	&1.1& 3.7&615.8& 11.3  $\pm$    0.2&	   16272183$-24$43356   	&	 133 	&	         III   	&	          GY253   	&	  I \\
56&	   16:27:22.1   	&	   $-24$:29:54.1   	&1.9&10.3&232.9&  2.8  $\pm$    0.1&	   16272180$-24$29533   	&	 134 	&	           I   	&	      WL6/GY254  	&	  I  \\
57&	   16:27:23.0   	&	   $-24$:48:08.5   	&2.6& 8.1&454.9&  1.8  $\pm$    0.09&	   16272297$-24$48071   	&	 -- 	&	               	&	            	&	  S      \\
58&	   16:27:24.4   	&	   $-24$:50:56.4   	&6.6&10.9&342.7&  0.5  $\pm$    0.06&	                      	&	 -- 	&	               	&	          	&	  U       \\
59&	   16:27:24.7   	&	   $-24$:29:39.0   	&2.5&10.6&219.5&  5.0  $\pm$    0.2&	   16272463$-24$29353   	&	 -- 	&	               	&	               	&	  S   \\
60&	   16:27:26.5   	&	   $-24$:39:23.3   	&1.0& 2.3&614.9&  4.08  $\pm$    0.1&	   16272648$-24$39230   	&	 140 	&	          II   	&	          GY262  	&	  I \\
61&	   16:27:27.0   	&	   $-24$:32:17.4   	&2.7& 8.2&274.5&  2.7  $\pm$    0.1&	   16272706$-24$32175   	&	 -- 	&	               	&	    GY266         	&	  S     \\
62&	   16:27:27.1   	&	   $-24$:40:51.9   	&0.9& 2.5&644.4& 22.0  $\pm$    0.2&	   16272693$-24$40508   	&	 141 	&	           I   	&	    IRS43/GY265  	&	  I  \\
63&	   16:27:27.4   	&	   $-24$:31:17.1   	&2.1& 9.2&271.3&  5.6  $\pm$    0.2&	   16272738$-24$31165   	&	 142 	&	          II   	&	   VSSG25/GY267  	&	  I \\
64&	   16:27:28.1   	&	   $-24$:39:34.7   	&1.1& 2.6&617.5& 13.7  $\pm$    0.2&	   16272802$-24$39335   	&	 143 	&	           I   	&	    IRS44/GY269  	&	  I \\
65&	   16:27:28.3   	&	   $-24$:27:21.3   	&3.2&13.0& 57.5&  1.5  $\pm$    0.2&	   16272844$-24$27210   	&	 144 	&	          II   	&	    IRS45/GY273  	&	  I \\
66&	   16:27:28.5   	&	   $-24$:54:34.3   	&2.5&14.7&236.2&  3.9  $\pm$    0.2&	   16272873$-24$54317   	&	 -- 	&	               	&	           	&	  S       \\
67&	   16:27:29.5   	&	   $-24$:39:17.0   	&1.4& 3.0&612.4&  0.36  $\pm$    0.03&	   16272943$-24$39161   	&	 145 	&	           I   	&	    IRS46/GY274  	&	  I \\
68&	   16:27:30.0   	&	   $-24$:27:42.8   	&2.8&12.8& 20.2&  0.9  $\pm$    0.4&	   16272960$-24$27419   	&	 -- 	&	               	&	            	&	  S      \\
69&	   16:27:30.0   	&	   $-24$:33:37.2   	&1.7& 7.2&353.6&  2.9  $\pm$    0.1&	   16272996$-24$33365   	&	 146 	&	        III?   	&	          GY278  	&	  I  \\
70&	   16:27:30.6   	&	   $-24$:52:09.7   	&3.7&12.4&315.0&  3.3  $\pm$    0.1&	                      	&	 -- 	&	               	&	           	&	  S    \\
71&	   16:27:30.9   	&	   $-24$:47:27.7   	&1.1& 8.0&446.8& 12.7  $\pm$    0.2&	   16273084$-24$47268   	&	 149 	&	         III   	&	            	&	  S       \\
72&	   16:27:31.2   	&	   $-24$:34:04.3   	&2.0& 6.9&366.2&  1.7  $\pm$    0.09&	   16273105$-24$34032   	&	 148 	&	        III?   	&	          GY283  	&	  I \\
73&	   16:27:31.3   	&	   $-24$:31:11.1   	&2.3& 9.5&283.2&  0.52  $\pm$    0.06&	                      	&	 -- 	&	               	&	         	&	  S        \\
74&	   16:27:32.6   	&	   $-24$:44:59.6   	&3.6& 6.0&561.5&  0.21  $\pm$    0.04&	   16273272$-24$45004   	&	 -- 	&	               	&	         	&	  S        \\
75&	   16:27:32.8   	&	   $-24$:33:24.8   	&1.7& 7.6&343.0&  3.2  $\pm$    0.1&	   16273267$-24$33239   	&	 152 	&	         III   	&	          GY289  	&	  I \\
76&	   16:27:32.9   	&	   $-24$:32:35.5   	&2.2& 8.4&318.9&  4.7  $\pm$    0.2&	   16273285$-24$32348   	&	 154 	&	          II   	&	          GY291  	&	  I \\
77&	   16:27:33.4   	&	   $-24$:41:15.8   	&1.0& 3.9&477.4& 26.4  $\pm$    0.3&	   16273311$-24$41152   	&	 155 	&	          II   	&	          GY292   	&	  I \\
78&	   16:27:35.6   	&	   $-24$:38:34.1   	&2.0& 4.5&401.8&  1.5  $\pm$    0.09&	   16273526$-24$38334   	&	 156 	&	        III?   	&	          GY295  	&	  I \\
79&	   16:27:35.8   	&	   $-24$:45:34.1   	&1.8& 6.9&519.2&  1.5  $\pm$    0.07&	   16273566$-24$45325   	&	 157 	&	         III   	&	          GY296  	&	  I \\
80&	   16:27:37.4   	&	   $-24$:42:39.6   	&1.9& 5.3&443.8&  1.5  $\pm$    0.08&	   16273724$-24$42380   	&	 161 	&	          II   	&	          GY301  	&	  I \\
81&	   16:27:38.2   	&	   $-24$:30:43.1   	&3.2&10.6&263.6&  1.7  $\pm$    0.1&	   16273812$-24$30429   	&	 -- 	&	               	&	   IRS50/GY306    	&	  S \\
82&	   16:27:38.3   	&	   $-24$:37:01.6   	&1.1& 5.8&514.1& 11.8  $\pm$    0.2&	   16273832$-24$36585   	&	 163 	&	          II   	&	    IRS49/GY308    	&	  I \\
83&	   16:27:38.7   	&	   $-24$:38:39.5   	&1.7& 5.2&543.5&  0.64  $\pm$    0.05&	   16273863$-24$38391   	&	 164 	&	          II   	&	         GY310  	&	  I  \\
84&	   16:27:39.0   	&	   $-24$:40:19.6   	&1.8& 5.0&302.6&  0.73  $\pm$    0.07&	   16273894$-24$40206   	&	 165 	&	          II   	&	          GY312   	&	  I \\
85&	   16:27:39.1   	&	   $-24$:47:21.6   	&3.2& 8.8&403.6&  3.6  $\pm$    0.1&	                      	&	 -- 	&	               	&	             	&	  S     \\
86&	   16:27:39.3   	&	   $-24$:39:14.9   	&0.9& 5.2&396.4& 43.0  $\pm$    0.4&	   16273942$-24$39155   	&	 166 	&	          II   	&	          GY314  	&	  I \\
87&	   16:27:39.9   	&	   $-24$:43:13.6   	&1.3& 6.1&426.8&  8.84  $\pm$    0.2&	   16273982$-24$43150   	&	 167 	&	           I   	&	    IRS51/GY315 	&	  I  \\
88&	   16:27:40.6   	&	   $-24$:46:55.3   	&3.5& 8.7&444.5&  0.23  $\pm$    0.03&	                      	&	 -- 	&	               	&	          	&	  U       \\
89&	   16:27:41.6   	&	   $-24$:35:37.8   	&4.0& 7.2&362.8&  0.64  $\pm$    0.07&	   16274149$-24$35376   	&	 169 	&	        III?   	&	          GY322  	&	  I \\
90&	   16:27:42.8   	&	   $-24$:38:51.6   	&2.5& 6.0&521.6&  0.63  $\pm$    0.05&	   16274270$-24$38506   	&	 172 	&	          II   	&	          GY326  	&	  I \\
91&	   16:27:43.7   	&	   $-24$:31:27.7   	&2.9&10.6&269.3&  2.7  $\pm$    0.1&	                      	&	 -- 	&	               	&	            	&	  S      \\
92&	   16:27:44.4   	&	   $-24$:48:57.4   	&5.1&10.8&319.9&  1.1  $\pm$    0.08&	                      	&	 -- 	&	               	&	           	&	  S       \\
93&	   16:27:47.1   	&	   $-24$:45:35.4   	&1.5& 8.8&430.2&  3.77  $\pm$    0.1&	   16274709$-24$45350   	&	 177 	&	          II   	&	          GY352  	&	  I \\
94&	   16:27:49.2   	&	   $-24$:39:07.2   	&3.2& 7.4&448.3&  0.81  $\pm$    0.06&	                      	&	 -- 	&	               	&	          	&	  S        \\
95&	   16:27:50.3   	&	   $-24$:31:47.2   	&4.6&11.3&258.8&  1.1  $\pm$    0.1&	                      	&	 -- 	&	               	&	            	&	  X     \\
96&	   16:27:51.7   	&	   $-24$:47:45.3   	&7.0&11.0&382.7&  0.41  $\pm$    0.06&	                      	&	 -- 	&	               	&	          	&	  U       \\
97&	   16:27:51.9   	&	   $-24$:31:45.4   	&2.4&11.6&252.8&  0.41  $\pm$    0.06&	   16275180$-24$31455   	&	 182 	&	           I   	&	    IRS54/GY378  	&	  I  \\
98&	   16:27:52.0   	&	   $-24$:46:30.2   	&1.4&10.2&409.9&  6.18  $\pm$    0.1&	   16275191$-24$46296   	&	 183 	&	         III   	&	          GY377  	&	  I \\
99&	   16:27:52.2   	&	   $-24$:40:51.3   	&1.1& 8.1&284.0& 158    $\pm$    1&	   16275209$-24$40503   	&	 184 	&	         III   	&	    IRS55/GY380  	&	  I \\
100&	   16:27:55.8   	&	   $-24$:44:51.1   	&3.5&10.1&417.9&  0.88  $\pm$    0.08&	   16275565$-24$44509   	&	 186 	&	        III?   	&	      GY398  	&	  I  \\
101&	   16:27:57.9   	&	   $-24$:40:02.4   	&1.4& 9.3&418.9&  5.99  $\pm$    0.1&	   16275782$-24$40017   	&	 188 	&	         III   	&	          GY410  	&	  I  \\
102&	   16:27:60.0   	&	   $-24$:48:20.8   	&1.3&12.8&327.6&  8.51  $\pm$    0.2&	   16275996$-24$48193   	&	 -- 	&	               	&	             	&	  S     \\
103&	   16:27:60.0   	&	   $-24$:42:41.5   	&8.3&10.1&375.2&  0.8  $\pm$    0.08&	                      	&	 -- 	&	               	&	            	&	  S      \\
104&	   16:28:02.0   	&	   $-24$:49:55.7   	&6.5&14.2&263.8&  0.7  $\pm$    0.09&	                      	&	 -- 	&	               	&	            	&	  S      \\
105&	   16:28:04.7   	&	   $-24$:34:55.8   	&1.6&12.1&306.7& 16.9  $\pm$    0.3&	   16280464$-24$34560   	&	 191 	&	        III?   	&	          GY463  	&	  I  \\
106&	   16:28:04.8   	&	   $-24$:37:10.5   	&2.3&11.3&301.1&  3.9  $\pm$    0.1&	   16280478$-24$37100   	&	 -- 	&	               	&	           	&	  S       \\
107&	   16:28:05.0   	&	   $-24$:50:18.7   	&9.1&14.9& 51.4&  1.0  $\pm$    0.2&	                      	&	 -- 	&	               	&	               	&	  S        \\
108&	   16:28:07.0   	&	   $-24$:48:29.9   	&3.9&14.1&127.1&  1.4  $\pm$    0.1&	                      	&	 -- 	&	               	&	             	&	  S     \\
109&	   16:28:08.0   	&	   $-24$:45:10.9   	&3.5&12.7&235.9&  0.48  $\pm$    0.08&	   16280810$-24$45121   	&	 -- 	&	               	&	           	&	  S      \\
110&	   16:28:08.8   	&	   $-24$:48:41.5   	&5.5&14.6&122.9&  0.58  $\pm$    0.1&	                      	&	 -- 	&	               	&	          	&	  U      \\
111&	   16:28:10.0   	&	   $-24$:39:31.7   	&5.5&12.1&328.9&  0.28  $\pm$    0.05&	                      	&	 -- 	&	               	&	           	&	  S      \\

\end{longtable}
\noindent Notes. \newline
$^a$ Identifier from \citep{Bontemps01}. \newline
$^b$:{\sc I} indicates a match in the ISOCAM survey Bo01. 
IRS~50 (Src. num 81) is listed in Bo01, but undetected in the ISOCAM survey. 
{\sc S} indicates a match with one or more {\em Spitzer} objects. 
{\sc X} indicates a match only with X-ray sources detected in the surveys
\citet{Flaccomio06} or \citet{Ozawa05}. 
{\sc U} indicates a source without any X-ray, optical and IR counterpart.
\end{landscape}
}

\longtab{2}{
\begin{landscape}
\small
\begin{longtable}{l l l l l l l l l l l l l l l l l}
\caption{\label{t1fit}  X-ray Parameters from model fit to source spectra. 
The errors are quoted at the 90\% confidence region of best fit. 
Unabsorbed X-ray fluxes and luminosities are calculated in the 0.3-8 keV energy band. }\\
\hline\hline
DROXO Src. num. & CCDs & Instr& SNR & Model & N$_\mathrm H$ & kT 1 & log E. M. 1 & kT 2 & log E. M. 2& kT 3 & E. M. 3 & log $F_\mathrm X$ & log $L_\mathrm X$ & $\chi ^2$ & D.o.F. & Prob    \\
   & & & & & 10$^{22}$ cm$^{-2}$  & keV &  cm$^{-3}$ & keV & cm$^{-3}$ & keV & cm$^{-3}$ & erg s$^{-1}$ cm$^{-2}$  & erg s$^{-1}$ & & &  \\ 
\hline
\endfirsthead
\caption{continued.}\\
\hline\hline
DROXO src. num. & CCDs & Instr& SNR & Model & N$_\mathrm H$ & kT 1 & log E. M. 1 & kT 2 & log E. M. 2& kT 3 & log E. M. 3 & log $F_\mathrm X$ & log $L_\mathrm X$ & $\chi ^2$ & D.o.F. & Prob    \\
   & & & & & 10$^{22}$ cm$^{-2}$  & keV & cm$^{-3}$ & keV & cm$^{-3}$ & keV &  cm$^{-3}$ & erg s$^{-1}$ cm$^{-2}$  & erg s$^{-1}$ & & &  \\ 
\hline
\endhead
\hline
\endfoot
1	 & 	M2PN	 & 	ALL	 & 	L	 & 	1T	 & 	$ 0.14^{0.29}_{0.06} $	 & 	$ 0.77^{0.94}_{0.63} $	 & 	$ 51.42^{51.61}_{51.28} $	 & 	--	 & 	--	 & 	--	 & 	--	 & 	1.8e-14	 & 	28.5	 & 	48	 & 	37	 & 	0.11  \\
2	 & 	M1PN	 & 	ALL	 & 	L	 & 	1T	 & 	$ 0.8^{1.4}_{0.5} $	 & 	$ 0.7^{1.3}_{0.3} $	 & 	$ 51.67^{52.64}_{0} $	 & 	--	 & 	--	 & 	--	 & 	--	 & 	3.4e-14	 & 	28.8	 & 	14.5	 & 	10	 & 	0.15  \\
3	 & 	M1M2PN	 & 	PN	 & 	H	 & 	1T	 & 	$ 0.93^{0.97}_{0.89} $	 & 	$ 3.3^{3.5}_{3.1} $	 & 	$ 53.23^{53.25}_{53.21} $	 & 	--	 & 	--	 & 	--	 & 	--	 & 	1.4e-12	 & 	30.4	 & 	262.3	 & 	221	 & 	0.03  \\
4	 & 	M1M2PN	 & 	PN	 & 	H	 & 	1T	 & 	$ 8.4^{9.0}_{7.9} $	 & 	$ 4.5^{5.0}_{3.8} $	 & 	$ 53.58^{53.65}_{53.53} $	 & 	--	 & 	--	 & 	--	 & 	--	 & 	3.5e-12	 & 	30.8	 & 	132.4	 & 	125	 & 	0.31  \\
5	 & 	M1M2PN	 & 	ALL	 & 	L	 & 	1T	 & 	$ 0.09^{0.59}_{0.0} $	 & 	$ >10 $			 & 	$ 50.88^{51.00}_{0} $	 & 	--	 & 	--	 & 	--	 & 	--	 & 	7.3e-15	 & 	28.1	 & 	1.9	 & 	3	 & 	0.6  \\
6	 & 	M1M2PN	 & 	ALL	 & 	L	 & 	1T	 & 	$ 3.9^{4.8}_{2.9} $	 & 	$ 9.1^{46.0}_{4.9} $	 & 	$ 52.23^{52.35}_{52.15} $	 & 	--	 & 	--	 & 	--	 & 	--	 & 	1.8e-13	 & 	29.5	 & 	67.7	 & 	90	 & 	0.96  \\
8	 & 	M1M2PN	 & 	ALL	 & 	L	 & 	1T	 & 	$ 9.1^{28.0}_{4.4} $	 & 	$ >10 $	 		 & 	$ 51.78^{52.02}_{51.56} $	 & 	--	 & 	--	 & 	--	 & 	--	 & 	5.8e-14	 & 	29	 & 	9.7	 & 	8	 & 	0.28  \\
9	 & 	M1M2PN	 & 	ALL	 & 	H	 & 	2T	 & 	$ 0.38^{0.41}_{0.35} $	 & 	$ 0.7^{6.4}_{2.5} $	 & 	$ 52.21^{52.27}_{52.15} $	 & 	$ 0.70^{0.74}_{0.67} $	 & 	$ 52.10^{52.16}_{52.04} $	 & 	--	 & 	--	 & 	2e-13	 & 	29.5	 & 	168.2	 & 	144	 & 	0.082  \\
10	 & 	M1M2PN	 & 	ALL	 & 	H	 & 	1T	 & 	$ 0.44^{0.62}_{0.24} $	 & 	$ 0.55^{0.76}_{0.38} $	 & 	$ 51.66^{51.99}_{51.38} $	 & 	--	 & 	--	 & 	--	 & 	--	 & 	3.2e-14	 & 	28.7	 & 	21.3	 & 	12	 & 	0.046  \\
11	 & 	M2PN	 & 	ALL	 & 	H	 & 	1T	 & 	$ 0.7^{1.1}_{0.5} $	 & 	$ 14^{ }_{ } $	 & 	$ 51.76^{51.87}_{51.69} $	 & 	--	 & 	--	 & 	--	 & 	--	 & 	6.4e-14	 & 	29	 & 	11.4	 & 	9	 & 	0.25  \\
12	 & 	M2PN	 & 	ALL	 & 	L	 & 	1T	 & 	$ 2.5^{17}_{0.7} $	 & 	$ >10 $	 		 & 	$ 51.36^{51.94}_{0} $	 & 	--	 & 	--	 & 	--	 & 	--	 & 	2.3e-14	 & 	28.6	 & 	1.7	 & 	2	 & 	0.43  \\
13	 & 	M1M2PN	 & 	ALL	 & 	H	 & 	1T	 & 	$ 0.23^{0.35}_{0.16} $	 & 	$ 1.6^{2.1}_{1.3} $	 & 	$ 51.45^{51.55}_{51.36} $	 & 	--	 & 	--	 & 	--	 & 	--	 & 	1.9e-14	 & 	28.5	 & 	21.4	 & 	13	 & 	0.065  \\
14	 & 	M1M2PN	 & 	ALL	 & 	H	 & 	1T	 & 	$ 1.1^{1.7}_{0.3} $	 & 	$ 0.5^{1.3}_{0.2} $	 & 	$ 51.94^{53.26}_{0} $	 & 	--	 & 	--	 & 	--	 & 	--	 & 	6e-14	 & 	29	 & 	10.6	 & 	9	 & 	0.3  \\
15	 & 	PN	 & 	PN	 & 	H	 & 	1T	 & 	$ 7.7^{8.3}_{7.1} $	 & 	$ 6.0^{7.5}_{5.1} $	 & 	$ 53.39^{53.45}_{53.33} $	 & 	--	 & 	--	 & 	--	 & 	--	 & 	2.5e-12	 & 	30.6	 & 	85.4	 & 	74	 & 	0.17  \\
17	 & 	M1M2PN	 & 	ALL	 & 	L	 & 	1T	 & 	$ 1.1^{2.1}_{0.5} $	 & 	$ 2.2^{6.3}_{0.9} $	 & 	$ 51.37^{51.86}_{51.08} $	 & 	--	 & 	--	 & 	--	 & 	--	 & 	1.7e-14	 & 	28.5	 & 	6.2	 & 	11	 & 	0.86  \\
19	 & 	M2PN	 & 	PN	 & 	L	 & 	1T	 & 	$ 0.05^{0.0}_{0.45} $	 & 	$ 0.3^{0.41}_{0.18} $	 & 	$ 50.88^{50.66}_{52.15} $	 & 	--	 & 	--	 & 	--	 & 	--	 & 	5.1e-15	 & 	27.9	 & 	7.5	 & 	6	 & 	0.28  \\
22	 & 	M1M2PN	 & 	ALL	 & 	L	 & 	1T	 & 	$ 3.0^{4.2}_{2.0} $	 & 	$ 3.1^{7.4}_{1.9} $	 & 	$ 51.84^{52.11}_{51.58} $	 & 	--	 & 	--	 & 	--	 & 	--	 & 	5.5e-14	 & 	29	 & 	49.1	 & 	34	 & 	0.045  \\
23	 & 	M2PN	 & 	PN	 & 	L	 & 	1T	 & 	$ 0.05^{0.1}_{0.01} $	 & 	$ 2.7^{5.7}_{1.8} $	 & 	$ 51.11^{51.2}_{51.01} $	 & 	--	 & 	--	 & 	--	 & 	--	 & 	9.8e-15	 & 	28.2	 & 	18.8	 & 	17	 & 	0.34  \\
25	 & 	M1M2PN	 & 	ALL	 & 	H	 & 	1T	 & 	$ 0.95^{0.99}_{0.91} $	 & 	$ 2.6^{2.8}_{2.5} $	 & 	$ 52.94^{52.96}_{52.92} $	 & 	--	 & 	--	 & 	--	 & 	--	 & 	6.5e-13	 & 	30	 & 	402.5	 & 	367	 & 	0.098  \\
27	 & 	M1M2PN	 & 	M1PN	 & 	H	 & 	1T	 & 	$ 13.0^{17.0}_{8.8} $	 & 	$ 2.7^{5.2}_{1.8} $	 & 	$ 52.79^{53.21}_{44.24} $	 & 	--	 & 	--	 & 	--	 & 	--	 & 	4.8e-13	 & 	29.9	 & 	18.4	 & 	13	 & 	0.14  \\
28	 & 	M1M2PN	 & 	ALL	 & 	L	 & 	1T	 & 	$ 1.0^{2.0}_{0.4} $	 & 	$ 5.9^{ }_{} $		 & 	$ 51.47^{51.83}_{51.3} $	 & 	--	 & 	--	 & 	--	 & 	--	 & 	2.9e-14	 & 	28.7	 & 	27.4	 & 	20	 & 	0.12  \\
30	 & 	M1M2PN	 & 	PN	 & 	H	 & 	1T	 & 	$ 0.70^{0.75}_{0.67} $	 & 	$ 2.0^{2.1}_{1.9} $	 & 	$ 52.63^{52.66}_{52.61} $	 & 	--	 & 	--	 & 	--	 & 	--	 & 	2.9e-13	 & 	29.7	 & 	157.8	 & 	126	 & 	0.029  \\
31	 & 	M1M2PN	 & 	ALL	 & 	H	 & 	1T	 & 	$ 0.8^{0.86}_{0.75} $	 & 	$ 2.0^{2.1}_{1.9} $	 & 	$ 52.52^{52.55}_{52.49} $	 & 	--	 & 	--	 & 	--	 & 	--	 & 	2.3e-13	 & 	29.6	 & 	182.9	 & 	154	 & 	0.056  \\
33	 & 	M1M2	 & 	ALL	 & 	H	 & 	1T	 & 	$ 0.40^{0.74}_{0.25} $	 & 	$ 2.6^{4.3}_{1.6} $	 & 	$ 51.48^{51.66}_{51.35} $	 & 	--	 & 	--	 & 	--	 & 	--	 & 	2.2e-14	 & 	28.6	 & 	9	 & 	9	 & 	0.43  \\
34	 & 	M1M2PN	 & 	ALL	 & 	H	 & 	1T	 & 	$ 2.7^{4.3}_{1.5} $	 & 	$ 5.2^{}_{} $		 & 	$ 51.61^{51.92}_{51.41} $	 & 	--	 & 	--	 & 	--	 & 	--	 & 	3.9e-14	 & 	28.8	 & 	8	 & 	9	 & 	0.53  \\
35	 & 	M2PN	 & 	PN	 & 	H	 & 	1T	 & 	$ 2.8^{3.3}_{2.3} $	 & 	$ 6.0^{9.4}_{4.2} $	 & 	$ 52.4^{52.51}_{52.31} $	 & 	--	 & 	--	 & 	--	 & 	--	 & 	2.5e-13	 & 	29.6	 & 	34.7	 & 	27	 & 	0.15  \\
36	 & 	M1M2	 & 	ALL	 & 	L	 & 	1T	 & 	$ 1^{1.2}_{0.9} $	 & 	$ 3.8^{5.2}_{3.0} $	 & 	$ 52.03^{52.11}_{51.96} $	 & 	--	 & 	--	 & 	--	 & 	--	 & 	9.3e-14	 & 	29.2	 & 	91.2	 & 	94	 & 	0.56  \\
37	 & 	M1M2PN	 & 	ALL	 & 	L	 & 	1T	 & 	$ 5.3^{12}_{2.3} $	 & 	$ 1.2^{2.8}_{0.5} $	 & 	$ 52.04^{53.61}_{0} $	 & 	--	 & 	--	 & 	--	 & 	--	 & 	7.2e-14	 & 	29.1	 & 	10	 & 	8	 & 	0.27  \\
38	 & 	M1M2PN	 & 	M2	 & 	H	 & 	1T	 & 	$ 6.4^{5.8}_{7.1} $	 & 	$ 4.4^{3.8}_{5.3} $	 & 	$ 52.96^{52.87}_{53.04} $	 & 	--	 & 	--	 & 	--	 & 	--	 & 	1.1e-12	 & 	30.3	 & 	71.3	 & 	48	 & 	0.016  \\
39	 & 	M1M2PN	 & 	M1M2	 & 	H	 & 	1T	 & 	$ 4.1^{5.1}_{3.1} $	 & 	$ 6.6^{25.0}_{4.1} $	 & 	$ 53.2^{52.34}_{52.08} $	 & 	--	 & 	--	 & 	--	 & 	--	 & 	1.6e-13	 & 	29.4	 & 	14.8	 & 	14	 & 	0.39  \\
40	 & 	M1M2PN	 & 	ALL	 & 	L	 & 	1T	 & 	$ 10^{12}_{9} $	 	& 	$ 3.7^{4.8}_{3.0} $	 & 	$ 52.73^{52.86}_{52.60} $	 & 	--	 & 	--	 & 	--	 & 	--	 & 	4.6e-13	 & 	29.9	 & 	225.8	 & 	209	 & 	0.2  \\
43	 & 	M2PN	 & 	PN	 & 	H	 & 	3T	 & 	$ 0.26^{0.30}_{0.24} $	 	& 	$ 2.40^{2.66}_{2.17} $	 & 	$ 52.49^{52.52}_{52.46} $	 & 	$ 0.18^{0.21}_{0.17} $	 & 	$ 52.21^{52.41}_{52.00}$	 & 	$ 0.61^{0.64}_{0.58} $ 	 & 	$52.57^{52.63}_{52.52}  $	 & 	5.9e-13	 & 	30.0	 & 	284.3	 & 	245	 & 	0.02  \\
44	 & 	M2PN	 & 	ALL	 & 	H	 & 	1T	 & 	$ 4.5^{6.7}_{3.1} $	 & 	$ 2.8^{5.0}_{1.9} $	 & 	$ 52.24^{52.55}_{51.99} $	 & 	--	 & 	--	 & 	--	 & 	--	 & 	1.3e-13	 & 	29.4	 & 	18	 & 	9	 & 	0.035  \\
45	 & 	M2PN	 & 	ALL	 & 	L	 & 	1T	 & 	$ 7.2^{23.0}_{2.7} $	 & 	$ 5.4^{ }_{ } $		 & 	$ 51.81^{52.99}_{0} $	 	 & 	--	 & 	--	 & 	--	 & 	--	 & 	6.3e-14	 & 	29	 & 	19.5	 & 	12	 & 	0.078  \\
46	 & 	M1M2PN	 & 	PN	 & 	H	 & 	1T	 & 	$ 2.2^{2.3}_{2.1} $	 & 	$ 3.0^{3.2}_{2.8} $	 & 	$ 53.02^{53.05}_{53} $	  	 & 	--	 & 	--	 & 	--	 & 	--	 & 	8.3e-13	 & 	30.2	 & 	251.1	 & 	203	 & 	0.012  \\
47	 & 	M2PN	 & 	ALL	 & 	H	 & 	1T	 & 	$ 7.4^{9.4}_{6.1} $	 & 	$ 3.9^{5.5}_{2.7} $	 & 	$ 52.71^{52.91}_{52.56} $	 & 	--	 & 	--	 & 	--	 & 	--	 & 	4.4e-13	 & 	29.9	 & 	31	 & 	24	 & 	0.15  \\
49	 & 	M2	 & 	ALL	 & 	H	 & 	1T	 & 	$ 6.5^{7.1}_{6.0} $	 & 	$ 4.5^{5.5}_{3.7} $	 & 	$ 53.65^{53.74}_{53.58} $	 & 	--	 & 	--	 & 	--	 & 	--	 & 	4.1e-12	 & 	30.8	 & 	61.4	 & 	59	 & 	0.39  \\
50	 & 	M2PN	 & 	ALL	 & 	H	 & 	1T	 & 	$ 0.04^{0.07}_{0.02} $	 & 	$ 0.91^{0.97}_{0.75} $	 & 	$ 51.57^{51.61}_{51.52} $	 & 	--	 & 	--	 & 	--	 & 	--	 & 	2.6e-14	 & 	28.7	 & 	21.1	 & 	19	 & 	0.33  \\
51	 & 	M1M2PN	 & 	ALL	 & 	L	 & 	1T	 & 	$ 11^{14}_{8} $	 	& 	$ 3.8^{6.4}_{2.5} $	 & 	$ 52.31^{52.62}_{52.09} $	 & 	--	 & 	--	 & 	--	 & 	--	 & 	1.8e-13	 & 	29.5	 & 	71.9	 & 	65	 & 	0.26  \\
52	 & 	M2	 & 	ALL	 & 	H	 & 	1T	 & 	$ 3.6^{4.5}_{2.9} $	 & 	$ 2.0^{2.6}_{1.6} $	 & 	$ 52.96^{53.15}_{52.80} $	 & 	--	 & 	--	 & 	--	 & 	--	 & 	6.3e-13	 & 	30	 & 	15	 & 	8	 & 	0.059  \\
53	 & 	M1M2PN	 & 	PN	 & 	H	 & 	3T	 & 	$ 0.090^{0.100}_{0.087} $	& $ 0.41^{0.42}_{0.40} $ & 	$ 52.98^{53.03}_{52.93} $	 & 	$ 3.5^{4.3}_{3.1} $	 & 	$ 52.50^{52.55}_{52.40} $	 & 	$ 0.91^{0.94}_{0.88} $	 & 	$ 52.94^{52.99}_{52.9} $	 & 	8e-13	 & 	30.1	 & 	587.8	 & 	433	 & 	1e-06  \\
54	 & 	M1M2PN	 & 	M1PN	 & 	H	 & 	1T+3T	 & 	$ 3.5^{4.3}_{2.9} $	 & 	$ 2.5^{3.3}_{2.0} $	 & 	$ 52.26^{52.41}_{52.11} $	 & 	--	 & 	--	 & 	--	 & 	--	 & 	1.3e-13	 & 	29.4	 & 	29.9	 & 	33	 & 	0.62  \\
55	 & 	M1M2PN	 & 	PN	 & 	L	 & 	1T	 & 	$ 3.4^{3.6}_{3.2} $	 & 	$ 2.8^{3.0}_{2.5} $	 & 	$ 52.98^{53.04}_{52.95} $	 & 	--	 & 	--	 & 	--	 & 	--	 & 	7.4e-13	 & 	30.1	 & 	509.2	 & 	531	 & 	0.74  \\
56	 & 	M2PN	 & 	ALL	 & 	H	 & 	1T	 & 	$ 8.2^{9.4}_{7.1} $	 & 	$ 5.8^{8.1}_{4.5} $	 & 	$ 52.81^{52.91}_{52.72} $	 & 	--	 & 	--	 & 	--	 & 	--	 & 	6.4e-13	 & 	30	 & 	53.2	 & 	37	 & 	0.041  \\
57	 & 	M2PN	 & 	ALL	 & 	H	 & 	1T	 & 	$ 0.29^{0.39}_{0.22} $	 & 	$ 1.6^{1.9}_{1.3} $	 & 	$ 51.42^{51.5}_{51.34} $	 & 	--	 & 	--	 & 	--	 & 	--	 & 	1.7e-14	 & 	28.5	 & 	6.8	 & 	9	 & 	0.66  \\
59	 & 	M2PN	 & 	ALL	 & 	H	 & 	1T	 & 	$ 2.6^{3.1}_{2.1} $	 & 	$ 5.2^{9.1}_{3.8} $	 & 	$ 52.39^{52.49}_{52.28} $	 & 	--	 & 	--	 & 	--	 & 	--	 & 	2.4e-13	 & 	29.6	 & 	25.7	 & 	22	 & 	0.26  \\
60	 & 	M1M2PN	 & 	M2PN	 & 	H	 & 	1T	 & 	$ 3.3^{3.4}_{3.1} $	 & 	$ 4.3^{4.7}_{4.0} $	 & 	$ 53.01^{53.04}_{52.98} $	 & 	--	 & 	--	 & 	--	 & 	--	 & 	9.2e-13	 & 	30.2	 & 	164.6	 & 	165	 & 	0.5  \\
61	 & 	M1M2PN	 & 	PN	 & 	L	 & 	2NH	 & 	$ 2.5^{3.2}_{2.1} $	 & 	$ 1.8^{2.4}_{1.4} $	 & 	$ 52.33^{52.5}_{52.19} $	 & 	--	 & 	--	 & 	--	 & 	--	 & 	1.4e-13	 & 	29.4	 & 	41.5	 & 	51	 & 	0.83  \\
62	 & 	M1M2PN	 & 	PN	 & 	L	 & 	1T	 & 	$ 4.4^{4.6}_{4.3} $	 & 	$ 3.2^{3.3}_{3.0} $	 & 	$ 53.29^{53.32}_{53.26} $	 & 	--	 & 	--	 & 	--	 & 	--	 & 	1.6e-12	 & 	30.4	 & 	835.5	 & 	854	 & 	0.67  \\
63	 & 	M2PN	 & 	ALL	 & 	H	 & 	1T	 & 	$ 0.75^{0.89}_{0.65} $	 & 	$ 2.7^{3.3}_{2.2} $	 & 	$ 52.17^{52.24}_{52.12} $	 & 	--	 & 	--	 & 	--	 & 	--	 & 	1.1e-13	 & 	29.3	 & 	48	 & 	34	 & 	0.057  \\
64	 & 	M1M2PN	 & 	PN	 & 	H	 & 	1T	 & 	$ 7.2^{7.4}_{7} $	 & 	$ 5.8^{6.2}_{5.4} $	 & 	$ 53.4^{53.42}_{53.37} $	 & 	--	 & 	--	 & 	--	 & 	--	 & 	2.5e-12	 & 	30.6	 & 	393.2	 & 	354	 & 	0.074  \\
65	 & 	M2	 & 	ALL	 & 	L	 & 	1T	 & 	$ 1.0^{1.9}_{0.3} $	 & 	$ 4.8^{ }_{ } $	 	 & 	$ 51.61^{52}_{51.36} $	 	 & 	--	 & 	--	 & 	--	 & 	--	 & 	3.8e-14	 & 	28.8	 & 	4.3	 & 	4	 & 	0.36  \\
66	 & 	M2PN	 & 	M2	 & 	H	 & 	1T	 & 	$ \sim 0 $	 & 	$ 0.6^{0.74}_{0.47} $	 	 & 	$ 51.3^{51.43}_{51.23} $	 & 	--	 & 	--	 & 	--	 & 	--	 & 	1.4e-14	 & 	28.4	 & 	5	 & 	4	 & 	0.29  \\
67	 & 	M1M2PN	 & 	PN	 & 	L	 & 	1T	 & 	$ 6.1^{9.5}_{4.4} $	 & 	$ 6.6^{14.0}_{3.7} $	 & 	$ 51.82^{52.07}_{51.67} $	 & 	--	 & 	--	 & 	--	 & 	--	 & 	6.8e-14	 & 	29.1	 & 	38.3	 & 	25	 & 	0.043  \\
68	 & 	M2	 & 	M2	 & 	H	 & 	1T	 & 	$ 1.9^{2.8}_{1.3} $	 & 	$ 2.2^{4.1}_{1.3} $	 & 	$ 52.3^{52.57}_{52.06} $	 & 	--	 & 	--	 & 	--	 & 	--	 & 	1.4e-13	 & 	29.4	 & 	0.1	 & 	3	 & 	0.99  \\
69	 & 	M2PN	 & 	ALL	 & 	H	 & 	1T	 & 	$ 6.3^{7.4}_{5.6} $	 & 	$ 5.0^{6.5}_{3.8} $	 & 	$ 52.64^{52.76}_{52.56} $	 & 	--	 & 	--	 & 	--	 & 	--	 & 	4.2e-13	 & 	29.9	 & 	53.7	 & 	36	 & 	0.029  \\
71	 & 	M1M2PN	 & 	ALL	 & 	H	 & 	1T	 & 	$ 0.81^{0.85}_{0.77} $	 & 	$ 2.7^{2.9}_{2.5} $	 & 	$ 52.67^{52.69}_{52.65} $	 & 	--	 & 	--	 & 	--	 & 	--	 & 	3.5e-13	 & 	29.8	 & 	220.2	 & 	197	 & 	0.12  \\
72	 & 	M2PN	 & 	ALL	 & 	H	 & 	1T	 & 	$ 0.82^{1.1}_{0.63} $	 & 	$ 2.2^{3.0}_{1.6} $	 & 	$ 51.9^{52.03}_{51.8} $	 	 & 	--	 & 	--	 & 	--	 & 	--	 & 	5.6e-14	 & 	29	 & 	10.2	 & 	14	 & 	0.75  \\
75	 & 	M2PN	 & 	ALL	 & 	H	 & 	1T	 & 	$ 1.9^{2.2}_{1.7} $	 & 	$ 3.2^{3.9}_{2.6} $	 & 	$ 52.35^{52.43}_{52.28} $	 & 	--	 & 	--	 & 	--	 & 	--	 & 	1.8e-13	 & 	29.5	 & 	35.2	 & 	30	 & 	0.24  \\
76	 & 	M2PN	 & 	ALL	 & 	H	 & 	1T	 & 	$ 2.1^{2.4}_{1.8} $	 & 	$ 3.0^{3.8}_{2.4} $	 & 	$ 52.39^{52.48}_{52.31} $	 & 	--	 & 	--	 & 	--	 & 	--	 & 	1.9e-13	 & 	29.5	 & 	28	 & 	29	 & 	0.52  \\
77	 & 	M1M2PN	 & 	M1M2	 & 	H	 & 	2T	 & 	$ 1.9^{2.0}_{1.8} $	 & 	$ 0.74^{0.86}_{0.63} $	 & 	$ 53.22^{53.35}_{53.09} $	 & 	$ 3.0^{3.3}_{2.7} $	 & 	$ 53.08^{53.13}_{53} $	 & 	--	 & 	--	 & 	2.1e-12	 & 	30.6	 & 	270.6	 & 	253	 & 	0.21  \\
78	 & 	M1M2PN	 & 	M1M2	 & 	H	 & 	1T	 & 	$ 0.32^{0.45}_{0.23} $	 & 	$ 2.6^{3.7}_{2.0} $	 & 	$ 51.65^{51.73}_{51.56} $	 & 	--	 & 	--	 & 	--	 & 	--	 & 	3.3e-14	 & 	28.8	 & 	12.3	 & 	10	 & 	0.27  \\
79	 & 	M1M2PN	 & 	ALL	 & 	L	 & 	1T	 & 	$ 1.0^{1.1}_{0.9} $	 & 	$ 0.7^{0.8}_{0.6} $	 & 	$ 52.03^{52.13}_{51.91} $	 & 	--	 & 	--	 & 	--	 & 	--	 & 	7.7e-14	 & 	29.1	 & 	91.9	 & 	82	 & 	0.21  \\
80	 & 	M1M2PN	 & 	M2	 & 	L	 & 	1T	 & 	$ 2.6^{3.3}_{2.0} $	 & 	$ 3.7^{6.4}_{2.5} $	 & 	$ 52.14^{52.29}_{51.98} $	 & 	--	 & 	--	 & 	--	 & 	--	 & 	1.2e-13	 & 	29.3	 & 	24.2	 & 	29	 & 	0.72  \\
81	 & 	M2PN	 & 	ALL	 & 	H	 & 	1T	 & 	$ 1.3^{1.6}_{1.0} $	 & 	$ 2.1^{3.0}_{1.5} $	 & 	$ 52.14^{52.29}_{52} $	 	 & 	--	 & 	--	 & 	--	 & 	--	 & 	9.5e-14	 & 	29.2	 & 	15.5	 & 	12	 & 	0.21  \\
82	 & 	M1M2PN	 & 	ALL	 & 	H	 & 	1T	 & 	$ 1.6^{1.7}_{1.6} $	 & 	$ 2.7^{2.9}_{2.6} $	 & 	$ 52.86^{52.89}_{52.83} $	 & 	--	 & 	--	 & 	--	 & 	--	 & 	5.5e-13	 & 	30	 & 	291.4	 & 	239	 & 	0.012  \\
83	 & 	M1M2PN	 & 	M2PN	 & 	H	 & 	1T	 & 	$ 0.7^{0.9}_{0.6} $	 & 	$ 3.2^{4.7}_{2.3} $	 & 	$ 51.78^{51.87}_{51.69} $	 & 	--	 & 	--	 & 	--	 & 	--	 & 	4.8e-14	 & 	28.9	 & 	17.9	 & 	15	 & 	0.27  \\
84	 & 	M1M2PN	 & 	ALL	 & 	H	 & 	1T	 & 	$ 4.5^{ }_{ } $	 		& $ 27^{ }_{ } $	 & 	$ 51.88^{ }_{ } $	  	 & 	--	 & 	--	 & 	--	 & 	--	 & 	8.4e-14	 & 	29.2	 & 	31	 & 	14	 & 	0.0055  \\
85	 & 	M1M2PN	 & 	PN	 & 	H	 & 	1T	 & 	$ 2.5^{3.2}_{2.1} $	 & 	$ >10 $	 		 & 	$ 52.19^{52.24}_{52.15} $	 & 	--	 & 	--	 & 	--	 & 	--	 & 	1.7e-13	 & 	29.5	 & 	17.7	 & 	19	 & 	0.55  \\
86	 & 	M1M2PN	 & 	M1M2	 & 	H	 & 	2T	 & 	$ 1.7^{1.8}_{1.6} $	 & 	$ 0.95^{1.20}_{0.68} $	 & 	$ 52.75^{52.96}_{52.45} $	 & 	$ 3.0^{3.2}_{2.8} $	 & 	$ 53.28^{53.31}_{53.20} $	 & 	--	 & 	--	 & 	1.9e-12	 & 	30.5	 & 	379.4	 & 	322	 & 	0.15  \\
87	 & 	M1M2PN	 & 	M1	 & 	H	 & 	1T	 & 	$ 3^{3.4}_{2.7} $	 & 	$ 3.8^{4.6}_{3.2} $	 & 	$ 52.86^{52.94}_{52.79} $	 & 	--	 & 	--	 & 	--	 & 	--	 & 	6.3e-13	 & 	30	 & 	40.4	 & 	37	 & 	0.32  \\
88	 & 	M1M2PN	 & 	ALL	 & 	L	 & 	1T	 & 	$ 4.6^{46}_{1.6} $	 & 	$ >10 $			 & 	$ 51.38^{54.28}_{51.14} $	 & 	--	 & 	--	 & 	--	 & 	--	 & 	2.7e-14	 & 	28.7	 & 	5.5	 & 	6	 & 	0.48  \\
91	 & 	M2PN	 & 	PN	 & 	L	 & 	1T	 & 	$ 4.5^{5.5}_{3.1} $	 	& 	$ >10 $		 & 	$ 52.25^{52.34}_{52.18} $	 & 	--	 & 	--	 & 	--	 & 	--	 & 	2e-13	 & 	29.5	 & 	48.5	 & 	45	 & 	0.33  \\
92	 & 	M1M2PN	 & 	ALL	 & 	L	 & 	1T	 & 	$ 2.0^{3.3}_{1.3} $	 & 	$ 17 $	 		 & 	$ 51.60^{51.81}_{51.52} $	 & 	--	 & 	--	 & 	--	 & 	--	 & 	4.5e-14	 & 	28.9	 & 	21.8	 & 	27	 & 	0.75  \\
93	 & 	M1M2PN	 & 	ALL	 & 	H	 & 	1T	 & 	$ 1.8^{2.0}_{1.6} $	 & 	$ 3.9^{4.8}_{3.2} $	 & 	$ 52.27^{52.34}_{52.21} $	 & 	--	 & 	--	 & 	--	 & 	--	 & 	1.6e-13	 & 	29.4	 & 	48.8	 & 	42	 & 	0.22  \\
94	 & 	M1M2PN	 & 	ALL	 & 	L	 & 	1T	 & 	$ 1.8^{2.6}_{1.3} $	 & 	$ >10 $	 		& 	$ 51.70^{51.77}_{51.59} $	 & 	--	 & 	--	 & 	--	 & 	--	 & 	4.8e-14	 & 	28.9	 & 	52.8	 & 	32	 & 	0.012  \\
95	 & 	M2PN	 & 	ALL	 & 	L	 & 	1T	 & 	$ 3.6^{7.0}_{2.0} $	 & 	$ >10 $	 		& 	$ 51.75^{52.19}_{51.59} $	 & 	--	 & 	--	 & 	--	 & 	--	 & 	5.4e-14	 & 	29	 & 	21.6	 & 	16	 & 	0.16  \\
96	 & 	M1M2PN	 & 	ALL	 & 	L	 & 	1T	 & 	$ 2.5^{12.0}_{0.3} $	 & 	$ 4.2 $			& 	$ 51.33^{53.47}_{0} $	 & 	--	 & 	--	 & 	--	 & 	--	 & 	1.9e-14	 & 	28.5	 & 	0.6	 & 	3	 & 	0.9  \\
97	 & 	M2PN	 & 	ALL	 & 	L	 & 	1T	 & 	$ 19^{39}_{6} $	 	& 	$ 2.5 $			 & 	$ 52.49^{53.88}_{0} $	 & 	--	 & 	--	 & 	--	 & 	--	 & 	2.3e-13	 & 	29.6	 & 	2	 & 	6	 & 	0.92  \\
98	 & 	M1M2PN	 & 	ALL	 & 	H	 & 	1T	 & 	$ 1.5^{1.6}_{1.4} $	 & 	$ 2.2^{2.4}_{2.0} $	 & 	$ 52.57^{52.62}_{52.53} $	 & 	--	 & 	--	 & 	--	 & 	--	 & 	2.6e-13	 & 	29.7	 & 	83.2	 & 	75	 & 	0.24  \\
99	 & 	M1M2PN	 & 	M1	 & 	H	 & 	2T	 & 	$ 0.9^{1}_{0.8} $	 & 	$ 0.8^{4.6}_{3.2} $	 & 	$ 53.02^{53.23}_{52.79} $	 & 	$ 0.77^{0.89}_{0.69} $	 & 	$ 53.43^{53.45}_{53.41} $	 & 	--	 & 	--	 & 	3e-12	 & 	30.7	 & 	245.8	 & 	209	 & 	0.041  \\
100	 & 	M1M2PN	 & 	ALL	 & 	H	 & 	1T	 & 	$ 0.19^{0.29}_{0.09} $	 & 	$ 3.1^{7.4}_{1.9} $	 & 	$ 51.14^{51.24}_{51.02} $	 & 	--	 & 	--	 & 	--	 & 	--	 & 	1.1e-14	 & 	28.3	 & 	2.8	 & 	8	 & 	0.95  \\
101	 & 	M1M2PN	 & 	ALL	 & 	L	 & 	1T	 & 	$ 1.5^{1.6}_{1.4} $	 & 	$ 2.6^{2.9}_{2.4} $	 & 	$ 52.72^{52.76}_{52.68} $	 & 	--	 & 	--	 & 	--	 & 	--	 & 	3.9e-13	 & 	29.8	 & 	443.7	 & 	474	 & 	0.84  \\
102	 & 	M1M2PN	 & 	ALL	 & 	H	 & 	2T	 & 	$ 0.47^{0.57}_{0.42} $	 & 	$ 0.7^{7.4}_{1.9} $	 & 	$ 52.04^{52.16}_{51.93} $	 & 	$ 0.73^{0.80}_{0.58} $	 & 	$ 52.20^{52.26}_{52.15} $	 & 	--	 & 	--	 & 	1.9e-13	 & 	29.5	 & 	127.7	 & 	100	 & 	0.032  \\
104	 & 	M1M2PN	 & 	ALL	 & 	L	 & 	1T	 & 	$ 0.9^{2.0}_{0.5} $	 & 	$ >10 $	 & 	$ 51.47^{51.58}_{51.27} $	 & 	--	 & 	--	 & 	--	 & 	--	 & 	2.8e-14	 & 	28.7	 & 	15.8	 & 	12	 & 	0.2  \\
105	 & 	M1M2PN	 & 	ALL	 & 	L	 & 	1T	 & 	$ 2.6^{2.7}_{2.5} $	 & 	$ 5.9^{6.4}_{5.5} $	 & 	$ 53.36^{53.38}_{53.34} $	 & 	--	 & 	--	 & 	--	 & 	--	 & 	2.3e-12	 & 	30.6	 & 	1521.9	 & 	1391	 & 	0.0078  \\
106	 & 	M1M2PN	 & 	ALL	 & 	L	 & 	1T	 & 	$ 1.2^{1.5}_{1..0} $	 & 	$ 3.3^{4.4}_{2.6} $	 & 	$ 52.13^{52.22}_{52.05} $	 & 	--	 & 	--	 & 	--	 & 	--	 & 	1.1e-13	 & 	29.3	 & 	121	 & 	107	 & 	0.17  \\
107	 & 	M1	 & 	ALL	 & 	L	 & 	1T	 & 	$ 1.5^{4.2}_{0.4} $	 & 	$ 2.2 $			 & 	$ 51.75^{52.47}_{0} $	 & 	--	 & 	--	 & 	--	 & 	--	 & 	4e-14	 & 	28.8	 & 	0.6	 & 	2	 & 	0.76  \\
108	 & 	M1M2	 & 	ALL	 & 	H	 & 	1T	 & 	$ 5.3^{10.0}_{2.4} $	 & 	$ 3.9^{0}_{1.6} $	 & 	$ 52.17^{52.79}_{0} $	 & 	--	 & 	--	 & 	--	 & 	--	 & 	1.3e-13	 & 	29.4	 & 	4.4	 & 	5	 & 	0.5  \\
109	 & 	M1M2PN	 & 	ALL	 & 	L	 & 	1T	 & 	$ 0.27^{1.40}_{0.01} $	 & 	$ 8.7 $			 & 	$ 51.05^{51.69}_{50.84} $	 & 	--	 & 	--	 & 	--	 & 	--	 & 	1.2e-14	 & 	28.3	 & 	6.2	 & 	7	 & 	0.52  \\
110	 & 	M1M2	 & 	ALL	 & 	L	 & 	1T	 & 	$ 1.4^{3.1}_{0.5} $	 & 	$ 9.1 $	 		& 	$ 51.51^{51.89}_{51.33} $	 & 	--	 & 	--	 & 	--	 & 	--	 & 	3.5e-14	 & 	28.8	 & 	5	 & 	6	 & 	0.54  \\

\end{longtable}
\noindent Notes. \newline 
Source 38 (Elias 29): Fit with variable abundance (Z/Z$_{COUP} = 3.2 \sim Z_\odot$). \newline
Source 53 (Sr12A):P$(\chi^2 > \chi^2_0)$ is low but the overall spectrum shape is well 
modeled with 3 thermal components. Abundance was a free parameter: $Z= 0.3 Z_{COUP}$ \newline
Source 54: spectrum contamination from source nr. 53, see Sect. \ref{src54} for details. \newline
Source 61: the best fit model is the sum of two differently absorbed {\sc APEC} models, 
the parameters in the table refer to the hot component. 
The soft component has parameters: N$_\mathrm H = 3_0^{34}\cdot 10^{20}$ cm$^{-2}$, 
kT $= 0.47_{0.23}^{0.67} $, log E.M. $= 50.47_{50.24}^{51.28}$ cm$^{-3}$,  
absorbed flux: f$_\mathrm X$ ($0.3-10.$ keV band) $= 1.67\cdot 10^{-15}$ erg cm$^{-2}$ s$^{-1}$,
unabsorbed flux: f$_\mathrm X$ ($0.3-10.$ keV band) $= 1.99\cdot 10^{-15}$ erg cm$^{-2}$ s$^{-1}$,
unabsorbed luminosity: L$_\mathrm X = 3.4\cdot10^{27}$ erg s$^{-1}$.

\end{landscape}
}

\longtab{3}{
\begin{longtable}{l l l l l l l l }
\caption{\label{upl}  Upper limits to count rate, X-ray flux and luminosity in the  0.3--10 keV for the 
ISOCAM YSOs \citep{Bontemps01} undetected in the DROXO field of view. Count rates are scaled to MOS detector
units as in Table \ref{tabdet} (see Sect. \ref{pwxdet}). {  Class III candidates from Bo01 are indicated 
by ''III?". Fluxes and luminosities are calculated by assuming a conversion factor from count rates of $cf = 6.1\cdot10^{-11}$ and a distance of 120~pc.
For ISOCAM nr. 92 (WL~16) we have taken into account the large absorption toward this star ($A_\mathrm V \sim 30$ mag, cf. Sect. 6.2) }}\\
\hline\hline
R.A.  & Dec. & ISOCAM nr. & Class & Lim. rate & Exposure time & log $F_\mathrm X$ &  log L$_\mathrm X$  \\
 J2000  & J2000  &  &  & ct/ks & ks & erg s$^{-1}$ cm$^{-2}$ & erg s$^{-1}$   \\ 
\hline
\endfirsthead
\caption{continued.}\\
\hline\hline
R.A.  & Dec. & ISOCAM nr. & Class & Lim. rate & Exposure time & log $F_\mathrm X$ &  log $L_\mathrm X$  \\
 J2000  & J2000  &  &  & ct/ks & ks & erg s$^{-1}$ cm$^{-2}$ &  erg s$^{-1}$   \\ 
\hline
\endhead
\hline
\endfoot
16:26:30.9	 & 	-24:31:07	 & 	47	 & 	III?	 & 	0.798	 & 	114.8	 & 	4.87e-14	 & 	28.92  \\
16:26:40.7	 & 	-24:30:53	 & 	55	 & 	III?	 & 	0.563	 & 	186.3	 & 	3.44e-14	 & 	28.77  \\
16:26:41.6	 & 	-24:40:15	 & 	56	 & 	II	 & 	0.292	 & 	350.6	 & 	1.78e-14	 & 	28.49  \\
16:26:42	 & 	-24:33:24	 & 	58	 & 	III	 & 	0.393	 & 	267.3	 & 	2.4e-14	 & 	28.62  \\
16:26:42.1	 & 	-24:31:02	 & 	59	 & 	II	 & 	0.456	 & 	225	 & 	2.78e-14	 & 	28.68  \\
16:26:52.1	 & 	-24:30:39	 & 	75	 & 	 II	 & 	0.41	 & 	254.7	 & 	2.5e-14	 & 	28.63  \\
16:26:53.6	 & 	-24:32:36	 & 	76	 & 	 II	 & 	0.348	 & 	305.8	 & 	2.12e-14	 & 	28.56  \\
16:26:56.7	 & 	-24:28:38	 & 	82	 & 	III?	 & 	0.514	 & 	188.6	 & 	3.14e-14	 & 	28.73  \\
16:26:58.3	 & 	-24:37:40	 & 	85	 & 	 II	 & 	0.328	 & 	300.9	 & 	2e-14	 & 	28.54  \\
16:27:02.5	 & 	-24:37:30	 & 	92	 & 	II	 & 	0.222	 & 	490.7	 & 	2.31e-13	 & 	29.60 \\
16:27:04.1	 & 	-24:28:33	 & 	95	 & 	II	 & 	1.163	 & 	67.7	 & 	7.09e-14	 & 	29.09  \\
16:27:05.4	 & 	-24:36:31	 & 	99	 & 	I	 & 	0.218	 & 	481.6	 & 	1.33e-14	 & 	28.36  \\
16:27:05.7	 & 	-24:40:12	 & 	100	 & 	III?	 & 	0.214	 & 	470.4	 & 	1.31e-14	 & 	28.35  \\
16:27:07.9	 & 	-24:40:27	 & 	104	 & 	III?	 & 	0.192	 & 	581.6	 & 	1.17e-14	 & 	28.31  \\
16:27:09.6	 & 	-24:29:55	 & 	109	 & 	III?	 & 	0.413	 & 	239.6	 & 	2.52e-14	 & 	28.64  \\
16:27:10.3	 & 	-24:33:22	 & 	111	 & 	III?	 & 	0.308	 & 	329	 & 	1.88e-14	 & 	28.51  \\
16:27:12.1	 & 	-24:34:48	 & 	115	 & 	 II	 & 	0.265	 & 	371.1	 & 	1.62e-14	 & 	28.45  \\
16:27:14.6	 & 	-24:26:55	 & 	118	 & 	 II	 & 	0.803	 & 	57.3	 & 	4.9e-14	 & 	28.93  \\
16:27:15.7	 & 	-24:26:46	 & 	120	 & 	II	 & 	1.953	 & 	38.8	 & 	1.19e-13	 & 	29.31  \\
16:27:24.3	 & 	-24:41:46	 & 	136	 & 	III?	 & 	0.215	 & 	651.6	 & 	1.31e-14	 & 	28.35  \\
16:27:24.8	 & 	-24:41:03	 & 	137	 & 	I	 & 	0.224	 & 	651.6	 & 	1.37e-14	 & 	28.37  \\
16:27:26.2	 & 	-24:42:45	 & 	139	 & 	 II	 & 	0.244	 & 	340.4	 & 	1.49e-14	 & 	28.41  \\
16:27:36.3	 & 	-24:28:34	 & 	158	 & 	III?	 & 	0.672	 & 	74.1	 & 	4.1e-14	 & 	28.85  \\
16:27:41.8	 & 	-24:46:45	 & 	170	 & 	 II	 & 	0.255	 & 	462.1	 & 	1.56e-14	 & 	28.43  \\
16:27:41.9	 & 	-24:43:37	 & 	171	 & 	 II	 & 	0.222	 & 	497	 & 	1.35e-14	 & 	28.37  \\
16:27:43.7	 & 	-24:43:07	 & 	173	 & 	III?	 & 	0.231	 & 	441.8	 & 	1.41e-14	 & 	28.39  \\
16:27:45.9	 & 	-24:37:60	 & 	174	 & 	III?	 & 	0.215	 & 	482.4	 & 	1.31e-14	 & 	28.35  \\
16:27:46	 & 	-24:44:52	 & 	175	 & 	 II	 & 	0.236	 & 	469	 & 	1.44e-14	 & 	28.39  \\
16:27:46.2	 & 	-24:31:40	 & 	176	 & 	 II	 & 	0.5	 & 	122.3	 & 	3.05e-14	 & 	28.72  \\
16:27:50	 & 	-24:44:15	 & 	179	 & 	III?	 & 	0.263	 & 	418.4	 & 	1.6e-14	 & 	28.44  \\
16:27:50.3	 & 	-24:39:01	 & 	181	 & 	III?	 & 	0.542	 & 	247.7	 & 	3.31e-14	 & 	28.76  \\
16:27:57.8	 & 	-24:36:01	 & 	189	 & 	III?	 & 	0.287	 & 	350.3	 & 	1.75e-14	 & 	28.48  \\
16:28:05.5	 & 	-24:33:55	 & 	192	 & 	III?	 & 	0.367	 & 	266.5	 & 	2.24e-14	 & 	28.59  \\
16:28:16.8	 & 	-24:37:04	 & 	196	 & 	II	 & 	0.963	 & 	49.2	 & 	5.88e-14	 & 	29.01  \\
16:28:22.1	 & 	-24:42:49	 & 	197	 & 	 II	 & 	0.848	 & 	93.5	 & 	5.17e-14	 & 	28.95  \\

\end{longtable}
}
\longtab{4}{\small
\begin{landscape}
\begin{longtable}{l l l l l l l l l l l}
\caption{\label{spitz} Spitzer IRAC fluxes of IR counterparts to 
DROXO X-ray sources from the catalog by \citet{Evans03}. 
The $\alpha$ index is fitted from K to Spitzer MIPS1 band ($2.2 - 24 \mu$m). 
{  The classification given in the last column is that given in \citet{Evans03}.}} \\
\hline\hline
DROXO src. num. & C2D id & Name & $\alpha$     &  IRAC Ch1           & Ch2                 &  Ch3                & Ch4   & Mips 1 & Mips 2           & Classification \\
    &        &      &              & $3.6\ \mu\mathrm m$ & $4.5\ \mu\mathrm m$ & $5.8\ \mu\mathrm m$ & $8.0\ \mu\mathrm m$&                24 $\mu$m & 70 $\mu$m &  \\
    &        &      &              &  mJy                &         mJy         &         mJy         & mJy    & mJy & mJy        &       \\
\hline
\endfirsthead
\caption{continued.}\\
\hline\hline
DROXO src. num. & C2D id & Name & $\alpha$     &  IRAC Ch1           & Ch2                 &  Ch3                & Ch4   & Mips 1 & Mips 2           & Classification \\
    &        &      &              & $3.6\ \mu\mathrm m$ & $4.5\ \mu\mathrm m$ & $5.8\ \mu\mathrm m$ & $8.0\ \mu\mathrm m$&                24 $\mu$m & 70 $\mu$m &  \\
    &        &      &              &  mJy                &         mJy         &         mJy         & mJy    & mJy & mJy        &       \\
\hline
\endhead
\hline
\endfoot
    1 & J162619.5-243727 &                 &   -1.35 &        37.6   &        28.8   &        23.3   &        26.3   &        24.9   &            -- &  YSOc: star+dust(IR4) \\
   2 & J162621.9-244440 &             GY3 &  -0.990 &        28.8   &        26.6   &        23.8   &        26.4   &        33.6   &            -- &  YSOc: star+dust(IR1) \\
   3 & J162623.7-244314 &     DoAr25/GY17 &   -1.12 &        367.   &        292.   &        299.   &        258.   &        399.   &        1100.0 &  YSOc: star+dust(IR3) \\
   4 & J162627.5-244154 &            GY33 &  -0.510 &        72.2   &        59.0   &        49.0   &        50.9   &        403.   &        783.   &  YSOc: star+dust(IR4) \\
   6 & J162635.2-244236 &                 &   -2.02 &       0.123   &      0.0967   &      0.0794   &     -0.0121   &            -- &            -- &                  two \\
   6 & J162635.3-244239 &                 &   0.540 &       0.721   &        1.17   &        1.84   &        2.82   &        3.56   &            -- &                 Galc \\
   8 & J162644.2-243448 &      WL12/GY111 &    2.49 &        239.   &        744.   &        1610.0 &        2240.0 &            -- &        8120.0 &               rising \\
   9 & J162644.3-244314 &           GY112 &   -2.59 &        51.8   &        34.8   &        24.9   &        14.6   &        1.95   &            -- &                 star \\
  10 & J162644.4-244714 &                 &   -2.45 &        21.0   &        14.2   &        10.0   &        6.18   &        1.35   &            -- &                 star \\
  11 & J162645.1-245205 &                 &   0.310 &      0.0496   &      0.0676   &      0.0703   &      0.0676   &       0.263   &            -- &                  two \\
  13 & J162647.0-244430 &           GY122 &   -2.43 &        25.2   &        17.4   &        12.5   &        7.26   &       0.887   &            -- &                 star \\
  14 & J162648.1-244203 &                 &   -2.40 &        24.0   &        17.0   &        12.3   &        7.28   &       0.875   &            -- &                 star \\
  15 & J162648.5-242839 &       WL2/GY128 &  0.02   &        79.2   &        111.   &        128.   &        148.   &        446.   &        1100.0 &             YSOc: red \\
  17 & J162649.0-243825 &      WL18/GY129 &  -0.930 &        101.   &        135.   &        103.   &        86.7   &        122.   &            -- &                 YSOc \\
  18 & J162652.6-244332 &                 &   -2.54 &       0.256   &       0.178   &       0.123   &      0.0706   &       0.455   &            -- &                 star \\
  18 & J162652.9-244324 &                 &   -1.04 &      0.0511   &      0.0506   &      0.0109   &      0.0422   &            -- &            -- &                  two \\
  21 & J162656.5-244960 &                 &   0.4   &      0.0628   &       0.111   &       0.183   &       0.206   &       0.304   &            -- &             cup-down \\
  22 & J162657.3-243539 &      WL21/GY164 &  -0.660 &        7.70   &        12.3   &        11.9   &        10.7   &        12.1   &            -- &                 YSOc \\
  23 & J162657.5-244606 &                 &   -2.67 &        32.8   &        20.7   &        14.3   &        8.49   &            -- &            -- &                 star \\
  24 & J162657.6-244313 &                 &  -0.790 &      0.0505   &      0.0530   &      0.0716   &      0.0152   &            -- &            -- &                  two \\
  25 & J162658.5-244537 &            SR24 &  -0.920 &        990.   &       -30.6   &        351.   &        1040.0 &        1320.0 &            -- &       star+dust(IR4) \\
  25 & J162658.4-244532 &                 &  -0.450 &        1390.0 &        1440.0 &        1740.0 &        1950.0 &        2230.0 &        6380.0 &  YSOc: star+dust(IR1) \\
  27 & J162659.2-243459 &      WL22/GY174 &    2.36 &        143.   &        487.   &        564.   &        3230.0 &        4540.0 &            -- &               rising \\
  28 & J162659.0-243557 &      WL14/GY172 &   -1.01 &        14.2   &        13.7   &        11.6   &        10.4   &        20.0   &            -- &  YSOc: star+dust(IR4) \\
  28 & J162659.0-243557 &      WL14/GY172 &   -1.01 &        14.2   &        13.7   &        11.6   &        10.4   &        20.0   &            -- &  YSOc: star+dust(IR4) \\
  30 & J162704.5-244260 &           GY193 &   -2.46 &        48.8   &        35.1   &        25.3   &        15.3   &        1.79   &            -- &                 star \\
  31 & J162704.5-244214 &           GY194 &   -2.45 &        48.6   &        36.3   &        26.2   &        15.4   &        1.79   &            -- &                 star \\
  32 & J162705.2-244113 &                 &    1.07 &      0.0330   &      0.0946   &       0.160   &       0.177   &       0.279   &            -- &             cup-down \\
  33 & J162706.6-244149 &           GY204 &  -0.870 &        29.4   &        24.3   &        21.6   &        22.3   &        57.9   &            -- &  YSOc: star+dust(IR2) \\
  34 & J162706.8-243815 &      WL17/GY205 &   0.610 &        240.   &        416.   &        553.   &        695.   &        2790.0 &        6070.0 &             YSOc: red \\
  35 & J162709.1-243408 &      WL10/GY211 &  -0.910 &        259.   &        310.   &        272.   &        222.   &        339.   &        784.   &  YSOc: star+dust(IR2) \\
  36 & J162709.3-244320 &                 &   -2.13 &        21.8   &        17.1   &        13.1   &        7.79   &       0.858   &            -- &                 star \\
  37 & J162709.3-244022 &           GY213 & -0.09   &        44.1   &        53.7   &        65.1   &        92.9   &        202.   &            -- &  YSOc: star+dust(IR3) \\
  39 & J162711.2-244047 &           GY224 & -0.05   &        203.   &        305.   &        358.   &        367.   &        907.   &        908.   &                 YSOc \\
  40 & J162711.7-243832 &      WL19/GY227 &  -0.430 &        215.   &        354.   &        406.   &        328.   &        223.   &            -- &                 YSOc \\
  42 & J162713.8-244332 &           GY235 &  -0.220 &        56.2   &        60.7   &        70.8   &        116.   &        309.   &        577.   &  YSOc: star+dust(IR2) \\
  43 & J162715.1-245139 &                 &   -1.06 &        82.5   &        72.8   &        58.6   &        52.3   &        139.   &            -- &  YSOc: star+dust(IR2) \\
  43 & J162714.5-245133 &                 &   -2.53 &        26.4   &        17.3   &        11.9   &        7.24   &        1.28   &            -- &                 star \\
  44 & J162715.5-243054 &     IRS35/GY238 & 0.010   &        19.0   &        31.4   &        41.1   &        50.7   &        75.4   &            -- &                 YSOc \\
  46 & J162715.7-243843 &      WL20/GY240 &  -0.430 &        82.1   &        13.0   &        130.   &        59.2   &        993.   &            -- &                  red \\
  46 & J162715.9-243843 &      WL20/GY240 &  -0.7   &        127.   &        143.   &        140.   &        99.0   &        816.   &        15700. &       star+dust(IR2) \\
  46 & J162715.9-243843 &      WL20/GY240 &  -0.7   &        127.   &        143.   &        140.   &        99.0   &        816.   &        15700. &       star+dust(IR2) \\
  47 & J162716.4-243114 &                 &   0.160 &        19.7   &        52.7   &        66.5   &        50.7   &        129.   &            -- &                 YSOc \\
  49 & J162718.2-242853 &       WL5/GY246 &  -0.520 &        209.   &        297.   &        298.   &        163.   &            -- &            -- &             cup-down \\
  50 & J162718.4-245454 &                 &   -1.24 &       -999.   &        51.2   &       -999.   &        38.2   &        39.0   &            -- &       star+dust(IR2) \\
  51 & J162718.4-243915 &           GY245 &   0.180 &        43.8   &        72.0   &        91.3   &        107.   &        251.   &            -- &                 YSOc \\
  52 & J162718.5-242906 &       WL4/GY247 &  -0.590 &        173.   &        195.   &        232.   &        252.   &        269.   &            -- &  YSOc: star+dust(IR1) \\
  52 & J162719.2-242844 &       WL3/GY249 & -0.03   &        98.6   &        163.   &        208.   &        204.   &        325.   &            -- &                 YSOc \\
  53 & J162719.5-244140 &      SR12/GY250 &   -2.56 &        140.   &        94.7   &        68.3   &        39.9   &        5.24   &            -- &                 star \\
  54 & J162721.5-244143 &     IRS42/GY252 & -0.03   &        1060.0 &        1630.0 &        2100.0 &        2980.0 &        3450.0 &        2940.0 &                 YSOc \\
  55 & J162721.8-244336 &           GY253 &   -1.89 &        64.7   &        60.4   &        53.3   &        30.9   &        3.18   &            -- &                 star \\
  56 & J162721.8-242953 &       WL6/GY254 &   0.720 &        467.   &        925.   &        1440.0 &        1730.0 &        4360.0 &        5110.0 &                 YSOc \\
  57 & J162723.0-244807 &                 &   -1.04 &        101.   &        89.1   &        84.8   &        104.   &        91.7   &            -- &  YSOc: star+dust(IR2) \\
  59 & J162724.6-242935 &                 &   -1.45 &        13.0   &        12.4   &        10.9   &        6.42   &        4.26   &            -- &  YSOc: star+dust(MP1) \\
  60 & J162726.5-243923 &           GY262 &  -0.460 &        182.   &        237.   &        222.   &        210.   &        257.   &            -- &               rising \\
  61 & J162727.1-243217 &                 &   -1.54 &        11.6   &        10.7   &        9.18   &        5.37   &        1.04   &            -- &                 star \\
  62 & J162726.9-244051 &     IRS43/GY265 &    1.17 &        629.   &        1240.0 &        1790.0 &        2190.0 &            -- &        34400. &               rising \\
  63 & J162727.4-243117 &    VSSG25/GY267 &  -0.940 &        130.   &        115.   &        101.   &        97.6   &        196.   &            -- &  YSOc: star+dust(IR4) \\
  64 & J162728.0-243933 &     IRS44/GY269 &    2.29 &        731.   &        1830.0 &        2940.0 &        2320.0 &            -- &        34700. &               rising \\
  65 & J162728.4-242721 &     IRS45/GY273 & -0.03   &        187.   &        272.   &        382.   &        481.   &        712.   &        8790.0 &                 YSOc \\
  66 & J162728.7-245432 &                 &   -2.40 &       -999.   &        5.27   &       -999.   &        2.24   &       0.923   &            -- &                 star \\
  67 & J162729.4-243916 &     IRS46/GY274 &   0.180 &        172.   &        271.   &        402.   &        411.   &        639.   &            -- &               rising \\
  68 & J162730.2-242743 &     IRS47/GY279 &  -0.120 &        740.   &        1190.0 &        1580.0 &        2040.0 &        1720.0 &        7040.0 &                 YSOc \\
  69 & J162729.9-243336 &           GY278 &   -1.38 &        22.1   &        21.6   &        19.6   &        11.2   &       0.865   &            -- &                 YSOc \\
  70 & J162730.5-245211 &                 &   -2.41 &      0.0755   &      0.0541   &      0.0266   &       0.103   &      0.0906   &            -- &                  two \\
  70 & J162730.5-245208 &                 &  0.08   &       0.256   &       0.324   &       0.404   &       0.637   &        1.74   &            -- &                 Galc \\
  71 & J162730.8-244727 &  B162730-244726 &   -2.40 &        80.9   &        60.9   &        47.7   &        26.7   &        2.22   &            -- &                 star \\
  72 & J162731.0-243403 &           GY283 &   -2.34 &        35.6   &        26.8   &        19.8   &        11.7   &        1.51   &            -- &                 star \\
  73 & J162731.3-243110 &                 &   0.720 &      0.0933   &       0.163   &       0.187   &       0.290   &        1.18   &            -- &                  red \\
  74 & J162732.7-2445   &                 &   -1.48 &        6.26   &        6.05   &        4.98   &        3.10   &        1.97   &            -- &  YSOc: star+dust(MP1) \\
  75 & J162732.6-243323 &           GY289 &  -0.920 &        45.6   &        41.9   &        36.8   &        32.8   &        56.5   &            -- &  YSOc: star+dust(IR4) \\
  76 & J162732.8-243235 &           GY291 &  -0.460 &        54.2   &        69.5   &        84.8   &        108.   &        94.5   &            -- &  YSOc: star+dust(IR2) \\
  77 & J162733.1-244115 &           GY292 &  -0.880 &        571.   &        494.   &        473.   &        558.   &        721.   &            -- &  YSOc: star+dust(IR4) \\
  78 & J162735.3-243833 &           GY295 &   -2.47 &        54.8   &        39.3   &        27.9   &        17.0   &        1.69   &            -- &                 star \\
  79 & J162735.7-244532 &           GY296 &   -2.28 &        22.0   &        15.6   &        11.3   &        6.92   &       0.507   &            -- &                 star \\
  80 & J162737.2-244238 &           GY301 &   0.130 &        91.6   &        142.   &        162.   &        179.   &        561.   &        845.   &             YSOc: red \\
  81 & J162738.1-243043 &                 &   -1.89 &        72.0   &        53.8   &        43.3   &        13.5   &            -- &            -- &                 star \\
  82 & J162738.3-243658 &     IRS49/GY308 &  -0.730 &        375.   &        335.   &        331.   &        397.   &        688.   &        924.   &  YSOc: star+dust(IR3) \\
  83 & J162738.6-243839 &           GY310 &  -0.490 &        23.7   &        22.3   &        21.2   &        29.2   &        87.6   &            -- &  YSOc: star+dust(IR2) \\
  84 & J162738.9-244020 &           GY312 &   0.640 &        15.8   &        26.7   &        38.2   &        53.2   &        461.   &        950.   &  YSOc: star+dust(IR1) \\
  85 & J162739.0-244721 &                 &   0.4   &       0.170   &       0.265   &       0.430   &       0.602   &        1.63   &            -- &                 Galc \\
  86 & J162739.4-243915 &           GY314 &  -0.370 &        282.   &        284.   &        333.   &        427.   &        1140.0 &        1310.0 &  YSOc: star+dust(IR2) \\
  87 & J162739.8-244315 &     IRS51/GY315 &  -0.150 &        752.   &        916.   &        1000.0 &        1070.0 &        2730.0 &        3260.0 &                 YSOc \\
  89 & J162741.5-243538 &           GY322 &   -1.98 &        38.5   &        33.2   &        24.7   &        14.7   &        5.77   &            -- &  YSOc: star+dust(MP1) \\
  90 & J162742.7-243851 &           GY326 &   -1.28 &        31.4   &        25.4   &        21.4   &        20.1   &        24.5   &            -- &  YSOc: star+dust(IR4) \\
  91 & J162743.6-243127 &                 &    1.72 &      0.0248   &      0.0560   &       0.123   &       0.175   &            -- &            -- &             cup-down \\
  92 & J162744.4-244856 &                 &   -1.33 &      0.0636   &      0.0589   &      0.0455   &      0.0251   &       0.136   &            -- &                  two \\
  93 & J162747.1-244535 &           GY352 &  -0.720 &        41.1   &        47.8   &        52.5   &        49.9   &        57.1   &            -- &  YSOc: star+dust(IR2) \\
  94 & J162749.1-243906 &                 &   0.510 &       0.187   &       0.357   &       0.614   &       0.908   &        1.70   &            -- &                 Galc \\
  97 & J162751.8-243145 &     IRS54/GY378 &  0.03   &        525.   &        712.   &        931.   &        1010.0 &        3560.0 &        6500.0 &             YSOc: red \\
  98 & J162751.9-244630 &           GY377 &   -2.26 &        40.3   &        31.6   &        25.4   &        14.6   &        1.44   &            -- &                 star \\
  99 & J162752.1-244050 &     IRS55/GY380 &   -2.61 &        203.   &        145.   &        106.   &        62.7   &        7.40   &            -- &                 star \\
 1   & J162755.6-244451 &           GY398 &   -2.36 &        34.2   &        26.0   &        18.3   &        10.7   &        1.28   &            -- &                 star \\
 101 & J162757.8-244002 &           GY410 &   -2.42 &        54.6   &        40.3   &        30.0   &        17.9   &        1.69   &            -- &                 star \\
 102 & J162759.9-244819 &                 &   -2.56 &        72.8   &        52.3   &        36.7   &        22.1   &        2.16   &            -- &                 star \\
 103 & J162760.0-244239 &                 &   0.660 &      0.0625   &       0.102   &       0.159   &       0.209   &       0.415   &            -- &             cup-down \\
 104 & J162802.0-244953 &                 &   -1.14 &       0.103   &       0.101   &      0.0904   &       0.105   &       0.441   &            -- &                 star \\
 104 & J162802.1-244956 &                 &   -2.60 &       0.113   &      0.0776   &      0.0361   &      0.0274   &       0.356   &            -- &                  two \\
 105 & J162804.6-243456 &           GY463 &   -1.85 &        52.7   &        48.2   &        41.0   &        24.7   &        3.     &            -- &                 star \\
 106 & J162804.8-243710 &                 &   -2.13 &        43.2   &        33.4   &        25.6   &        14.3   &        4.72   &            -- &  YSOc: star+dust(MP1) \\
 107 & J162804.9-245014 &                 &   -2.73 &      0.0902   &      0.06     &     0.00124   &      0.0493   &       0.524   &            -- &                  two \\
 107 & J162805.3-245019 &                 &   -2.34 &      0.0542   &      0.0395   &     0.00871   &    0.000535   &       0.303   &            -- &                  two \\
 108 & J162806.8-244828 &                 &    0.   &       0.405   &       0.446   &       0.856   &       0.989   &        1.93   &            -- &                 Galc \\
 109 & J162808.1-244512 &                 &   -2.22 &        6.38   &        4.79   &        3.41   &        2.18   &       0.303   &            -- &                 star \\
 111 & J162810.2-243928 &                 &   -2.04 &       0.267   &       0.209   &      0.0903   &      0.0890   &            -- &            -- &                  two \\

\end{longtable}
\end{landscape}

}

\longtabL{5}{
\begin{longtable}{llllll}
\caption{\label{photprop} Effective temperatures and bolometric luminosities of ISOCAM objects in DROXO (see Sect. 
\ref{photos}).}\\
\hline\hline
 ISO & DROXO Src. & Class & $\log$ T$_\mathrm{eff}$ (K) &  $\log$ L/L$_\mathrm{bol}$ & $\log$ M/M$\odot$ \\
\hline
\endfirsthead
 \caption{continued.}\\
\hline\hline
 ISO & DROXO Src. & Class & $\log$ T$_\mathrm{eff}$ (K) &  $\log$ L/L$_\mathrm{bol}$ & $\log$ M/M$\odot$ \\
\hline
\endhead
\hline
\endfoot
1	 & 	 	 & 	II	 & 	3.66	 & 	0.3	 & 	-0.18  \\
2	 & 	 	 & 	II	 & 	3.59	 & 	-0.05	 & 	-0.41  \\
3	 & 	 	 & 	II	 & 	3.61	 & 	0.03	 & 	-0.35  \\
5	 & 	 	 & 	III	 & 	3.65	 & 	0.26	 & 	-0.2  \\
6	 & 	 	 & 	II	 & 	3.64	 & 	0.21	 & 	-0.23  \\
9	 & 	 	 & 	II	 & 	3.47	 & 	-0.92	 & 	-0.91  \\
11	 & 	 	 & 	III	 & 	3.57	 & 	-0.13	 & 	-0.49  \\
12	 & 	 	 & 	II	 & 	3.48	 & 	-0.86	 & 	-0.88  \\
13	 & 	 	 & 	II	 & 	3.58	 & 	-0.1	 & 	-0.45  \\
14	 & 	 	 & 	III	 & 	3.63	 & 	0.15	 & 	-0.26  \\
16	 & 	 	 & 	III	 & 	3.68	 & 	0.61	 & 	0.04  \\
17	 & 	 	 & 	II	 & 	3.68	 & 	0.65	 & 	0.07  \\
18	 & 	 	 & 	III	 & 	3.64	 & 	0.2	 & 	-0.23  \\
19	 & 	 	 & 	II	 & 	3.68	 & 	0.67	 & 	0.08  \\
20	 & 	 	 & 	II	 & 	3.63	 & 	0.13	 & 	-0.28  \\
23	 & 	 	 & 	II	 & 	3.43	 & 	-1.6	 & 	-1.32  \\
24	 & 	 	 & 	II	 & 	3.67	 & 	0.39	 & 	-0.12  \\
26	 & 	 	 & 	II	 & 	3.54	 & 	-0.31	 & 	-0.62  \\
28	 & 	 	 & 	III	 & 	3.67	 & 	0.48	 & 	-0.06  \\
30	 & 	 	 & 	II	 & 	3.44	 & 	-1.2	 & 	-1.07  \\
32	 & 	2	 & 	II	 & 	3.43	 & 	-1.47	 & 	-1.24  \\
33	 & 	 	 & 	II	 & 	3.38	 & 	-2.93	 & 	-2.26  \\
35	 & 	 	 & 	II	 & 	3.46	 & 	-0.96	 & 	-0.93  \\
36	 & 	 	 & 	II	 & 	3.69	 & 	0.81	 & 	0.17  \\
37	 & 	 	 & 	II	 & 	3.51	 & 	-0.6	 & 	-0.77  \\
38	 & 	3	 & 	II	 & 	3.63	 & 	0.15	 & 	-0.27  \\
39	 & 	 	 & 	II	 & 	3.76	 & 	1.51	 & 	0.47  \\
40	 & 	 	 & 	II	 & 	3.68	 & 	0.65	 & 	0.06  \\
41	 & 	 	 & 	II	 & 	3.53	 & 	-0.43	 & 	-0.69  \\
43	 & 	4	 & 	II	 & 	3.63	 & 	0.16	 & 	-0.26  \\
44	 & 	 	 & 	III	 & 	3.6	 & 	0	 & 	-0.36  \\
46	 & 	 	 & 	II	 & 	3.61	 & 	0.04	 & 	-0.34  \\
47	 & 	 	 & 	III	 & 	3.55	 & 	-0.23	 & 	-0.57  \\
51	 & 	 	 & 	II	 & 	3.52	 & 	-0.49	 & 	-0.72  \\
52	 & 	 	 & 	II	 & 	3.51	 & 	-0.6	 & 	-0.77  \\
53	 & 	 	 & 	II	 & 	3.51	 & 	-0.62	 & 	-0.78  \\
55	 & 	 	 & 	III	 & 	3.43	 & 	-1.57	 & 	-1.3  \\
56	 & 	 	 & 	II	 & 	3.53	 & 	-0.39	 & 	-0.66  \\
58	 & 	 	 & 	III	 & 	3.68	 & 	0.7	 & 	0.1  \\
59	 & 	 	 & 	II	 & 	3.55	 & 	-0.23	 & 	-0.57  \\
59	 & 	 	 & 	II	 & 	3.57	 & 	-0.13	 & 	-0.48  \\
62	 & 	 	 & 	II	 & 	3.68	 & 	0.66	 & 	0.07  \\
63	 & 	 	 & 	II	 & 	3.51	 & 	-0.55	 & 	-0.75  \\
64	 & 	 	 & 	III	 & 	3.64	 & 	0.19	 & 	-0.24  \\
66	 & 	9	 & 	III	 & 	3.52	 & 	-0.48	 & 	-0.72  \\
67	 & 	 	 & 	II	 & 	3.66	 & 	0.34	 & 	-0.15  \\
68	 & 	 	 & 	II	 & 	3.67	 & 	0.43	 & 	-0.09  \\
69	 & 	13	 & 	III	 & 	3.49	 & 	-0.75	 & 	-0.83  \\
70	 & 	15	 & 	II	 & 	3.38	 & 	-3.01	 & 	-2.41  \\
72	 & 	17	 & 	II	 & 	3.5	 & 	-0.67	 & 	-0.8  \\
73	 & 	 	 & 	III	 & 	3.68	 & 	0.68	 & 	0.08  \\
74	 & 	 	 & 	III	 & 	3.58	 & 	-0.09	 & 	-0.45  \\
75	 & 	 	 & 	II	 & 	3.49	 & 	-0.71	 & 	-0.82  \\
76	 & 	 	 & 	II	 & 	3.52	 & 	-0.5	 & 	-0.72  \\
78	 & 	 	 & 	II	 & 	3.66	 & 	0.33	 & 	-0.16  \\
79	 & 	 	 & 	II	 & 	3.44	 & 	-1.27	 & 	-1.11  \\
82	 & 	 	 & 	III	 & 	3.43	 & 	-1.73	 & 	-1.4  \\
83	 & 	 	 & 	II	 & 	3.62	 & 	0.11	 & 	-0.29  \\
84	 & 	22	 & 	II	 & 	3.46	 & 	-1.02	 & 	-0.96  \\
84	 & 	 	 & 	II	 & 	3.46	 & 	-1.01	 & 	-0.96  \\
86	 & 	 	 & 	II	 & 	3.53	 & 	-0.42	 & 	-0.68  \\
87	 & 	 	 & 	II	 & 	3.47	 & 	-0.9	 & 	-0.9  \\
88	 & 	25	 & 	II	 & 	3.67	 & 	0.5	 & 	-0.04  \\
88	 & 	 	 & 	II	 & 	3.67	 & 	0.53	 & 	-0.02  \\
89	 & 	28	 & 	II	 & 	3.5	 & 	-0.68	 & 	-0.8  \\
89	 & 	 	 & 	II	 & 	3.5	 & 	-0.66	 & 	-0.8  \\
91	 & 	 	 & 	III	 & 	3.69	 & 	0.75	 & 	0.13  \\
92	 & 	 	 & 	II	 & 	4.04	 & 	2.26	 & 	0.56  \\
93	 & 	 	 & 	II	 & 	3.54	 & 	-0.26	 & 	-0.59  \\
93	 & 	 	 & 	II	 & 	3.54	 & 	-0.26	 & 	-0.59  \\
94	 & 	 	 & 	II	 & 	3.43	 & 	-1.69	 & 	-1.38  \\
95	 & 	 	 & 	II	 & 	3.65	 & 	0.25	 & 	-0.21  \\
96	 & 	30	 & 	III	 & 	3.54	 & 	-0.29	 & 	-0.61  \\
97	 & 	31	 & 	III	 & 	3.57	 & 	-0.13	 & 	-0.48  \\
98	 & 	 	 & 	II	 & 	3.62	 & 	0.09	 & 	-0.31  \\
100	 & 	 	 & 	III	 & 	3.43	 & 	-1.5	 & 	-1.26  \\
102	 & 	33	 & 	II	 & 	3.44	 & 	-1.2	 & 	-1.07  \\
103	 & 	34	 & 	II	 & 	3.68	 & 	0.69	 & 	0.09  \\
104	 & 	 	 & 	III	 & 	3.45	 & 	-1.11	 & 	-1.01  \\
105	 & 	35	 & 	II	 & 	3.67	 & 	0.51	 & 	-0.04  \\
106	 & 	 	 & 	II	 & 	3.54	 & 	-0.31	 & 	-0.62  \\
107	 & 	37	 & 	II	 & 	3.54	 & 	-0.32	 & 	-0.63  \\
109	 & 	 	 & 	III	 & 	3.57	 & 	-0.12	 & 	-0.48  \\
110	 & 	 	 & 	II	 & 	3.68	 & 	0.62	 & 	0.04  \\
111	 & 	 	 & 	III	 & 	3.52	 & 	-0.44	 & 	-0.69  \\
112	 & 	39	 & 	II	 & 	3.67	 & 	0.54	 & 	-0.01  \\
113	 & 	 	 & 	III	 & 	3.59	 & 	-0.05	 & 	-0.41  \\
113	 & 	 	 & 	III	 & 	3.59	 & 	-0.05	 & 	-0.41  \\
114	 & 	40	 & 	III	 & 	3.9	 & 	1.88	 & 	0.51  \\
115	 & 	 	 & 	II	 & 	3.47	 & 	-0.94	 & 	-0.92  \\
116	 & 	 	 & 	II	 & 	3.65	 & 	0.26	 & 	-0.2  \\
117	 & 	42	 & 	II	 & 	3.54	 & 	-0.3	 & 	-0.62  \\
118	 & 	 	 & 	II	 & 	3.57	 & 	-0.12	 & 	-0.48  \\
119	 & 	44	 & 	II	 & 	3.57	 & 	-0.13	 & 	-0.48  \\
120	 & 	 	 & 	II	 & 	3.56	 & 	-0.19	 & 	-0.54  \\
121	 & 	46	 & 	II	 & 	3.6	 & 	-0.01	 & 	-0.38  \\
121	 & 	 	 & 	II	 & 	3.67	 & 	0.49	 & 	-0.05  \\
122	 & 	 	 & 	II	 & 	3.46	 & 	-0.94	 & 	-0.92  \\
123	 & 	 	 & 	II	 & 	3.45	 & 	-1.09	 & 	-1  \\
124	 & 	 	 & 	II	 & 	3.61	 & 	0.01	 & 	-0.36  \\
125	 & 	49	 & 	III	 & 	4.02	 & 	2.22	 & 	0.55  \\
126	 & 	 	 & 	III	 & 	3.61	 & 	0.03	 & 	-0.35  \\
127	 & 	51	 & 	II	 & 	3.62	 & 	0.07	 & 	-0.32  \\
128	 & 	52	 & 	II	 & 	3.67	 & 	0.4	 & 	-0.11  \\
129	 & 	 	 & 	II	 & 	3.66	 & 	0.32	 & 	-0.17  \\
130	 & 	53	 & 	III	 & 	3.62	 & 	0.08	 & 	-0.31  \\
132	 & 	54	 & 	II	 & 	3.68	 & 	0.69	 & 	0.09  \\
133	 & 	55	 & 	III	 & 	3.66	 & 	0.36	 & 	-0.14  \\
135	 & 	 	 & 	III	 & 	3.67	 & 	0.44	 & 	-0.09  \\
136	 & 	 	 & 	III	 & 	3.46	 & 	-1.02	 & 	-0.96  \\
138	 & 	 	 & 	II	 & 	3.43	 & 	-1.59	 & 	-1.32  \\
139	 & 	 	 & 	II	 & 	3.49	 & 	-0.76	 & 	-0.84  \\
139	 & 	 	 & 	II	 & 	3.49	 & 	-0.75	 & 	-0.84  \\
140	 & 	60	 & 	II	 & 	3.67	 & 	0.55	 & 	-0.01  \\
142	 & 	63	 & 	II	 & 	3.62	 & 	0.11	 & 	-0.29  \\
144	 & 	65	 & 	II	 & 	3.62	 & 	0.07	 & 	-0.32  \\
146	 & 	69	 & 	III	 & 	3.56	 & 	-0.16	 & 	-0.52  \\
147	 & 	68	 & 	II	 & 	3.68	 & 	0.63	 & 	0.05  \\
148	 & 	72	 & 	III	 & 	3.56	 & 	-0.19	 & 	-0.54  \\
149	 & 	71	 & 	III	 & 	3.61	 & 	0.04	 & 	-0.34  \\
151	 & 	 	 & 	II	 & 	3.54	 & 	-0.26	 & 	-0.59  \\
152	 & 	75	 & 	III	 & 	3.64	 & 	0.19	 & 	-0.24  \\
153	 & 	79	 & 	III	 & 	3.44	 & 	-1.3	 & 	-1.13  \\
154	 & 	76	 & 	II	 & 	3.64	 & 	0.21	 & 	-0.23  \\
155	 & 	77	 & 	II	 & 	3.68	 & 	0.69	 & 	0.09  \\
156	 & 	78	 & 	III	 & 	3.51	 & 	-0.58	 & 	-0.76  \\
157	 & 	79	 & 	III	 & 	3.47	 & 	-0.89	 & 	-0.89  \\
158	 & 	 	 & 	III	 & 	3.43	 & 	-1.35	 & 	-1.17  \\
160	 & 	 	 & 	II	 & 	3.43	 & 	-1.49	 & 	-1.25  \\
161	 & 	80	 & 	II	 & 	3.65	 & 	0.24	 & 	-0.21  \\
163	 & 	82	 & 	II	 & 	3.67	 & 	0.39	 & 	-0.12  \\
164	 & 	83	 & 	II	 & 	3.44	 & 	-1.24	 & 	-1.1  \\
165	 & 	84	 & 	II	 & 	3.45	 & 	-1.12	 & 	-1.02  \\
166	 & 	86	 & 	II	 & 	3.66	 & 	0.35	 & 	-0.15  \\
168	 & 	 	 & 	II	 & 	3.65	 & 	0.25	 & 	-0.2  \\
169	 & 	89	 & 	III	 & 	3.58	 & 	-0.1	 & 	-0.46  \\
169	 & 	 	 & 	III	 & 	3.51	 & 	-0.58	 & 	-0.76  \\
170	 & 	 	 & 	II	 & 	3.38	 & 	-3.01	 & 	-2.41  \\
171	 & 	 	 & 	II	 & 	3.52	 & 	-0.52	 & 	-0.74  \\
172	 & 	90	 & 	II	 & 	3.53	 & 	-0.42	 & 	-0.68  \\
173	 & 	 	 & 	III	 & 	3.61	 & 	0.04	 & 	-0.34  \\
174	 & 	 	 & 	III	 & 	3.49	 & 	-0.76	 & 	-0.84  \\
175	 & 	 	 & 	II	 & 	3.43	 & 	-1.61	 & 	-1.33  \\
176	 & 	 	 & 	II	 & 	3.46	 & 	-1.01	 & 	-0.95  \\
176	 & 	 	 & 	II	 & 	3.46	 & 	-0.99	 & 	-0.94  \\
177	 & 	93	 & 	II	 & 	3.54	 & 	-0.26	 & 	-0.59  \\
178	 & 	 	 & 	II	 & 	3.51	 & 	-0.55	 & 	-0.75  \\
179	 & 	 	 & 	III	 & 	3.5	 & 	-0.68	 & 	-0.81  \\
180	 & 	 	 & 	III	 & 	3.67	 & 	0.44	 & 	-0.09  \\
181	 & 	 	 & 	III	 & 	3.45	 & 	-1.08	 & 	-0.99  \\
183	 & 	98	 & 	III	 & 	3.6	 & 	-0.03	 & 	-0.39  \\
184	 & 	99	 & 	III	 & 	3.66	 & 	0.36	 & 	-0.14  \\
185	 & 	 	 & 	II	 & 	3.47	 & 	-0.88	 & 	-0.89  \\
186	 & 	100	 & 	III	 & 	3.47	 & 	-0.92	 & 	-0.91  \\
187	 & 	 	 & 	II	 & 	3.53	 & 	-0.39	 & 	-0.66  \\
188	 & 	101	 & 	III	 & 	3.6	 & 	-0.02	 & 	-0.38  \\
189	 & 	 	 & 	III	 & 	3.5	 & 	-0.67	 & 	-0.8  \\
190	 & 	 	 & 	II	 & 	3.43	 & 	-1.39	 & 	-1.19  \\
191	 & 	105	 & 	III	 & 	3.63	 & 	0.12	 & 	-0.29  \\
192	 & 	 	 & 	III	 & 	3.48	 & 	-0.85	 & 	-0.88  \\
193	 & 	 	 & 	II	 & 	3.46	 & 	-1.01	 & 	-0.95  \\
194	 & 	 	 & 	II	 & 	3.51	 & 	-0.54	 & 	-0.75  \\
195	 & 	 	 & 	II	 & 	3.61	 & 	0.05	 & 	-0.33  \\
196	 & 	 	 & 	II	 & 	3.52	 & 	-0.48	 & 	-0.72  \\
197	 & 	 	 & 	III	 & 	3.47	 & 	-0.9	 & 	-0.9  \\
198	 & 	 	 & 	III	 & 	3.69	 & 	0.83	 & 	0.19  \\
199	 & 	 	 & 	II	 & 	3.61	 & 	0.03	 & 	-0.35  \\

\end{longtable}
}

\clearpage

\section{\label{atlas} Atlas of DROXO sources.}
\onecolumn We show here an example of the atlas that we have produced for each DROXO source. 
It is reported the number of the source, the literature name as in \citet{Bontemps01}, 
the EPIC image, the J-band 2MASS image, the EPIC spectrum, and the light curve.  
The atlas is available only online at the following web address: 
{\sc http:www.astropa.unipa.it/~pilli/atlas\_droxo\_sources.pdf.}
\begin{figure}[!h]
\includegraphics[width=0.9\textwidth]{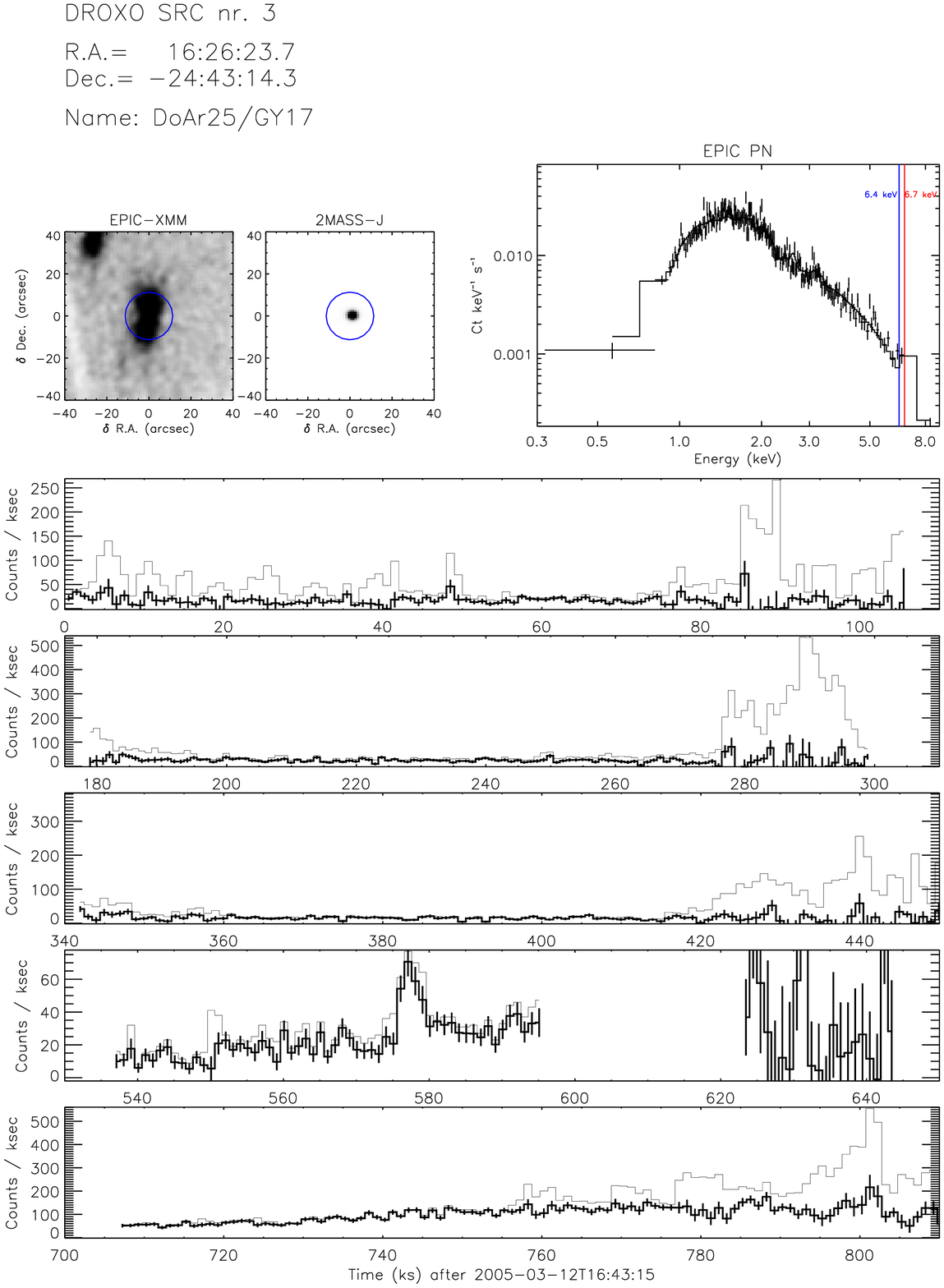}
\end{figure}

\end{appendix}
\end{document}